\documentclass[12pt]{article}
\usepackage{amsmath}
\usepackage{graphicx,psfrag,epsf}
\usepackage{enumerate}
\usepackage{natbib}
\usepackage{url} 
\usepackage[T1]{fontenc} 
\usepackage[utf8]{inputenc} 
\usepackage{mathtools} 
\usepackage[lofdepth,lotdepth]{subfig} 
\usepackage{mwe} 
\usepackage{amssymb}
\usepackage{hyperref}
\usepackage{xcolor}
\usepackage{adjustbox}
\usepackage{multirow}
\usepackage{pdflscape}
\usepackage[mathscr]{euscript}
\usepackage{etex}
\usepackage{amsthm}
\usepackage{tabulary}
\usepackage{varwidth}
\usepackage{array}
\usepackage{caption} 
\usepackage{tikz}
\usepackage{pgfplots}  
\pgfplotsset{compat=1.16} 
\usetikzlibrary{arrows}
\usetikzlibrary{patterns} 
\usetikzlibrary{shapes.arrows}
\usetikzlibrary{decorations.pathreplacing}
\usetikzlibrary{positioning}

\newcommand{\Xv}{\mathbf{X}}
\newcommand{\xv}{\mathbf{x}}
\newcommand{\thetav}{\boldsymbol{\theta}}
\newcommand{\sigmav}{\boldsymbol{\sigma}}
\newcommand{\Thetav}{\boldsymbol{\Theta}}

\newcommand{\E}[1]{{\color{black} #1}}
\definecolor{ao(english)}{rgb}{0.0, 0.5, 0.0}

\newlength{\starsize}
\newlength{\starspread}
\tikzset{starsize/.code={\setlength{\starsize}{#1}},
	starspread/.code={\setlength{\starspread}{#1}}}
\tikzset{starsize=1mm,
	starspread=3mm}
\pgfdeclarepatternformonly[\starspread,\starsize]
{custom fivepointed stars}
{\pgfpointorigin}
{\pgfqpoint{\starspread}{\starspread}}
{\pgfqpoint{\starspread}{\starspread}}
{
	\pgftransformshift{\pgfqpoint{\starsize}{\starsize}}
	\pgfpathmoveto{\pgfqpointpolar{18}{\starsize}}
	\pgfpathlineto{\pgfqpointpolar{162}{\starsize}}
	\pgfpathlineto{\pgfqpointpolar{306}{\starsize}}
	\pgfpathlineto{\pgfqpointpolar{90}{\starsize}}
	\pgfpathlineto{\pgfqpointpolar{234}{\starsize}}
	\pgfpathclose%
	\pgfusepath{fill}
}
\pgfdeclarepatternformonly{mydots1}{\pgfqpoint{-1pt}{-1pt}}{\pgfqpoint{1pt}{1pt}}{\pgfqpoint{2.4pt}{2.4pt}}
{%
	\pgfpathcircle{\pgfqpoint{0pt}{0pt}}{.5pt}%
	\pgfusepath{fill}%
}%

\pgfdeclarepatternformonly{mydots2}{\pgfqpoint{-1pt}{-1pt}}{\pgfqpoint{1pt}{1pt}}{\pgfqpoint{1.7pt}{1.7pt}}
{%
	\pgfpathcircle{\pgfqpoint{0pt}{0pt}}{.5pt}%
	\pgfusepath{fill}%
}%

\pgfdeclarepatternformonly{mydots3}{\pgfqpoint{-1pt}{-1pt}}{\pgfqpoint{1pt}{1pt}}{\pgfqpoint{1.2pt}{1.2pt}}
{%
	\pgfpathcircle{\pgfqpoint{0pt}{0pt}}{.5pt}%
	\pgfusepath{fill}%
}%
\newcommand{\blind}{0}

\begin{document}

\def\spacingset#1{\renewcommand{\baselinestretch}%
{#1}\small\normalsize} \spacingset{1}


\if0\blind
{
  \title{\bf Impact of the error structure on the design and analysis of enzyme kinetic models}
  \author{Elham Yousefi\thanks{
    The authors gratefully acknowledge \textit{support by project grants LIT-2017-4-SEE-001 funded by the Upper Austrian Government and Austrian Science Fund (FWF): I 3903-N32.}}\hspace{.2cm}\\
    Department of Applied Statistics, Johannes Kepler University Linz\\
    and \\
    Werner G. M{\"u}ller* \\
    Department of Applied Statistics, Johannes Kepler University Linz}
  \maketitle
} \fi

\if1\blind
{
  \bigskip
  \bigskip
  \bigskip
  \begin{center}
    {\LARGE\bf Title}
\end{center}
  \medskip
} \fi

\bigskip
\begin{abstract}
The statistical analysis of enzyme kinetic reactions usually involves models of the response functions which are well defined on the basis of Michaelis-Menten type equations. The error structure however is often without good reason assumed as additive Gaussian noise. This simple assumption may lead to undesired properties of the analysis, particularly when simulations are involved and consequently negative simulated reaction rates may occur. 
In this study we investigate the effect of assuming multiplicative lognormal errors instead. While there is typically little impact on the estimates, the experimental designs and their efficiencies are decisively affected, particularly when it comes to model discrimination problems.
\end{abstract}

\noindent%
\vfill

\newpage
\spacingset{1.45} 
\section{Introduction}
\label{sec-intro}

The experimental study of enzyme catalyzed reactions can help to provide valuable information for researches of a great range of specializations. Biotechnologists for instance study the principles of enzymology such as structure, kinetics, inhibition and classification quantified by rate and selectivity for determining steps that can result in increased product yield or suitable feeding strategies in fed-batch processes. \E{Studying the reversible interaction of drugs binding to their target enzyme is of high importance in pharmaceutical research. Also in drug discovery, (visual) inspection of concentration-response plots is important to diagnose non-ideal behavior and determination of $IC_{50}$ (see later parts of this section) or other similar quantitative measures.} Appropriate mechanistic and/or kinetic models, which itself might require challenging strategies to set and select, are instrumental in fulfilling those important goals.

Even though mechanistic models resulted from both generally accepted theories and empirical researches helped to understand many processes and making inferences in fields like biological sciences, chemical engineering, drug developments, etc., experimental data are still required to validate the proposed models. Collecting these experimental data requires 
{experimental effort reflected in} time, allocation of expenses, manpower and other costly factors. Optimal experimentation on the other hand can help reduce these expenses by providing high informative data according to the purpose of the experiment. Further, if the theory suggests more than one model, again optimal experimental design plays an important role to provide informative data for discrimination and/or model selection. 

Enzymes are organic catalysts that significantly speed up the rate of chemical and biochemical reactions that take place within cells. The molecules that an enzyme works with are called substrates. Products are the result of typical binding of substrates and enzymes on the active site of the enzymes. The standard two parameter Michaelis-Menten model is used to describe this type of reaction
\begin{equation} \label{formula-Michaelis}
	E[y]=\dfrac{\theta_{V}x_{S}}{\theta_{M} + x_{S} } ~~~~ x_{S}\in[a,b] , ~ a\geq 0,
\end{equation}
in which $E[y]$ is used to denote the expected reaction rate, when no inhibition is present. The design, controllable or independent variable, $x_{S}$, represents the substrate concentration supposed to be non-negative. Since there always exists at least an initial substrate concentration as the minimum value to start the enzymatic reactions; e.g. $x_{S}\geq 0$. The parameter $\theta_{V}$ is the maximum velocity the system could reach which should also be non-negatively varying according to physical definition of velocity and $\theta_{M}$ is the Michaelis-Menten constant, the value of $x_{S}$ at which the maximum velocity is half (\citealp{michaelis1913kinetics}). 
Note that according to these biochemical definitions for the parameters, the expected reaction rate of the system should be more than or equal to zero.

A number of substances known as inhibitors may cause a reduction in the rate of an enzyme catalyzed reaction. In such kinetic profiles more than one factor is controlled and the Michaelis-Menten model is extended to include the second controllable variable $x_{I}$, i.e. the inhibition concentration which is taken to be more than or equal to zero in a controlled experiment. Two of commonly used reaction rate equations of enzyme kinetics at the presence of inhibitors are competitive and non-competitive inhibition models \E{which are widely used in drug discovery (\citealp{copeland2005evaluation})} and have already been investigated by many authors \E{in optimal design} 
(\citealp{bogacka2011optimum,atkinson2012optimum,harman2020design})
.

\textbf{Competitive inhibition:} In this type of enzyme catalyzed reaction, \E{the inhibitor compete with the substrate for the pool of free enzyme molecules. Hence} binding of an inhibitor to the active site of an enzyme prevents the substrate binding and therefore no product is produced. In this case, the statistical model \E{which describes this influence on the reaction rate is}
\begin{equation} \label{formula-comp-model}
	y=\eta_{C}+\epsilon=\dfrac{\theta_{V}x_{S}}{\theta_{M}\left( 1+\frac{x_{I}}{\theta_{K}}\right) + x_{S} } + \epsilon,\\ 
\end{equation}
where \E{$\eta_{C}$ denotes the expected reaction rate for the competitive inhibition model, being used in later parts.} $\theta_{K}\geq 0$ denotes the inhibition constant, an indication of how potent an inhibitor is; it is the concentration required to produce half maximum inhibition. 
The independent random errors are normally distributed $\epsilon \sim \mathcal{N}(0,\sigma^2)$. The term statistical model is used instead of the model itself, since in practical studies, real observations are exposed to uncontrolled factors like random errors and therefore they are included in statistical models here.

\textbf{Non-competitive inhibition:}  This type of inhibition, models a system where the inhibitor and the substrate are both bound to the enzyme and form a complex in such a way that the enzyme is inactivated to form a product. The statistical model for the reaction rate $y$ in \E{the case where the inhibitor displays equal affinity for both the free enzyme and the enzyme-substrate complex}, is defined as

\begin{equation} \label{formula-noncomp-model} 
	y=\eta_{N}+\epsilon=\dfrac{\theta_{V}x_{S}}{\left(\theta_{M}+x_{S} \right) \left( 1+\frac{x_{I}}{\theta_{K}}\right) } + \epsilon,
\end{equation}
\E{where $\eta_{N}$ is similarly the representation for the expected reaction rate of the non-competitive inhibition model.} 

\textbf{Encompassing model:} \cite{atkinson2011some} suggested to combine the competitive and non-competitive inhibition models each having 3 parameters, to form a 4 parameter encompassing model. This model is similarly represented as

\begin{equation} \label{formula-combined-model}
	y=\eta_{E}+\epsilon=\dfrac{\theta_{V}x_{S}}{\theta_{M}\left( 1+\frac{x_{I}}{\theta_{K}}\right) + x_{S}\left( 1+\frac{(1-\lambda)x_{I}}{\theta_{K}}\right) } + \epsilon,
\end{equation}
\E{where $\eta_{E}$ represents the expected reaction rate of the encompassing model as before. Also $0\leq\lambda\leq 1$ is a non-negative parameter, where $\lambda=1$ corresponds to competitive model (\ref{formula-comp-model}) and $\lambda=0$ to non-competitive model (\ref{formula-noncomp-model}).}

\E{\textbf{Fractional activity and $IC_{50}$ determination:} In drug discovery terminology, at any concentration of inhibitor, the total concentration of enzyme in the sample is, by mass-balance, equal to the sum of the concentration of free enzyme molecules and the concentration of enzyme-inhibitor complex and therefore the fractional activity, the expected reaction rate of the free enzyme over the total enzyme concentration can be defined as $E_i(y)/E_0(y)$ (\citealp{copeland2005evaluation}). The fraction of enzyme occupied by the inhibitor, can also be shown by  $1-(E_i(y)/E_0(y))$ again by mass-balance and the $\%$ inhibition is accordingly equal to $100(1-(E_i(y)/E_0(y)))$. Therefore both plots of fractional velocity remaining as a function of inhibitor concentrations and the same behavior on a semilog plot (same plot on a different scaling and log transformation of data) will be decreasing functions of inhibitor concentrations. Finally the fractional velocity of $0.5$, corresponding to $50\%$ inhibition of the target enzyme which is basically referred to as inhibitor concentration at fractional activity of $0.5$ determines the $IC_{50}$ value. These calculations for the encompassing model (by comparing the expected reaction rates of the encompassing model and that of the Michaelis-Menten model at $x_{S}=\theta_{M}$ and using the definition of $IC_{50}$) results in 
	\begin{equation} \label{IC50}
		IC_{50}=2\theta_{K}/(2-\lambda)
	\end{equation}
	(\citealp{atkinson2012optimum}). Also for competitive and noncompetitive inhibition models $IC_{50}$ could be driven from eq. (\ref{IC50}) using their respective inhibition constants. These all can suggest the non-negativity of the third parameter, $\theta_{K}$ (of the encompassing model), and similarly those of competitive and non-competitive inhibition models.}

\cite{atkinson2012optimum} \E{computed} $D$-, $D_s$-, $T$- and so called compound $T$-optimal designs \E{(all being optimality criteria for estimation and discrimination which will be described in sections \ref{sec-optimaldesign-estimation} and \ref{sec-discrimination})} for competitive, non-competitive inhibition and the encompassing models. The same setting was used by \cite{harman2020design} to illustrate their genuinely symmetric discriminating design criterion, called $\delta$-optimality, based on linearization of the models and notion of flexible nominal sets.
However, according to biological definitions of parameters and the design or controllable variables (regressors), the modeled reaction rate needs to be positive, which is not necessarily the case for the additive normal error models used so far. To ensure nonnegative values we suggest instead working with logarithms of the models which assumes multiplicative log normal errors and investigate its effect on estimates and optimal designs. 

Enzyme kinetics is a frequent application field in the experimental design literature and Michaelis-Menten based models have become showcase examples, with recent references abound \cite{chen_standardized_2017},  \cite{schorning_optimal_2018} and \cite{marinas-collado_optimal_2019}.  
While those papers are concentrating on optimal design for parameter estimation, the present work adds to the literature by discussing the model transformation issue in deep. 
This aspect is also touched as a side issue in the recent paper by \cite{huang_optimal_2020-3} but only for parameter estimation, while we also put a focus on model discrimination.  

The rest of the paper proceeds as follows. After the introductory Section \ref{sec-estimation/seletion-120data}, initial estimation of the parameters for further use is conducted using 120 real observations \E{(being discussed in detail in later parts) from} \cite{bogacka2011optimum}. Also hypothesis testing is performed as an illustration of model selection at the end of this section. Section \ref{sec-optimaldesign-estimation} first provides calculation of optimal designs for precise estimation of the parameters in both the original and the log-transformed models. In the next section, optimal discriminating designs are derived by making use of compound $T$ ($CT$), $D_s$ and $\delta$ criteria. The discriminating performance of all exact optimal designs are compared with each other through a simulation study and contrasted to the results from the additive error case. Finally, we also calculate designs for discriminating between the two possible model specifications. \E{Discussions on the results plus an interpretable description, in terms of pharmacology, for one suggested optimal design is provided in the conclusions.}

\section{Statistical specification and estimation}\label{sec-estimation/seletion-120data}
\subsection{Parameter estimation}
\textbf{A standard statistical model:} All three models (\ref{formula-comp-model}), (\ref{formula-noncomp-model}) and (\ref{formula-combined-model}) above, could be formulated in terms of a general nonlinear statistical model of \E{$N$} observations, as
\begin{equation} \label{formula-generalmodel}
	y_i=\eta(\thetav,\xv_i)+\epsilon_i, \quad\quad i=1,\dots,N,
\end{equation}
where $\thetav=(\theta_{1},\dots,\theta_{m})^{T}$ is the vector of $m$ unknown parameters, $\thetav \in \Thetav \subseteq {\mathbb{R}_{+}^{m}}$, $\Thetav$ is a compact set of {all non negative} admissible parameter values. \E{$\xv_i=(x_{Si},x_{Ii})^{T}$} is the $i$th pair value of design variables \E{(which are the substrate and inhibition concentrations in the investigated models here).} \E{$\mathfrak{X}=\left[ [x_S]_{\min},[x_S]_{\max}\right] \times \left[[x_I]_{\min},[x_I]_{\max}\right]$} represents the rectangular design region which we assume to be the Cartesian product of the  set of acceptable values for the design variables, where $0\leq [x_S]_{\min} < [x_S]_{\max} $ and $0\leq [x_I]_{\min} < [x_I]_{\max} $ \E{(we may need to discretize the design region for computational purposes).}
Further $y_i$ denotes the ith observation and $\eta(\thetav,\xv_i)$ is the expected response for the ith observation, where $\eta: \Thetav\times \mathfrak{X} \rightarrow \mathbb{R}$ is a nonlinear function of the unknown parameters and the design variables.

As briefly noted in Section \ref{sec-intro}, following to the biochemical definitions for the parameters and the pair of design variables \E{$\xv=(x_{S},x_{I})^{T}$}, the reaction rate $y$ in all the above enzyme kinetic models should of course not be negative. This important issue is usually not taken into account by the common practice of simply assuming additive normal errors. It is evident that such errors could potentially lead to negative observations, if their variance is just large enough. Note that negativity of the reaction rate renders the likelihood estimation invalid. \cite{harman2020design} investigated the case to assume multiplicative log-normal errors instead of the additive normal ones to have liberty in inflating the error variance by any factor without producing faulty observations (eg. for simulation purposes). Now, we suggest to take the natural logarithms of the enzyme kinetics models assuming multiplicative log-errors. This way the errors are switched into additive normal and this process is fully matched with the assumptions under which the standard model is defined. Thus we defined the log-model as
\begin{equation} \label{formula-logmodel}
	\ln(y_i)=\ln(\eta(\thetav,\xv_i)\times\epsilon_i)=\ln(\eta(\thetav,\xv_i))+\ln(\epsilon_i), \quad\quad i=1,\dots,N
\end{equation}
where $\ln(\epsilon)\sim \mathcal{N}(0,\sigma^2)$.  Furthermore, there still remains the uncertainty about which enzyme kinetic model to be selected. Therefore, we would like to consider how the designs may differ under the assumption of the log models for enzyme kinetics, compared to their standard models using both estimation and discrimination criteria. The aim of this research is to investigate how the log models of enzyme kinetics and their error structure may influence the optimal design points produced.\\

To proceed further with optimal design for a nonlinear model we usually require some nominal values
 (see 
\citealp{chernoff1953locally}
), ideally estimated from data of previous experiments. 
For computation of these initial estimates in the models (\ref{formula-comp-model}), (\ref{formula-noncomp-model}) and (\ref{formula-combined-model}) we used data \E{from} \cite{bogacka2011optimum} which consists of $N=120$ triple values of \E{15 different concentrations of substrate (sertraline) \E{spanning a range of $[0,30]$ while being more dense in lower concentrations and more sparse in higher concentrations to provide reasonable substrate saturation as typically is used in \cite{copeland2005evaluation}} and 8 different inhibitor concentration (dextrometorphan) spanning a range of $[0,60]$ and the reaction rate $y$ for each combination of them,} resulted from an initial experiment on Dextrometorphan-Sertraline. \E{Note that the sample size $N$ is actually representing the number of observations.} All computations \E{of this part} were performed using the function {\sf nls} in {\sf R}.

\begin{figure}[htp]
	
	\begin{center}
		\begin{minipage}{.5\linewidth}
			
			\subfloat{\label{plot.res-vs-fit-bef/aft:(a)}\includegraphics[width= 8cm,height=5cm]{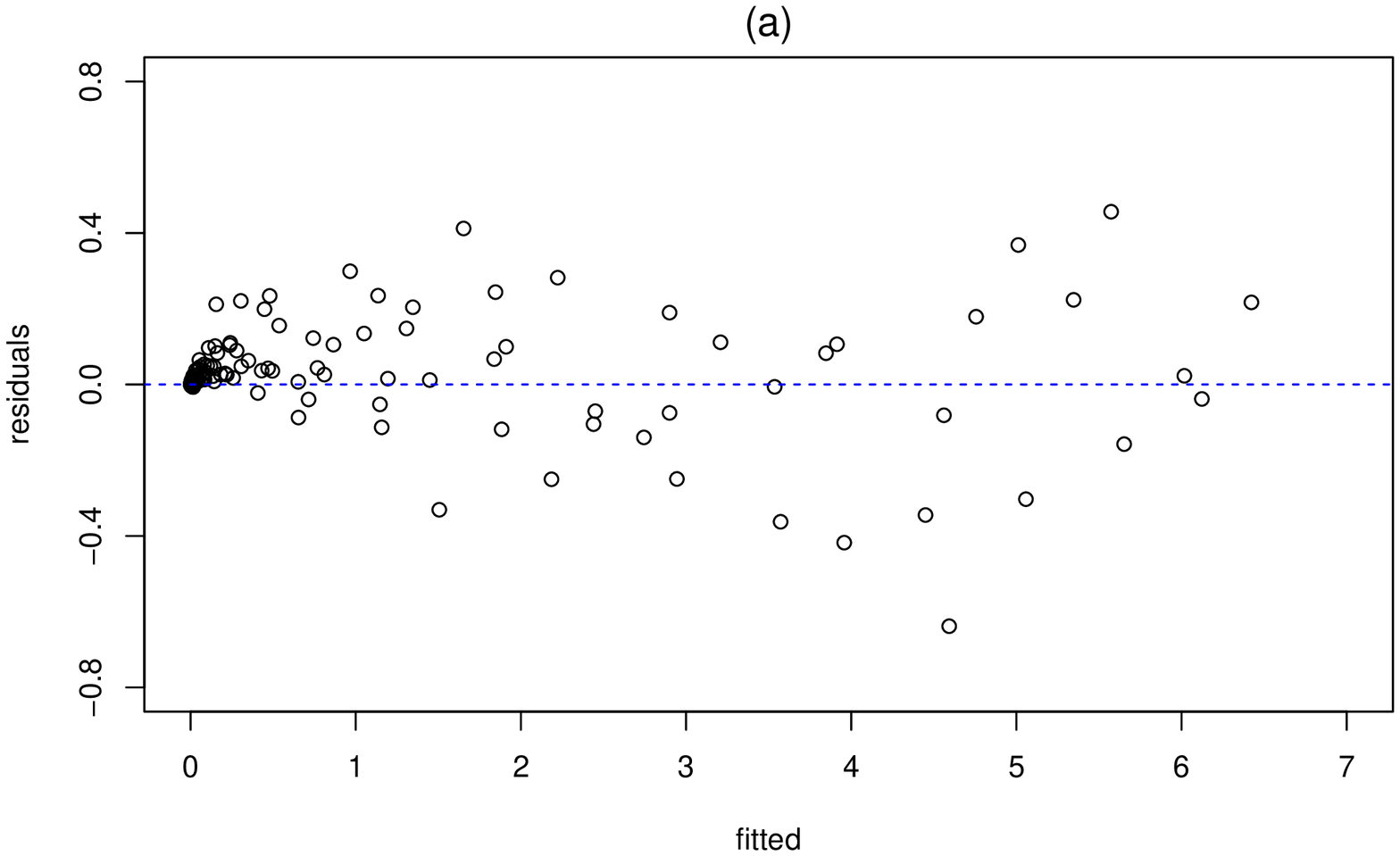}}
		\end{minipage}%
		\begin{minipage}{.5\linewidth}
			
			\subfloat{\label{plot.res-vs-fit-bef/aft:(b)}\includegraphics[width=8cm,height=5cm]{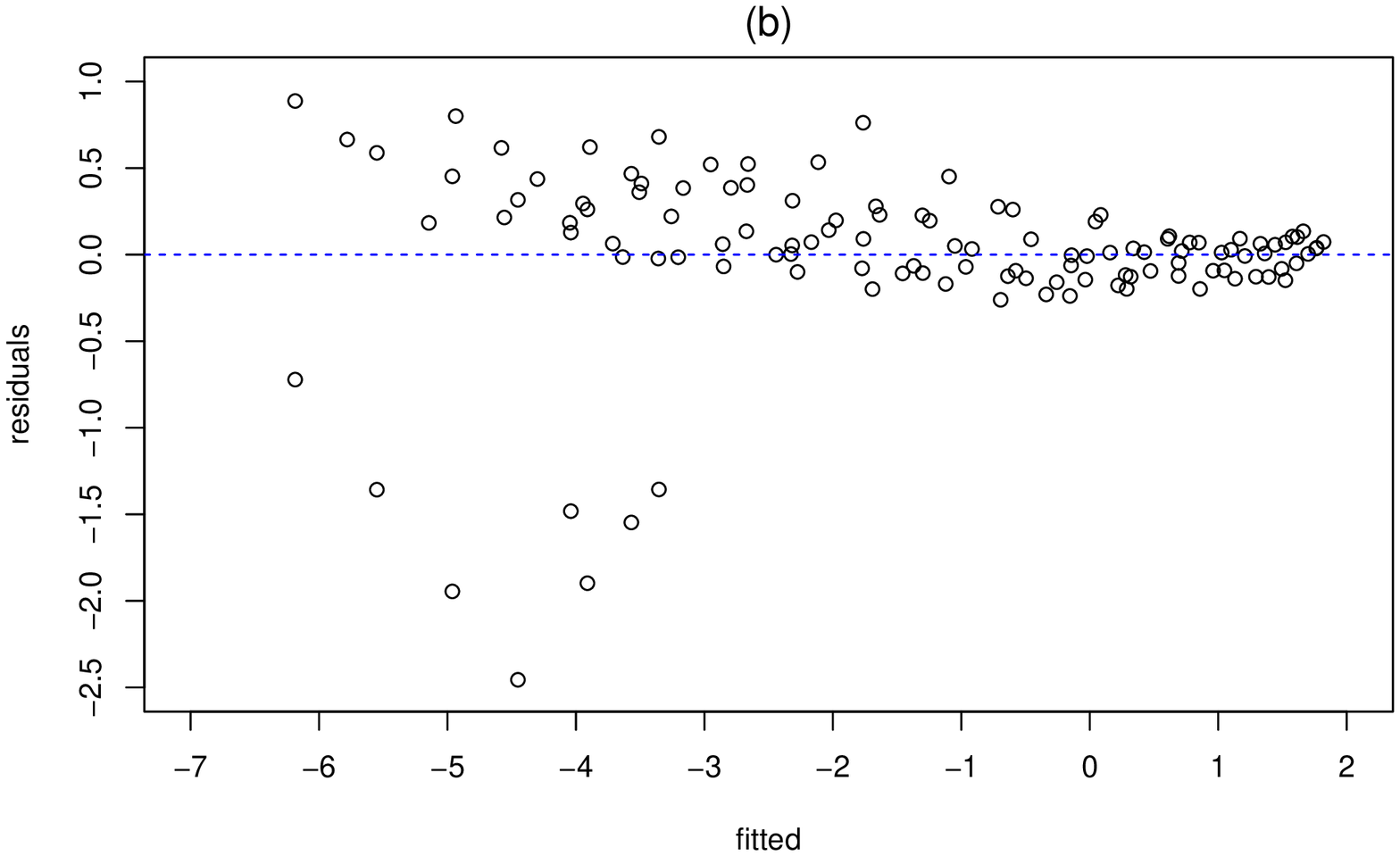}}
		\end{minipage}\par\medskip
		
		\begin{minipage}{\linewidth}
			\centering
			\subfloat{\label{plot.res-vs-fit-bef/aft:(c)}\includegraphics[width=8cm,height=5cm]{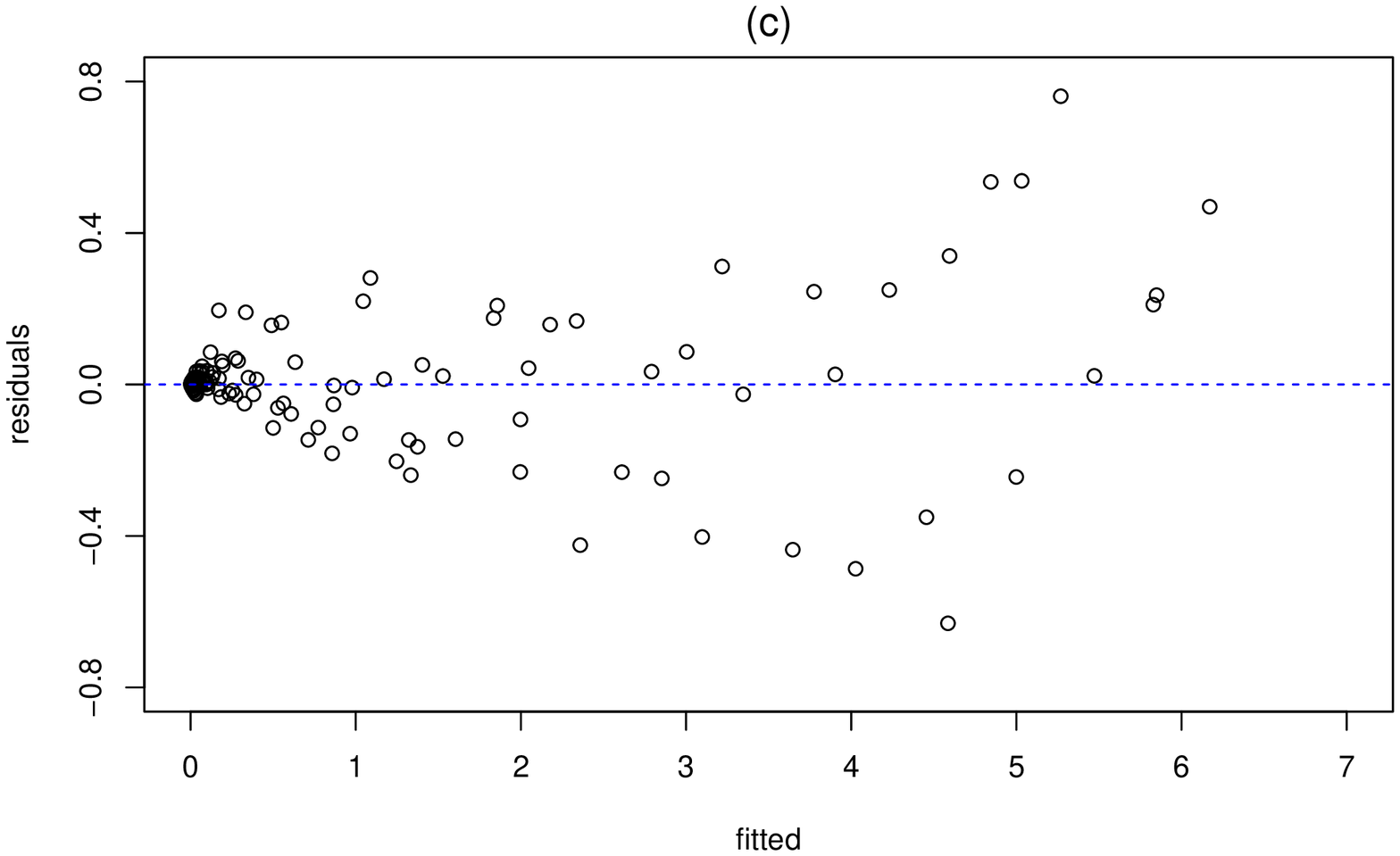}}
		\end{minipage}%
	\end{center}
	\caption{{Scatter plot of residuals versus fitted values for (a): standard case, (b): log case, (c): back-transformed case.}}
	\label{plot.res-vs-fit-bef/aft}
\end{figure}

The data contained some zero values of observed concentrations of substrate and the reaction rates, which cannot be log-transformed. We have thus chosen to replace these few zeros in the data set by some arbitrary small value $\varepsilon$.  For a small enough $\varepsilon$ there is no impact on the estimates in the original model and we have eventually chosen $\varepsilon=0.02$, which renders the smallest possible residual standard error in a back-transformed model (\ref{formula-combined-model}) (0.1870) compared to the same value in the standard case (0.1526) \E{(see Table \ref{table.est-se-log1-2} )}. 
The residual standard error equations for three different cases are 
\begin{align*} \label{formula-MSE-original-log(1)} 
	\mbox{SSE}&=\sum_{i=1}^{N} \left( y_i-\widehat{y_i} \right)^2,\quad \mbox{MSE}=\dfrac{\mbox{SSE}}{N-{m}} \tag{The standard case, model (\ref{formula-generalmodel}) }
	\\
	\mbox{SSE}_{l}^{*}&=\sum_{i=1}^{N} \left( \ln(y_i)-\widehat{\ln({y}_{i})} \right)^2, \quad \mbox{MSE}_{l}^{*}=\dfrac{\mbox{SSE}_{l}^{*}}{N-{m}}  \tag{The log case, model (\ref{formula-logmodel})}
	\\
	\mbox{SSE}_{b}^{*}&=\sum_{i=1}^{N} \left( y_i-\exp(\widehat{\ln({y}_{i})}) \right)^2,\quad \mbox{MSE}_{b}^{*}=\dfrac{\mbox{SSE}_{b}^{*}}{N-m}  \tag{The back-transformed case} 
\end{align*}
Here $N-m$ in each of the equations is the degree of freedom of the corresponding $\mbox{SSE}$. The scatter plot of residuals versus fitted values of $N=120$ observations for the standard case, the log case and the back-transformed case is displayed in Figure \ref{plot.res-vs-fit-bef/aft}. As we can observe from the panels \ref{plot.res-vs-fit-bef/aft:(a)} and \ref{plot.res-vs-fit-bef/aft:(c)}, the similarity of the fits is confirmed. Although the residual pattern for \E{ the standard case is a bit superior to the one for the back-transformed case from the perspective of being more spread around zero} 
, the advantage of not violating non-negativity motivates us to proceed further with the log-model. A robustness analysis was also performed, particularly on the eight observations in the lower left part of the scatter plot \ref{plot.res-vs-fit-bef/aft:(b)} which seem not to follow the trend. We looked into their reaction rate values and it was observed that their deletion would not have any noticeable effect on the initial estimates and they thus need not be discarded as outliers.


%
%
%
%
Table \ref{table.est-se-log1-2} represents the initial estimations for \E{competitive and non-competitive} models (\ref{formula-comp-model}) and (\ref{formula-noncomp-model}), in both the standard and the log case, \E{respectively}. Similar initial estimations for the \E{encompassing} model (\ref{formula-combined-model}) are represented in Table \ref{table.est-se-log3}. 
As it is observed from the Tables \ref{table.est-se-log1-2} and \ref{table.est-se-log3}, logarithmic transformations do not change the estimates \E{considerably} 
except for $\theta_{K}$ in the noncompetitive model. Note that there are some (slight) discrepancies between the estimates given here and to what \cite{atkinson2012optimum} used for some of his comparisons. For inner consistency we decided to only use the values from Tables \ref{table.est-se-log1-2} and  \ref{table.est-se-log3} throughout this paper.

\begin{table}[htp]
	\caption{The parameter estimates and their corresponding standard error estimates.}
	\label{table.est-se-log1-2}
	\begin{normalsize}
		\hfill{\renewcommand{\arraystretch}{1.4}
			\begin{center}
				\resizebox{\textwidth}{!}{
					\begin{tabular}{ccccccccccccccc}
						\hline
						\multicolumn{7}{c}{ Competitive model (\ref{formula-comp-model})}&&
						\multicolumn{7}{c}{ Non-competitive model (\ref{formula-noncomp-model}) }\\
						\cline{1-7} \cline{9-15}
						\multicolumn{3}{c}{ Standard case ($\hat{\sigma}=0.1553$)}&&\multicolumn{3}{c}{ Log case ($\hat{\sigma}=0.5160$) }&&\multicolumn{3}{c}{ Standard case ($\hat{\sigma}=0.2272$) }&&\multicolumn{3}{c}{ Log case ($\hat{\sigma}=0.5306$) }\\
						\cline{1-3} \cline{5-7} \cline{9-11} \cline{13-15}
						&\text{Estimate} $\hat{\thetav}$&\text{SE} $\hat{\sigmav}$&&&\text{Estimate} $\hat{\thetav}$&\text{SE} $\hat{\sigmav}$&&&\text{Estimate} $\hat{\thetav}$&\text{SE} $\hat{\sigmav}$&&&\text{Estimate} $\hat{\thetav}$&\text{SE} $\hat{\sigmav}$\\
						\cline{1-3} \cline{5-7} \cline{9-11} \cline{13-15}
						$\theta_{V}$&$7.2976$&$0.1143$&&$\theta_{V}$&$6.0645$&$0.9260$&&$\theta_{V}$&$8.6957$&$0.2227$&&$\theta_{V}$&$12.0125$&$2.0553$\\
						$\theta_{M}$&$4.3860$&$0.2333$&&$\theta_{M}$&$3.2799$&$0.7288$&&$\theta_{M}$&$8.0664$&$0.4880$&&$\theta_{M}$&$8.5359$&$1.5721$\\
						$\theta_{K}$&$2.5821$&$0.1454$&&$\theta_{K}$&$3.3153$&$0.6041$&&$\theta_{K}$&$12.0566$&$0.6709$&&$\theta_{K}$&$5.6638$&$0.8879$\\
						\hline
					\end{tabular}
				}
			\end{center}
		}
	\end{normalsize}
\end{table}

\begin{table}[htp]
	\caption{The parameter and their corresponding standard error estimates for the encompassing model (\ref{formula-combined-model}).}
	\label{table.est-se-log3}
	\begin{footnotesize}
		\hfill{\renewcommand{\arraystretch}{1.4}
			\begin{center}
				\begin{tabular}{ccccccc}
					\hline 
					\multicolumn{3}{c}{ Standard case ($\hat{\sigma}=0.1526$) }&&\multicolumn{3}{c}{ Log case ($\hat{\sigma}=0.5128$) }\\
					\cline{1-3} \cline{5-7} 
					&\text{Estimate} $\hat{\thetav}$&\text{SE} $\hat{\sigmav}$&&&\text{Estimate} $\hat{\thetav}$&\text{SE} $\hat{\sigmav}$\\
					\cline{1-3} \cline{5-7} 
					$\theta_{V}$&$7.4253$&$0.1298$&&$\theta_{V}$&$6.9897$&$1.3406$\\
					$\theta_{M}$&$4.6808$&$0.2724$&&$\theta_{M}$&$3.9799$&$1.0403$\\
					$\theta_{K}$&$3.0581$&$0.2815$&&$\theta_{K}$&$3.7380$&$0.7218$\\
					$\lambda$&$0.9636$&$0.0191$&&$\lambda$&$0.8737$&$0.1123$\\
					\hline
				\end{tabular}
			\end{center}
		}
	\end{footnotesize}
\end{table}

\subsection{Model discrimination and/or selection} 
Competitive and non-competitive inhibition models of enzyme kinetics, are two distinct models, none of which could be obtained from the other by implementing some restrictions on the parameters or through a limiting process. Therefore, in context of Cox's definition of models to do hypothesis testing (\citealp{cox1961tests, cox1962further}) these models are (separate) non-nested; although the encompassing model (\ref{formula-combined-model}) may be used further in the next sections in order to ease specification of methods. \E{This point is also mentioned in \cite{copeland2005evaluation} chapter 5, about the competitive and noncompetitive enzyme models to use some tests for validation of models if there is ambiguity about the model which best describes the data, although in some cases the model best describing the data may be clear.} 
One may observe a subtle differentiation between model discrimination and selection. While the former is considered strictly as a selection problem among two or more alternatives (of which one is considered the correct one), the latter is more concerned with decision making problems through using some statistical measures of fit and is oriented toward hypothesis testing problems where the same features are considered in the test statistics. Hence, since rejection of the null hypothesis does not necessarily imply acceptance of the alternative. This is specially problematic as in the case of nonnested models we are required to select an arbitrary candidate model as the null as contrasted to the nested models where the most parsimonious model is a natural choice
(\cite{pesaran2001non}).

\subsubsection{Utilizing the likelihood ratio test for model selection} \label{subsec:likelihood ratio test}
Thus, since in our present context of hypothesis testing the considered models  (\ref{formula-comp-model}), denoted by index $C$, and (\ref{formula-noncomp-model}), denoted by index $N$,  are attributed asymmetric status, we are required to examine two systems of hypothesis testing using fixed design points in order to potentially suggest one model at the end. 
However, 
instead of using the Cox proposals for hypothesis testing of non nested models (\citealp{cox1961tests,cox1962further}) we consider the Monte Carlo distribution of the log-likelihood ratio (\citealp{deldossi2019optimal}). Therefore
\E{similar to} 
\cite{pesaran2001non}, given \E{$N=120$ data from}  \cite{bogacka2011optimum} with \E{fixed design points}, which also were used in the parameter estimation part, we performed two likelihood ratio tests in each of which the null and the alternative hypothesis are defined as      
\[ \textrm{I})  \left\{ \begin{split}
	& H_0: y=\eta_{C}+\epsilon, \quad \epsilon \sim \mathcal{N}(0, \sigma^2)\\
	& H_1: y=\eta_{N}+\epsilon, \quad \epsilon \sim \mathcal{N}(0, \sigma^2)
\end{split} \right.
\]
and
\[ \textrm{II}) \left\{ \begin{split}
	& H_0: y=\eta_{N}+\epsilon, \quad \epsilon \sim \mathcal{N}(0, \sigma^2)\\
	& H_1: y=\eta_{C}+\epsilon, \quad \epsilon \sim \mathcal{N}(0, \sigma^2)
\end{split}  \right.
\]
respectively, \E{where in each of the tests $\sigma$ is estimated through the respective residual standard error for each model, presented in Table \ref{table.est-se-log1-2}. } Note that in both tests $\textrm{I}$ and $\textrm{II}$, $y$ follows the normal distribution with mean equal to its corresponding expected reaction rate and the covariance equal to the covariance of the corresponding errors. Two similar hypothesis tests can be defined for the log case with the only difference that in those cases the logarithm of errors follow normal distribution with the same parameter values defined in each of the tests above. For the case of nonnormal errors one could further refer to KL-optimality, see for example 
\cite{deldossi2019optimal}.

In order to implement the likelihood ratio \E{tests $\textrm{I}$ and $\textrm{II}$}, the log-likelihood
ratio test statistics are used as
\begin{equation} \label{log-likelihood test statistics}
	W_{CN}=\ell_{C}(\hat{\thetav}_{C})-\ell_{N}(\hat{\thetav}_{N}) \quad\text{and}\quad  W_{NC}=\ell_{N}(\hat{\thetav}_{N})-\ell_{C}(\hat{\thetav}_{C}),
\end{equation}
\E{respectively,} where $\ell_{C}$ and $\ell_{N}$ are the log-likelihood functions for \E{competitive model with $\eta_{C}$} and \E{non-competitive model with $\eta_{N}$} respectively and $\hat{\thetav}_{C}$ and $\hat{\thetav}_{N}$ are the corresponding maximum likelihood estimates of the parameters. Let $p_{CN}$ and $p_{NC}$ be the $p$ values of $W_{CN}$ and $W_{NC}$ respectively. Then, given a significance level $\alpha$, the hypothesis testings can lead to four different categories
\begin{enumerate}
	\item If $p_{CN} < \alpha$ and $p_{NC} \geq \alpha$, we reject the competitive and accept the non-competitive model;
	\item If $p_{CN} \geq \alpha$ and $p_{NC} < \alpha$, we accept the competitive and reject the non-competitive model;
	\item If $p_{CN} \geq \alpha$ and $p_{NC} \geq \alpha$, we accept the competitive if $p_{CN}> p_{NC}$ and accept the non-competitive if $p_{NC} > p_{CN}$;
	\item If $p_{CN} < \alpha$ and $p_{NC} < \alpha$, we reject the competitive if $p_{CN} < p_{NC}$ and reject the non-competitive if $p_{NC} < p_{CN}$.
\end{enumerate}

The Monte Carlo process of approximating the sample distribution of $W_{CN}$ and its corresponding approximated $p$-value $\hat{p}_{CN}$ \E{in test $\textrm{I}$, } is performed according to the following steps 
\begin{itemize}
	\item Generate $B=10,000$ samples \E{of the size $N=120$} from the competitive model under $H_{0}$ \E{starting with} the corresponding parameter estimates from  the table \ref{table.est-se-log1-2};
	\item At each of the Monte Carlo simulation steps for $b=1,2,\dots B$:
	\begin{itemize}
		\item Compute the maximum likelihood parameter estimates $\hat{\thetav}_{C}^b$ and $\hat{\thetav}_{N}^b$ by maximizing and $\ell_{C}(\thetav_{C}^b)$ and $\ell_{N}(\thetav_{N}^b)$ \E{for $\eta_{C}$ and $\eta_{N}$}, respectively;
		\item Compute $W_{CN}^b=\ell_{C}(\hat{\thetav}_{C}^b)-\ell_{N}(\hat{\thetav}_{N}^b)$;
	\end{itemize}
	\item Calculate the Monte Carlo $p$-value as
	$\hat{p}_{CN}=\sum_{b=1}^{B} I(W_{CN}^{b} < w_{CN})/B$ ,
\end{itemize} 
where \E{$w_{CN}=45.79$} is the computed value of $W_{CN}$ using the $N=120$ observations of $y$. 
\E{
Recall that in these tests the sample size is taken to be equal to $N=120$ and therefore the design points are also the same $15\times 8=120$ combinations of the substrate and inhibitor from the last section.}
A similar Monte Carlo process can be followed to obtain $W_{NC}$ and $\hat{p}_{NC}$ \E{for the second test} by reversing the role of competitive and non-competitive models such that \E{$w_{NC}=-45.67$ (this value is only slightly different from $w_{CN}=45.79$ which is due to the simulation error).
}Using the  Monte Carlo $p$-values in  Table \ref{hypo-test}, in the standard case we reject the non-competitive model according to the fourth category above\E{, on the significance level $\alpha=0.05$}. \E{Two similar hypothesis tests in the log case results in $w_{CN}=-w_{NC}=3.84$ } and finally the  consequence holds \E{to reject the non-competitive model (and to accept the competitive)} in the log case, according to the second category. 
\begin{table}[htp] 
	\caption{{$p$-value estimates of the likelihood ratio tests}}
	\label{hypo-test}
	\begin{footnotesize}
		\hfill{\renewcommand{\arraystretch}{1.4}
			\begin{center}
				\begin{tabular}{ccccccccc}
					\hline 
					&&\multicolumn{3}{c}{ Standard case }&&\multicolumn{3}{c}{ Log case }\\
					\cline{3-5} \cline{7-9}
					&&\multicolumn{3}{c}{ \E{Hypothesis test} }&&\multicolumn{3}{c}{ \E{Hypothesis test}  }\\
					\cline{3-5} \cline{7-9} 
					&& \E{$\textrm{I}$} && \E{$\textrm{II}$} && \E{$\textrm{I}$} && \E{$\textrm{II}$}\\
					\cline{3-5} \cline{7-9}
					$\hat{p}$ \E{(under $H_{0}$)}&&$0.0064$&&$0$&&$0.0605$&&$0.0017$\\
					\hline
				\end{tabular}
			\end{center}
		}
	\end{footnotesize}
\end{table}

In these tests we could make a decision about the models, using the $N=120$ fixed design. However, if we are interested in obtaining optimal designs (which will be defined formally in the next section) for the above procedure we would be required to solve a rather formidable multivariate nonconvex optimization problem. 
It is thus impossible to derive \E{optimal designs directly from this procedure}, in practice and we require to use some simplified criteria of optimality to derive the optimal design points for model discrimination (and parameter estimation), which will be described in the next section.    

\section{Optimal designs for estimation of parameters}\label{sec-optimaldesign-estimation}
In this section we implement optimal design criteria, specifically $D$ and $D_s$, for the model (\ref{formula-logmodel}) in general. The methods require initial estimates presented in Tables \ref{table.est-se-log1-2} and \ref{table.est-se-log3}. A thorough comparison of the resulted designs in the log case compared to the standard case is done using relative efficiencies. In all cases  optimality of the resulted designs in the log case is evaluated using the \E{equivalence theorems given in the next section}.

\subsection{D-optimality}\label{sec-D-optimality}
$D$-optimal designs, introduced by \cite{wald1943efficient}, are used when estimation of all parameters is of primary interest to the experimenter. In these situations we are faced with one model at a time. By a design we mean a set \E{of $n$ mutually distinct design points}, $\xv_1,\xv_2,\dots,\xv_n$, with their corresponding \E{proportion of replication of observations taken at each $\xv_i$ (weights, any real number between 0 and 1)} denoted by  $\omega_1,\omega_2,\dots, \omega_{n}$ which define a probability measure as $\xi=\left\lbrace (\xv_1{^T},\omega_1),(\xv_2{^T},\omega_2),\dots, (\xv_n{^T},\omega_n)\right\rbrace $, on design region $\mathfrak{X}$ \E{(being discretized in computations) such that $\sum_{i=1}^{n} \omega_i=1$. In order to obtain exact designs, $N_i=N\times \omega_i$, $i=1,\dots, n$ are rounded to integers such that $N=\sum_{i=1}^{n} N_i$ for all observations.} By an optimal design we mean a selection of some design $\xi^{*}$ which renders an optimum value of some criteria of optimality, according to the goal followed in designing an experiment. Therefore, in the context of enzyme kinetic models, the aim of this section is the optimal selection of pairs of substrates and inhibitors in each of the enzyme kinetic models also for their log-transformed cases \E{instead of screening experiments with quite large spans of substrate-inhibition titrations (with 96-, 384- or 1536-microwell plates being the typical ones, cf. \citealp{copeland2005evaluation}) which are the usual procedures in investigating the effect of these simultaneous titrations in response rates of enzyme kinetics in biopharmaceutical research.}

The information provided in a design $\xi$ is measured by its Fisher information matrix, defined below in (\ref{formula-Fisher-IM}), which essentially describes the amount of information provided in the data about the unknown parameters. For nonlinear models and independent observations the inverse of the Fisher information matrix is proportional to the asymptotic covariance matrix for the maximum likelihood estimates of the unknown parameters. \E{$D$-optimality performs well when the so-called parameter curvature is negligible, in the case of nonlinear models. For the discussion on the effects of parameter curvature please refer to later parts of this section.} Therefore, due to dependence of the Fisher information matrix on the unknown parameters, as discussed an initial estimate of them is needed to obtain the optimal designs which, in this case, are called locally optimal \cite{chernoff1953locally}. Consequently, we need to linearize each model at its respective initial estimate, \E{$\bar{\thetav}$}, as
\begin{equation} \label{formula-linearized model}
f(\xv_i,\bar{\thetav})=\dfrac{\partial \ln(\eta(\thetav,\xv_i))}{\partial \thetav}\big|_{\bar{\thetav}} ,	
\end{equation} 
where $f^{T}(\xv_i,\bar{\thetav})$ is the $m$ dimensional vector of partial derivatives for the $i$th 
\E{design point}. So the Fisher information matrix for 
\E{a design with $n$ support points} is  

\begin{equation} \label{formula-Fisher-IM}
M(\xi,\bar{\thetav})=\sum_{i=1}^{n}\omega_i f(\xv_i,\bar{\thetav})f^{T}(\xv_i,\bar{\thetav})=\mathbf{F}^{T}(\Xv,\bar{\thetav})\mathbf{W}\mathbf{F}(\Xv,\bar{\thetav}),
\end{equation} 
\E{in which $\Xv$ denotes the collection of all design points}.  $\mathbf{F}(\xv_i,\bar{\thetav})$ is the $n\times m$ dimensional matrix for \E{$n$ support points} with ith row $f^{T}(\xv_i,\bar{\thetav})$ and $\mathbf{W}$ is the diagonal matrix of $n$ weights $\omega_i$. 

Optimal designs for estimation of parameters are aimed to maximize a function $\Phi$ of the the Fisher information matrix, called the optimality criterion. Therefore, for the case of D-optimality the criterion is defined as 
\begin{equation} \label{D-criterion}
\Phi_{D}(\xi,\bar{\thetav})=\det\left\lbrace M(\xi,\bar{\thetav})\right\rbrace.
\end{equation}
Thus, a design is called $D$-optimal if it maximize the determinant of the information matrix (or similarly to minimize the determinant of the covariance matrix).   
\E{An analogue of the celebrated equivalence theorem (\citealp{kiefer1960equivalence}) which states equivalence of two extremum problems, approximate $D$-optimum and $G$-optimum (eq. (\ref{G-criterion})) designs, can be formulated for nonlinear models (\citealp{white1973extension}). Using this useful property, one can check whether a computed design is actually $D$-optimum. A $G$-optimum design minimize the maximum over $\xv$ of the sensitivity function and is defined as} 
\begin{equation} \label{G-criterion}
d(\xv,\xi,\bar{\thetav})=f^{T}(\xv,\bar{\thetav}) M^{-1}(\xi,\bar{\thetav})f(\xv,\bar{\thetav}).
\end{equation}
The equivalence theorem states that the following three conditions are equivalent:

\begin{enumerate}
	\item Design $\xi^{*}$ maximizes $\Phi_{D}(\xi,\bar{\thetav})$.
	\item Design $\xi^{*}$ minimizes $\max_{\mathfrak{X}} d(\xv,\xi,\bar{\thetav})$.
	\item $\max_{\mathfrak{X}} d(\xv,\xi^{*},\bar{\thetav})=m$, where $m$ is the number of parameters in the model and the maxima occur at the points of support of the optimal design, i.e. $d(\xv_{i}^{*},\xi^{*},\bar{\thetav})=m$.

	Therefore for any non-optimum design $\xi$,
	\item $\max_{\mathfrak{X}} d(\xv,\xi,\bar{\thetav})>m$.
	
\end{enumerate}
In order to compare any design to a $D$-optimum design, we used $D$-efficiency which is defined as

\begin{equation} \label{D-efficiency}
\text{Eff}_{D}(\xi)=\left[ \dfrac{\det\left\lbrace M(\xi,\bar{\thetav}\right\rbrace}{\det\left\lbrace M(\xi^{*},\bar{\thetav})\right\rbrace}\right]^{\frac{1}{m}}.
\end{equation}
\E{If a design $\xi$ with $n$ support points which has $\text{Eff}_{D}$ is used in an experiment, this means
	that the same accuracy in estimation could be achieved by performing only $n\times\text{Eff}_{D}$ trials
	under the $D$-optimal design $\xi^{*}$.}

\E{
Note that for nonlinear models the $D$-optimality criterion is only suitable when the so-called parameter curvature is negligible. \cite{hamilton1985quadratic} proposed to instead consider a quadratic design criterion based on second-order approximation of the volume of the parameter inference region, when the sample size is small. In order to investigate this parameter curvature effect we computed this quadratic design criterion 
for the encompassing model \ref{formula-combined-model} in both the standard and the log case. It is observed that for all our cases this effect is actually negligible and the new designs based on the proposed quadratic design criterion are essentially the same (with minor deviations in the weights) as the computed $D$-optimal designs and we thus refrain from reporting them for simplicity and brevity.}

\subsection{Ds-optimality}
$D_s$-optimality, introduced by \cite{atkinson1974planning}, is a special case of $D$-optimality, which is aimed to compute the optimal designs when the interest is in estimation of a subset of $s$ \E{(which is equal to one in our case)} parameters while the other $m-s$ parameters can be considered being nuisance. In this case the information matrix will be partitioned as
\[M(\xi,\bar{\thetav})=\begin{pmatrix} \label{partitioed FIM1}
	M_{11}(\xi,\bar{\thetav}) & M_{12}(\xi,\bar{\thetav})\\\
	M_{21}(\xi,\bar{\thetav}) & M_{22}(\xi,\bar{\thetav})
\end{pmatrix}.
\]
\E{where the block $M_{11}$ refers to the parameter(s) of interest.} A general equation for the partitions would be as
\begin{equation} \label{partitioed FIM2}
	M_{jk}(\xi,\bar{\thetav})=\sum_{i=1}^{n}\omega_i f_{j}(\xv_i,\bar{\thetav})f_{k}^{T}(\xv_i,\bar{\thetav})=\mathbf{F}_{j}^{T}(\Xv,\bar{\thetav})\mathbf{W}\mathbf{F}_{k}(\Xv,\bar{\thetav}), \quad j,k=1,2,
\end{equation}
where $f_{1}^{T}(\xv_i,\bar{\thetav})$ and $f_{2}^{T}(\xv_i,\bar{\thetav})$\E{, $i=1,\dots,n$,} are $s$ and $m-s$ dimensional vectors, which are similarly computed from Eq. (\ref{formula-linearized model}) with the difference that in these cases the partial derivatives are with respect to $\thetav_1$ and $\thetav_2$, being $s\times1$ and $(m-s)\times1$, respectively \E{such that $(f_{1}^{T}(\xv_i,\bar{\thetav}),f_{2}^{T}(\xv_i,\bar{\thetav}))=f^{T}(\xv_i,\bar{\thetav})$ }. Further $\mathbf{F}_{1}(\Xv,\bar{\thetav})$ and $\mathbf{F}_{2}(\Xv,\bar{\thetav})$ are $n\times s$ and $n\times (m-s)$ dimensional matrices each having the ith row as $f_{1}^{T}(\xv_i,\bar{\thetav})$ and $f_{2}^{T}(\xv_i,\bar{\thetav})$, respectively. Furthermore, $\mathbf{W}$ is a diagonal matrix of $n$ weights $\omega_i$, as before.

The covariance matrix for the \E{maximum likelihood} estimate of the parameter(s) of interest denoted by $Q^{-1}(\xi,\bar{\thetav})$, is the $s\times s$ upper left submatrix of $M^{-1}(\xi,\bar{\thetav})$  (see eg. \cite{atkinson2007optimum}, Chapter 10). So, using the partitioned matrix inversion we have
\begin{equation*} \label{inverse-P}
	Q^{-1}(\xi,\bar{\thetav})=\left\lbrace M_{11}(\xi,\bar{\thetav})-M_{12}(\xi,\bar{\thetav})M_{22}^{-1}(\xi,\bar{\thetav})M_{21}(\xi,\bar{\thetav})\right\rbrace^{-1},
\end{equation*}
\E{in which $M_{22}(\xi,\bar{\thetav})$ is assumed nonsingular. }A design will be $D_s$-optimal, if it minimizes the determinant of $Q^{-1}(\xi,\bar{\thetav})$ or similarly maximizes the determinant
\begin{align} \label{det-P}
	\Phi_{s}(\xi,\bar{\thetav})&=\det\left\lbrace Q(\xi,\bar{\thetav}) \right\rbrace=det\left\lbrace M_{11}(\xi,\bar{\thetav})-M_{12}(\xi,\bar{\thetav})M_{22}^{-1}(\xi,\bar{\thetav})M_{21}(\xi,\bar{\thetav})\right\rbrace \nonumber \\
	&=\dfrac{\det\left\lbrace M(\xi,\bar{\thetav})\right\rbrace }{\det\left\lbrace M_{22}(\xi,\bar{\thetav})\right\rbrace}.
\end{align}
Therefore, the similar equation to eq. (\ref{G-criterion}) for the \E{sensitivity} function in this case will be as 
\begin{equation} \label{Ds-variance}
	d_{s}(\xv,\xi,\bar{\thetav})=f^{T}(\xv,\bar{\thetav}) M^{-1}(\xi,\bar{\thetav})f(\xv,\bar{\thetav})-f_{2}^{T}(\xv,\bar{\thetav}) M_{22}^{-1}(\xi,\bar{\thetav})f_{2}(\xv,\bar{\thetav}).
\end{equation}
In order to check whether a computed design is actually $D_s$-optimal, we need to check if
\begin{equation}
	d_{s}(\xv,\xi^{*},\bar{\thetav})\leq s,
\end{equation}
with equality at points of support of the optimum design e.g. $d(\xv_{i}^{*},\xi^{*},\bar{\thetav})=s$.

In order to compare any design to a $D_s$-optimum design, $D_s$-efficiency is similarly defined as
\begin{equation} \label{Ds-efficiency}
	\text{Eff}_{Ds}(\xi)=\left[ \dfrac{\det\left\lbrace Q(\xi,\bar{\thetav}\right\rbrace}{\det\left\lbrace Q(\xi^{*},\bar{\thetav})\right\rbrace}\right]^{\frac{1}{s}}.
\end{equation}

Table \ref{table.D,Ds-designs-efficiencies} presents the $D$ and $D_s$-optimal designs consisting of recalculations of the designs for the standard case already presented by \cite{atkinson2012optimum} with the difference that here the design region is the \E{discretized} rectangular $\mathfrak{X}=[0,30]\times[0,60]$ and initial parameter estimations are taken from Tables \ref{table.est-se-log1-2} and \ref{table.est-se-log3} then followed by optimal design calculations for the log case. \E{As mentioned before,} for $D_s$-optimality we assume that $s=1$ meaning that we are interested in computation of optimal designs for estimation of a single parameter of interest, $\lambda$, in the encompassing model \E{such that a precise estimation of $\lambda$ test whether a simpler model is adequate and therefore is of high importance in enzyme kinetic models discussed in this work.} The design region used for the log case is the rectangular $\mathfrak{X}=[\varepsilon,30]\times[0,60]$ \E{constructing a grid of  $31\times61$ points  (note that a denser grid of the points does not affect the final resulted designs in all considered criteria of optimality in the log case and therefore speeds up the calculations).} Assumed parameter spaces can be \E{$\thetav \in \left( 0,\infty\right)$}, but sometimes for computational purposes we had to use nonrestrictive upper bounds.
Note that some discrepancies in the design recalculations of the standard case compared to the designs presented by \cite{atkinson2012optimum} are due to differences in the initial estimates and the designs space. 
Note that for computation of all $D$-optimal designs we used the package {\sf OptimalDesign} in {\sf R} 
and a linear programming simplex method \E{(\cite{harman2008computing})} was used for computation of $D_s$ optimal designs.  \E{\cite{harman2008computing} basically use the fact that a $c$-optimal design being the one which minimizes the variance for the best linear unbiased estimator of $c^{T}\thetav$, is equivalent to the desired $D_s$-optimal designs where $c^{T}=(0,0,0,1)$ suggest the interest in minimizing the variance for the unbiased estimator of $\lambda$. The computation is then handled through a linear programming simplex method.}

\begin{table}[htp] 
	\caption{$D$ and $Ds$-optimal designs}
	\label{table.D,Ds-designs-efficiencies}
	\begin{center}
		\begin{footnotesize}
		\begin{tabular}{lllcccccc}
			\hline
			&&&&&&&&\\
			
			&&Design&&$x_S$&$x_I$&$\omega$&&\\
			\hline
			Standard case &&{\scriptsize $4D_{N}$}&&$30.000$& $0.000$ & $0.25$&&\\ 
			&&&&$5.223$& $0.000$ & $0.25$&\\ 
			&&&&$30.000$& $12.045$& $0.25$&\\
			&&&&$5.223$& $12.045$& $0.25$&\\
			&&{\scriptsize $4D_{C}$}&&$30.000$& $0.000$ & $0.25$&&\\ 
			&&&&$3.348$& $0.000$ & $0.25$&\\ 
			&&&&$30.000$& $20.297$& $0.25$&\\
			&&&&$7.902$ & $7.137$& $0.25$&\\
			&&{\scriptsize $4D_{E}$}&&$30.000$& $0.000$ & $0.25$&&\\ 
			&&&&$3.616$& $0.000$ & $0.25$&\\ 
			&&&&$30.000$& $18.290$& $0.25$&\\
			&&&&$7.500$ & $7.584$& $0.25$&\\
			&&{\scriptsize $Ds_{N}$}&&$30.000$& $0.000$ & $0.086$&&\\ 
			&&&&$3.884$& $0.000$ & $0.208$&\\ 
			&&&&$30.000$& $16.952$& $0.206$&\\
			&&&&$3.884$ & $16.952$& $0.500$&\\
			&&{\scriptsize $Ds_{C}$}&&$30.000$& $0.000$ & $0.027$&&\\ 
			&&&&$2.545$& $0.000$ & $0.088$&\\ 
			&&&&$30.000$& $28.550$& $0.371$&\\
			&&&&$7.098$ & $8.253$& $0.514$&\\
			\hline 
			Log-case &&{\scriptsize $4D_{N},3D_{N},4D_{C},4D_{E},Ds_{N}$}&&$\varepsilon$& $0$ & $0.25$&& \\ 
			&&&&$30$ & $0$ & $0.25$&\\ 
			&&&&$\varepsilon$& $60$& $0.25$&\\
			&&&&$30$ & $60$& $0.25$&\\
			&&{\scriptsize $3D_C$}&&$\varepsilon$& $0$ & $1/3$&&\\ 
			&&&&$30$ & $0$ & $1/3$&\\ 
			&&&&$\varepsilon$& $60$& $1/3$&\\
			
			&&{\scriptsize $Ds_{C}$}&&$\varepsilon$& $0$ & $0.017$&&\\ 
			&&&&$30$ & $0$ & $0.173$&\\ 
			&&&&$\varepsilon$& $60$& $0.327$&\\
			&&&&$30$ & $60$ & $0.483$&\\ 
			\hline
		\end{tabular}
		\end{footnotesize}
	\end{center}
\end{table}

It is remarkable that all the optimal designs for the log-model are concentrated at the corners of the design region with the interpretation that
the best designs for precise estimation of parameters are the most extreme pair concentrations of substrate and inhibition which makes them easy to use in practice. Also they are robust to the choice of initial estimates, which indicates that they behave much like linear models over a wide region of the parameter space, another attractive feature.
Note that $4D_{N}$ and $4D_{C}$ stand for $D$-optimal designs for estimation of \E{four} parameters of the encompassing model using the initial estimates for the non competitive and competitive models in table \ref{table.est-se-log1-2} and $\lambda=0$ or $\lambda=1$, respectively. $4D_{E}$ is the $D$-optimal design for the \E{four parameter} encompassing model using the initial estimates in table \ref{table.est-se-log3}. Further, $3D_{N}$ denotes the $D$-optimal design for the three parameter non-competitive model, which surprisingly has four points of support and $3D_{C}$ is similarly computed for the \E{three parameter} competitive model. \E{Recall that in $D_s$ optimal designs $s=1$ meaning that we are interested in estimation of the parameter $\lambda$. } Therefore $Ds_{N}$ and $Ds_{C}$ are $D_s$-optimum for estimation of $\lambda$ in the encompassing model for two different cases of $\lambda=0$ and $\lambda=1$, respectively. Similar recalculations of designs for the standard case shows that in all these cases optimal designs are more spread over the rectangular design region and not completely located in the extremes. Note that in the standard case, omitted from the table $3D_{N}$ and $3D_{C}$ are the first three support points of their corresponding designs $4D_{N}$ and $4D_{C}$  with weights of $1/3$ each. 
Comparing the difference in optimal resulting designs from both the log and standard cases once again highlight the importance to know which error structure to use in an experiment.

This is even emphasized by looking at Table \ref{table.D,Ds-efficiencies}, which presents a comprehensive comparison of all the $D$ and $D_s$ designs of the standard and log case using relative $D$ and $D_s$ efficiencies. The upper part of the table are the efficiencies of all designs relative to the designs of the standard case, the lower parts are relative to the log case designs. We are using the symbol of $-$ to indicate that due to not having enough support points the information matrices are not full rank and therefore the designs are singular.

\begin{figure}[htp]
	
	\begin{minipage}{.5\linewidth}
		\centering
		\subfloat{\label{fig:variance(equivalence)D-optim:a}\includegraphics[scale=.5]{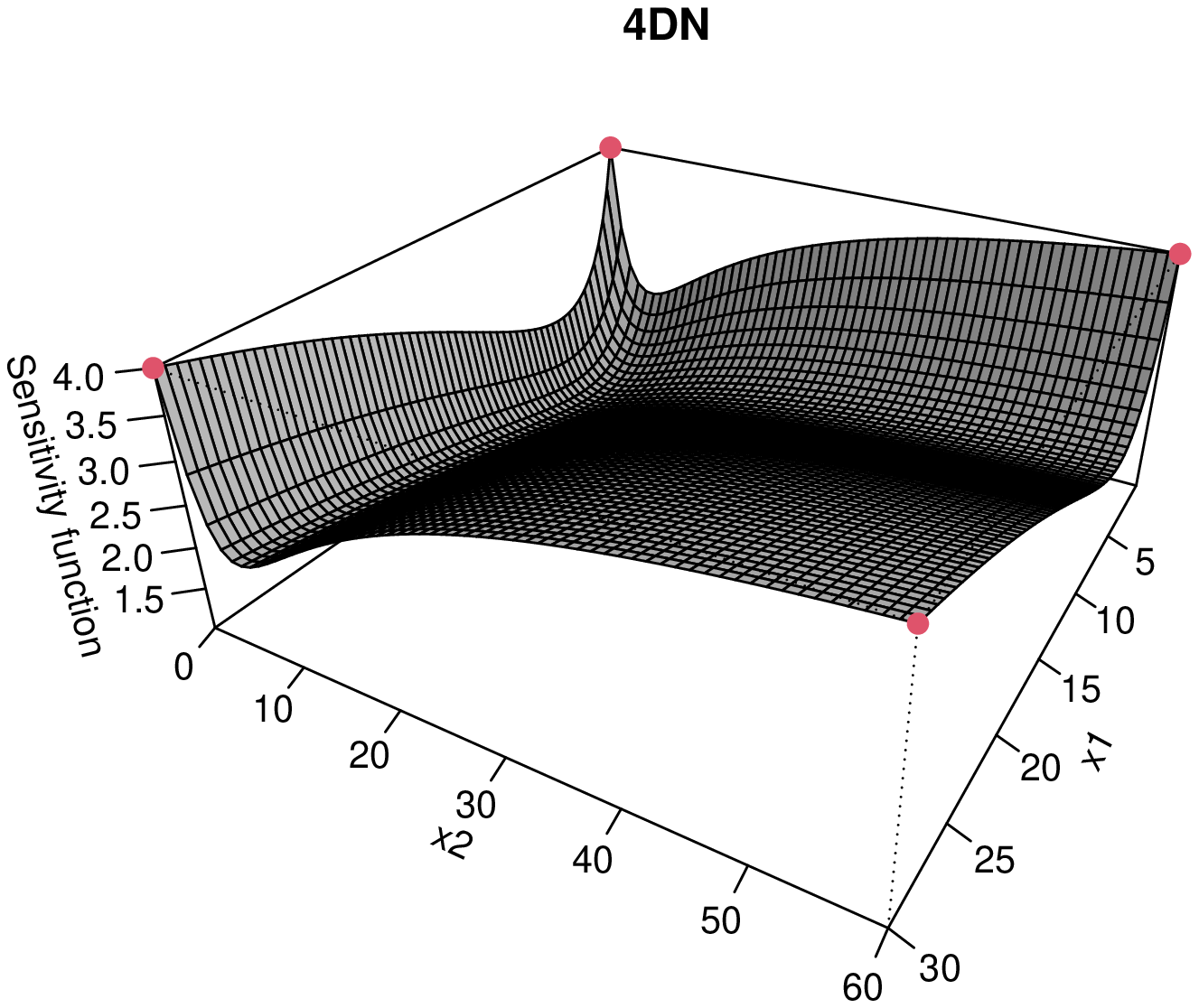}}
	\end{minipage}%
	\begin{minipage}{.5\linewidth}
		\centering
		\subfloat{\label{fig:variance(equivalence)D-optim:b}\includegraphics[scale=.5]{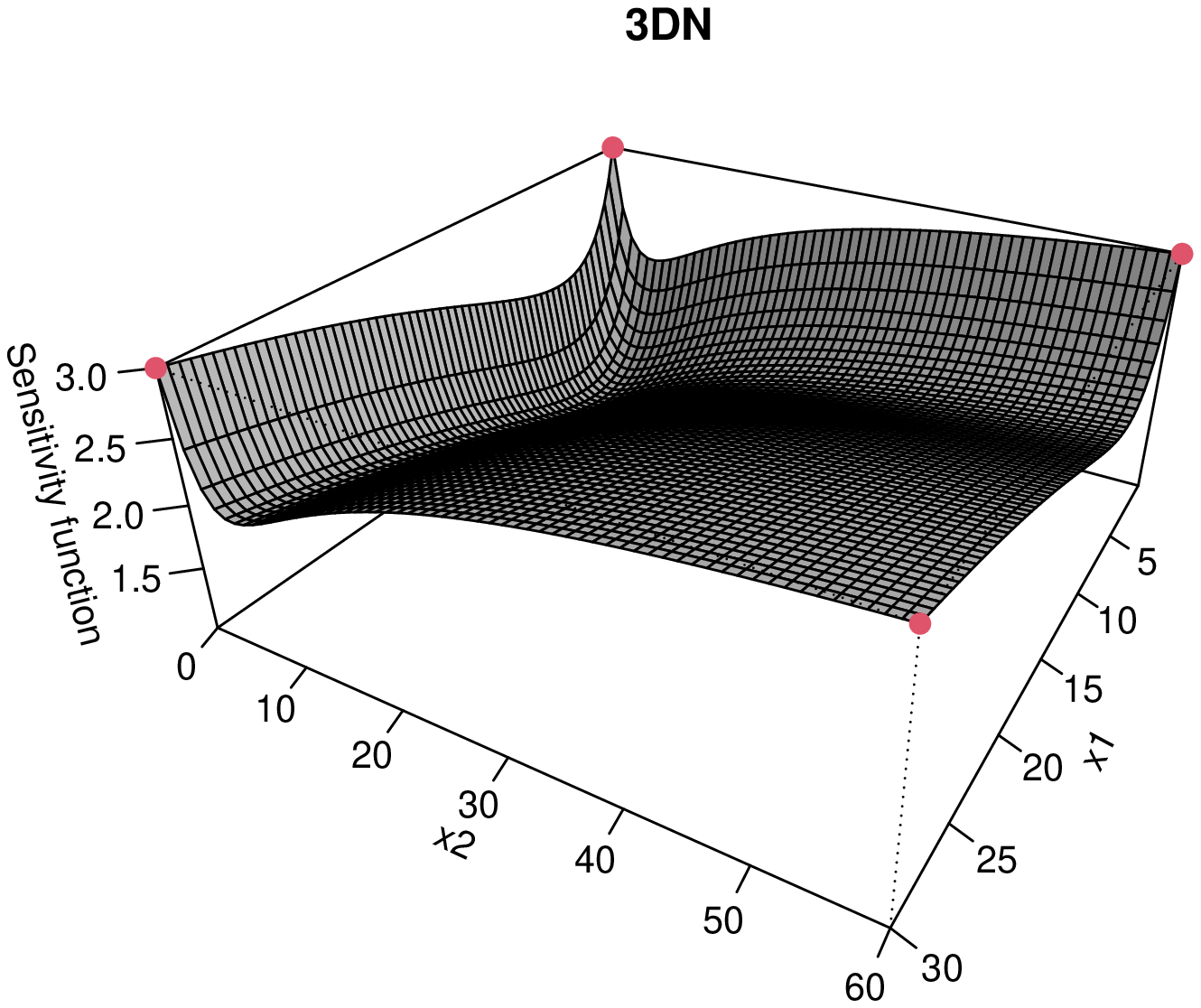}}
	\end{minipage}\par\medskip
	
	\begin{minipage}{.5\linewidth}
		\centering
		\subfloat{\label{fig:variance(equivalence)D-optim:c}\includegraphics[scale=.5]{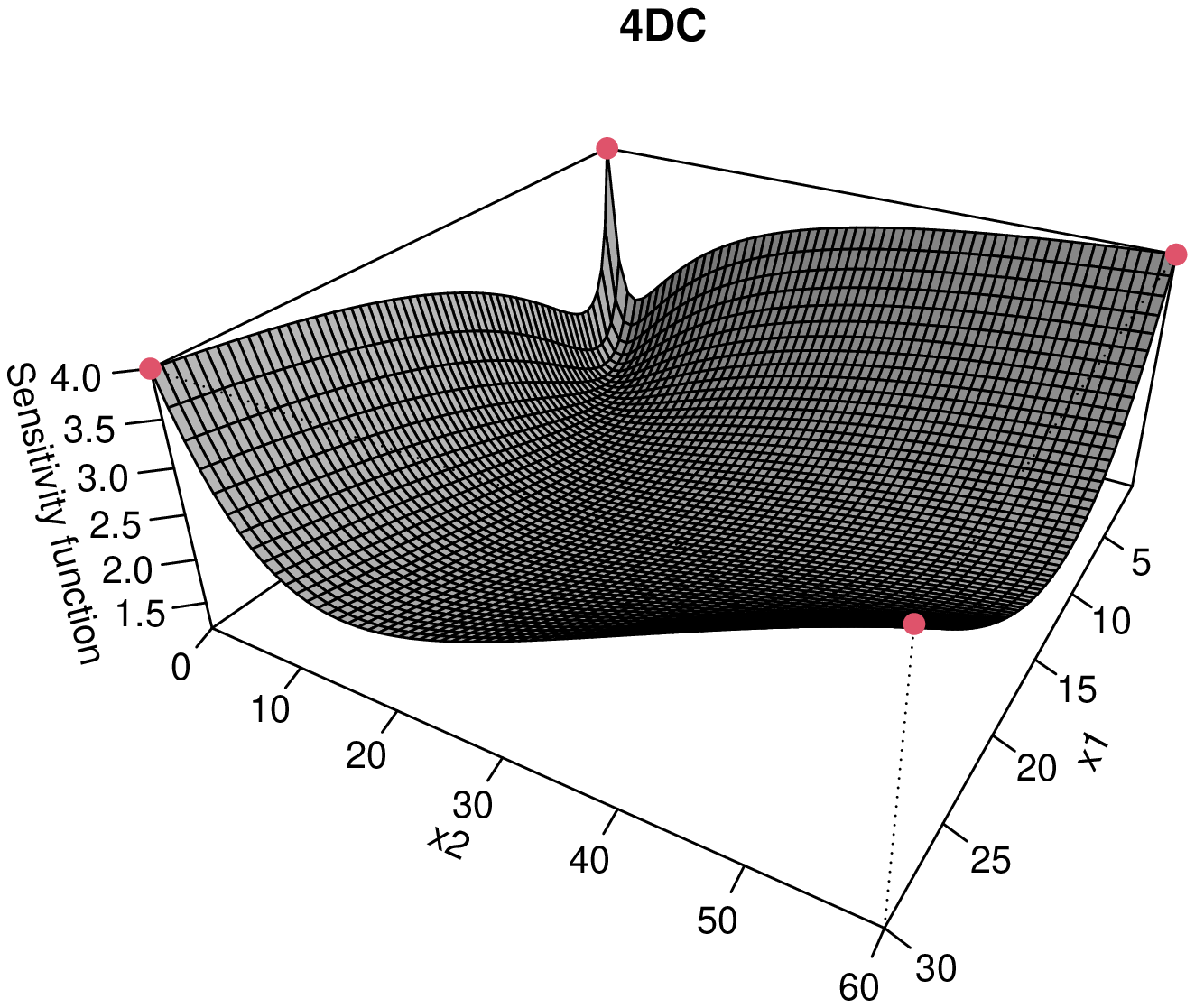}}
	\end{minipage}%
	\begin{minipage}{.5\linewidth}
		\centering
		\subfloat{\label{fig:variance(equivalence)D-optim:d}\includegraphics[scale=.5]{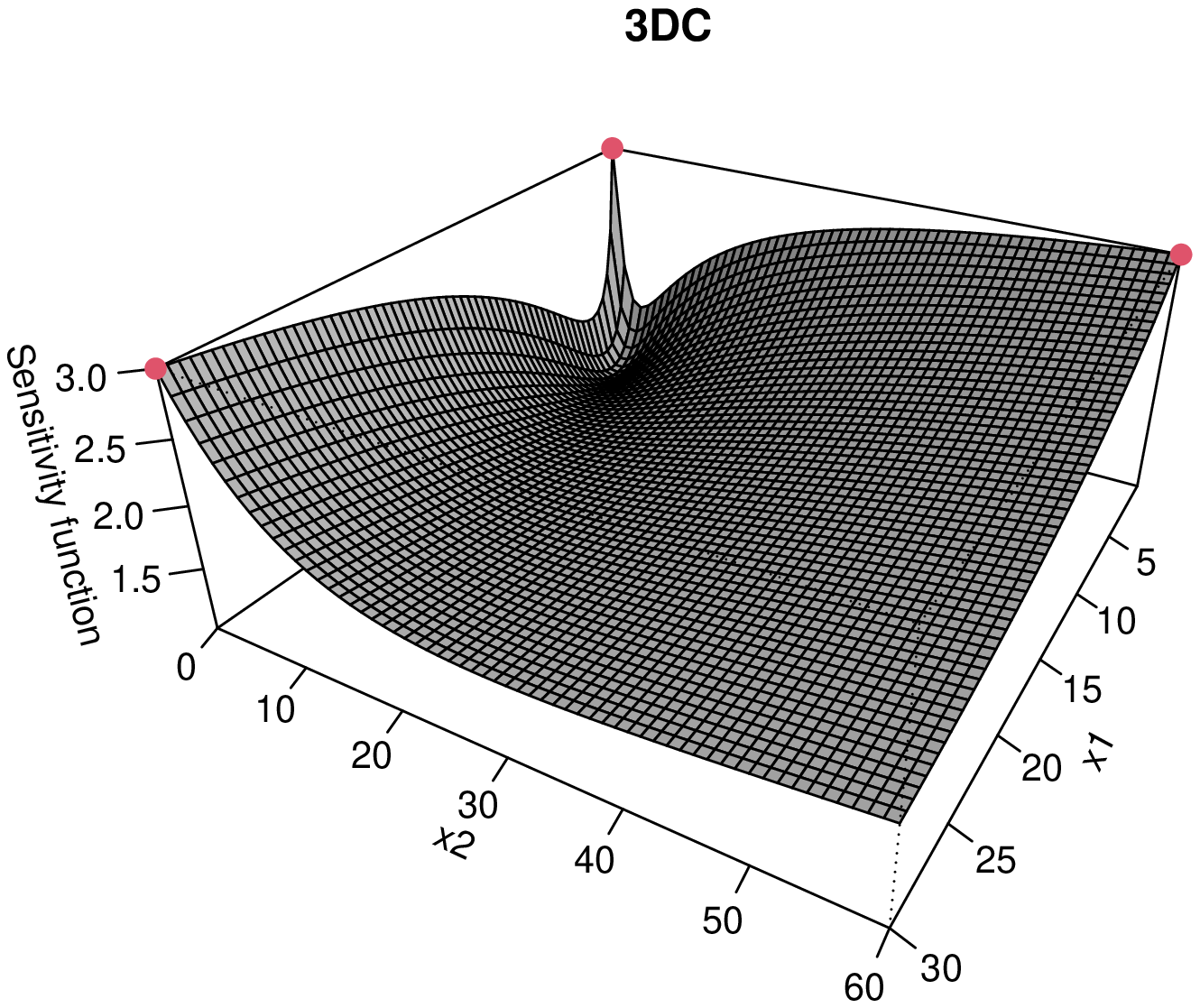}}
	\end{minipage}\par\medskip
	
	\begin{minipage}{\linewidth}
		\centering
		\subfloat{\label{fig:variance(equivalence)D-optim:e}\includegraphics[scale=.5]{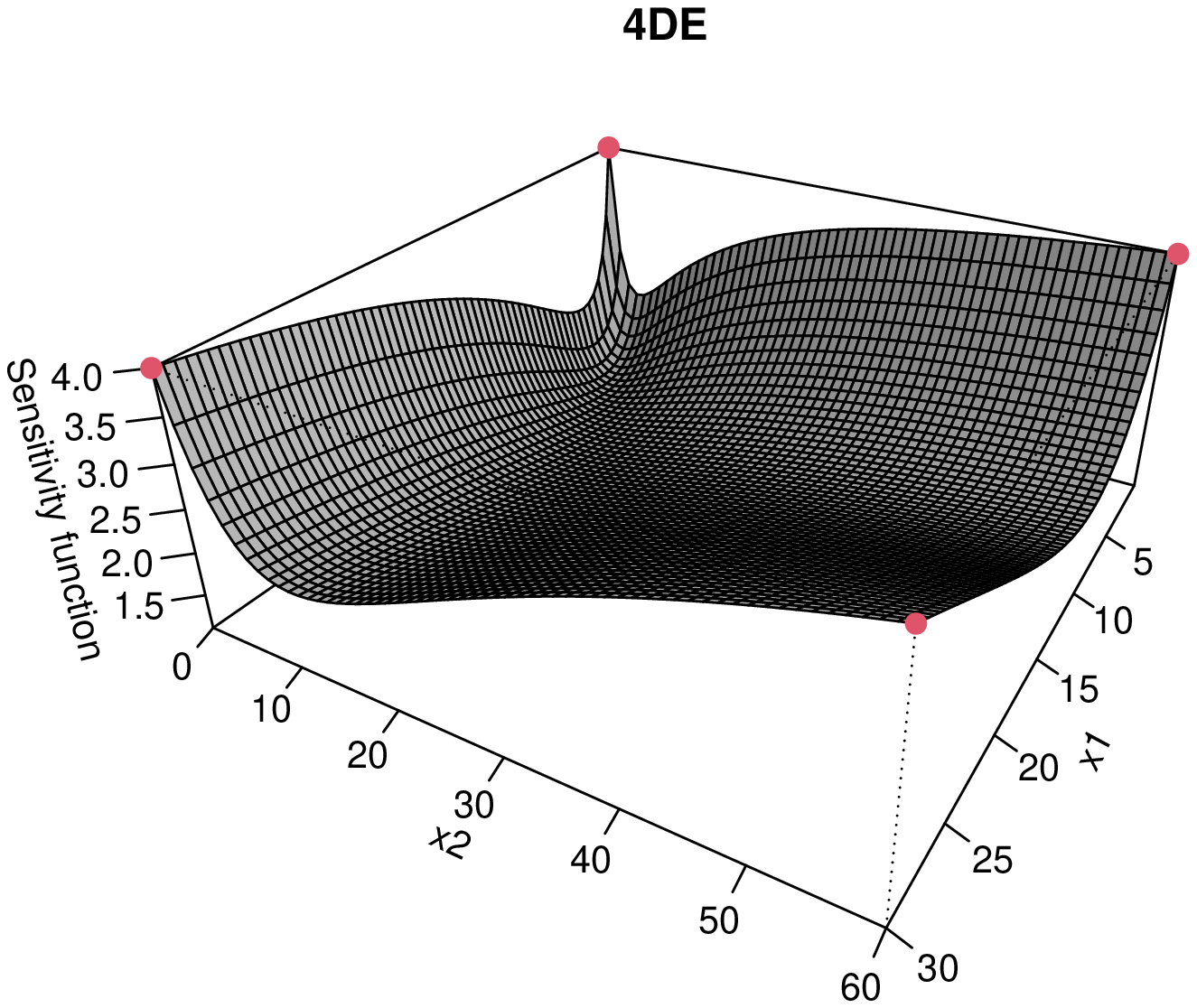}}
	\end{minipage}%
	\caption{{Plot of sensitivity function for D-optimal designs}}
	\label{fig:variance(equivalence)D-optim}
\end{figure}

\begin{figure}[htp]
	
	\begin{minipage}{.5\linewidth}
		\centering
		\subfloat{\label{fig:variance(equivalence)Ds-optim:c}\includegraphics[scale=.5]{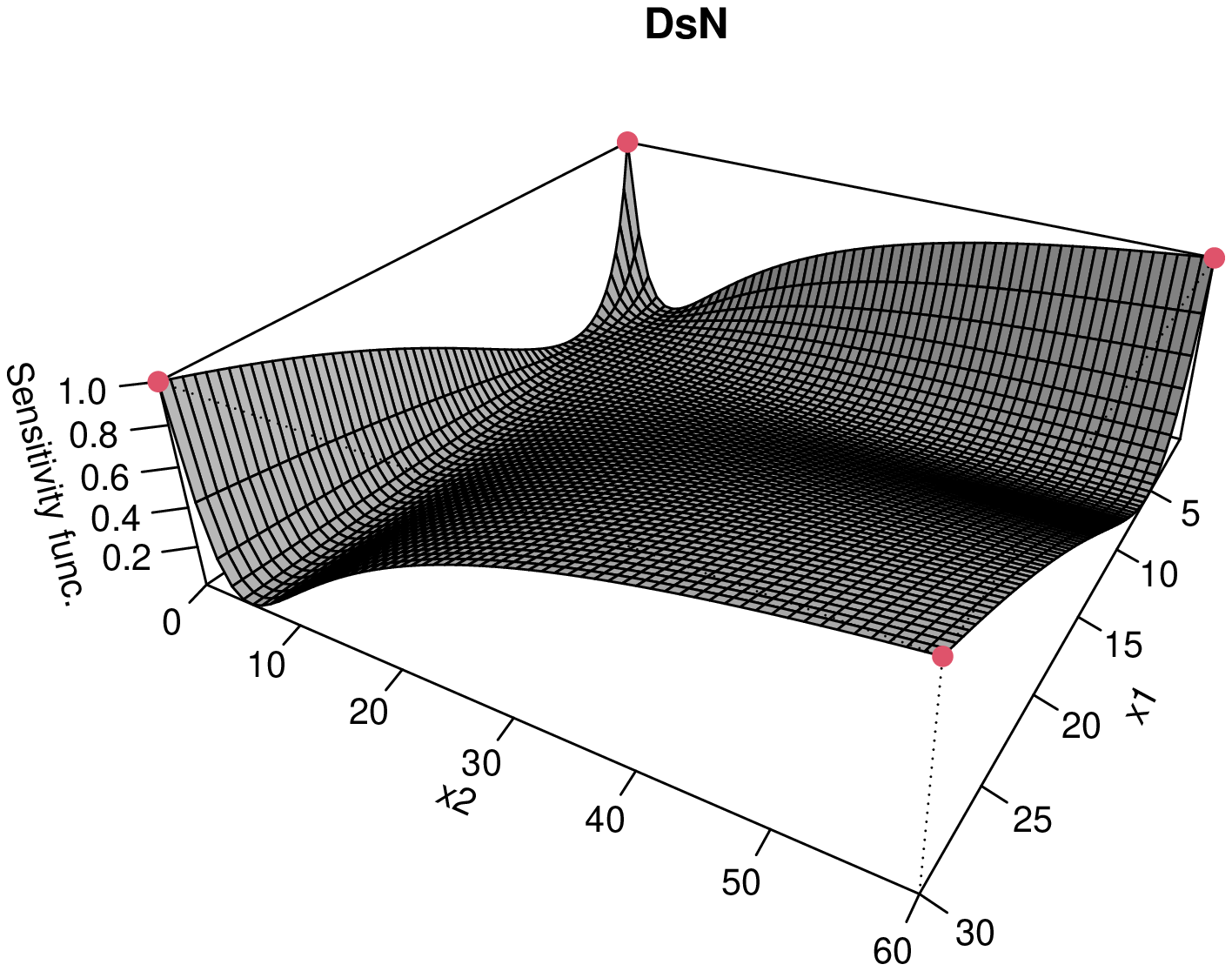}}
	\end{minipage}%
	\begin{minipage}{.5\linewidth}
		\centering
		\subfloat{\label{fig:variance(equivalence)Ds-optim:d}\includegraphics[scale=.5]{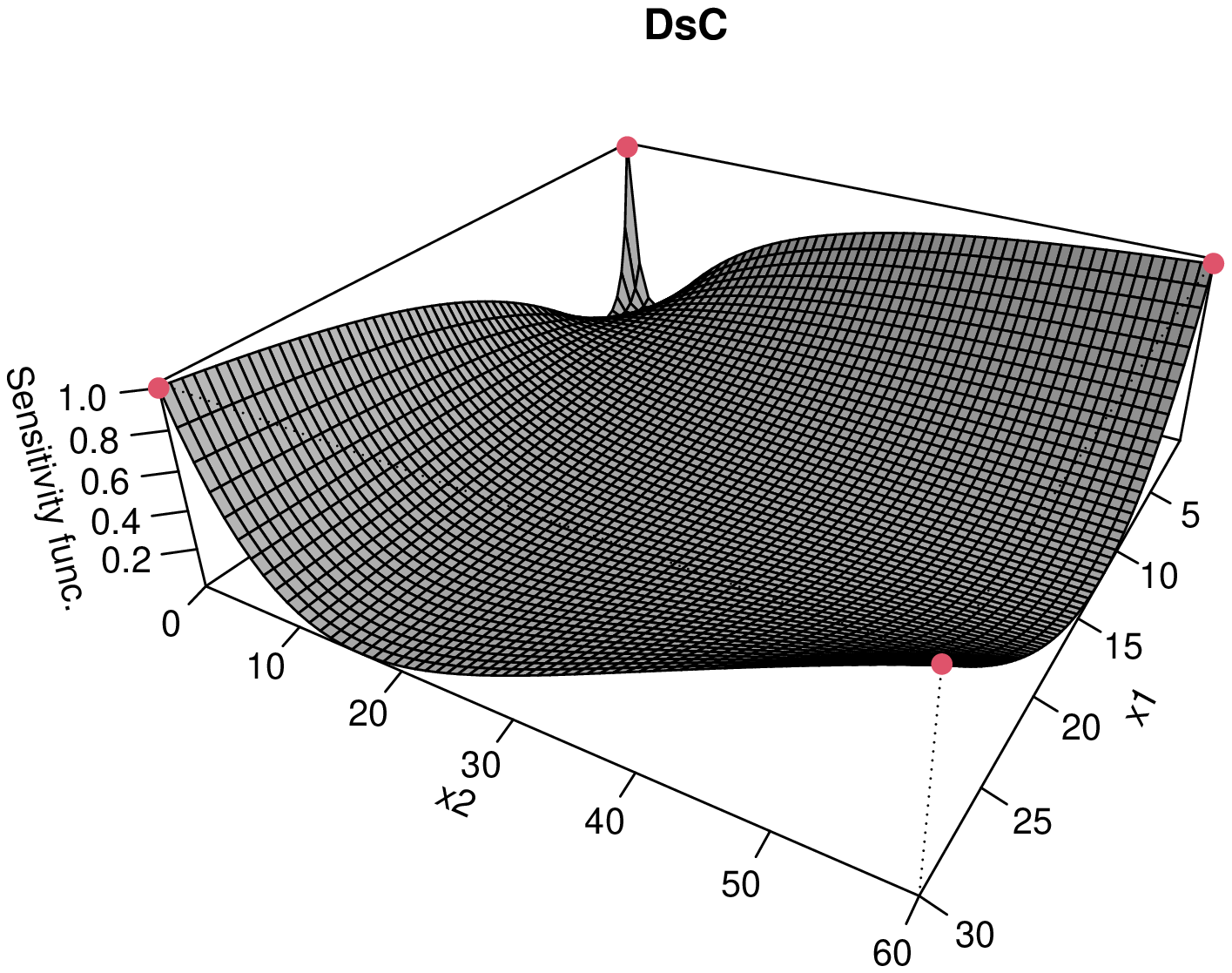}}
	\end{minipage}\par\medskip
	
	\caption{{Plot of sensitivity function for Ds-optimal designs}}
	\label{fig:variance(equivalence)Ds-optim}
\end{figure}

The following conclusions may be drawn from the table: 
\begin{itemize}
	\item Naturally, higher efficiencies are observed whenever similar cases are relatively compared; i.e. when designs of the standard case are relative to designs of the standard case or the designs of the log case are relative to the log case designs.
	\item For the case of the standard model considered as the reference (i.e. the model in the denominator of relative efficiency)
	we typically see that the efficiencies are always higher when the designs are compared to the encompassing rather than the pure models  (except for $Ds_{C}$ in the standard and log case compared to $4D_{C}$ and $3D_{C}$ in the lower part). For example, notice the $D$-efficiencies $100$ ($38.26$) and $87.01$ ($52.70$) in the first row of the table. The situation is exactly reverse for the log case. 
	
	\item Smaller efficiencies are observed when the log case designs are relative to standard designs and the other way around. Higher defects are observed in designs of the log case relative to the standard case designs. Notice the values in the last three rows of the upper part of the table with the first seven rows of the lower part.
\end{itemize}
The latter observation indicates that while the designs for the log case are robust to misspecification of nominal values they are much less so for misspecification of the error structure. It seems that when an experimenter is unsure about that it is much safer to use the additive normal error specification. 
\begin{center} 
	\begin{table}[htp] 
		\caption{The $D$ and $Ds$ efficiencies for all $D$ and $Ds$ designs}
		\label{table.D,Ds-efficiencies}
		\resizebox{\textwidth}{!}
		{
			\begin{small}
				\begin{tabular}{lclccccccccc}
					\hline 
					&&&&\multicolumn{8}{c}{Standard case, {reference model}}\\ 
					\cline{5-12}&&&&\multicolumn{5}{c}{Eff$_{D}(\%)$}&&\multicolumn{2}{c}{Eff$_{Ds}(\%)$}\\
					\cline{5-9}\cline{11-12}&&Design &&$4D_{N}$&$3D_{N}$&$4D_{C}$&$3D_{C}$&$4D_{E}$&&
					$Ds_{N}$&$Ds_{C}$\\
					\hline 
					Standard case
					&&$4D_{N}$&&$100$&$87.01$&$85.12$&$76.95$&$88.58$&&$72.19$&$47.37$\\
					&&$3D_{N}$&&$-$&$100$&$-$&$91.76$&$-$&&$-$&$-$\\
					&&$4D_{C}$&&$87.36$&$83.30$&$100$&$87.55$&$ 99.67$&&$47.91$&$61.25$\\
					&&$3D_{C}$&&$-$&$91.41$&$-$&$100$&$-$&&$-$&$-$\\
					&&$4D_{E}$&&$90.74$&$85.13$&$99.69$&$ 87.04$&$100$&&$52.25$&$61.57$\\
					&&$Ds_{N}$&&$78.04$&$56.08$&$63.67$&$51.63$&$66.46$&&$100$&$49.10$\\
					&&$Ds_{C}$&&$47.35$&$37.37$&$56.39$&$34.64$&$55.85$&&$45.78$&$100$\\ 
					\hline
					Log case
					&&$4D_{N},3D_{N},4D_{C},4D_{E},Ds_{N}$&&$0.70$&$2.76$&$0.49$&$4.69$&$0.50$&&$0.01$&$0.00$\\
					&&$3D_{C}$&&$-$&$0.08$&$-$&$0.07$&$-$&&$-$&$-$\\
					&&$Ds_{C}$&&$0.41$&$1.42$&$0.29$&$2.11$&$0.30$&&$0.00$&$0.00$\\
					
					\hline
					&&&&\multicolumn{8}{c}{Log case, {reference model}}\\ 
					\cline{5-12}&&&&\multicolumn{5}{c}{Eff$_{D}(\%)$}&&\multicolumn{2}{c}{Eff$_{Ds}(\%)$}\\
					\cline{5-9}\cline{11-12}&&Design &&$4D_{N}$&$3D_{N}$&$4D_{C}$&$3D_{C}$&$4D_{E}$&&
					$Ds_{N}$&$Ds_{C}$\\
					\hline
					Standard case
					&&$4D_{N}$&&$38.26$&$52.70$&$25.92$&$30.84$&$30.08$&&$14.63$&$6.96$\\
					&&$3D_{N}$&&$-$&$44.26$&$-$&$19.97 $&$-$&&$-$&$-$\\
					&&$4D_{C}$&&$35.81$&$52.82$&$27.98$&$30.88$&$30.86$&&$11.15$&$9.42$\\
					&&$3D_{C}$&&$-$&$56.22$&$-$&$31.38$&$-$&&$-$&$-$\\
					&&$4D_{E}$&&$36.04$&$52.32$&$27.43$&$30.33$&$30.56$&&$11.78$&$9.18$\\
					&&$Ds_{N}$&&$40.80$&$55.00$&$30.47$&$35.89$&$34.22$&&$16.65$&$8.44$\\
					&&$Ds_{C}$&&$24.43$&$31.82$&$21.20$&$17.57$&$22.48$&&$11.05$&$16.85$\\
					\hline
					Log case
					&&$4D_{N},3D_{N},4D_{C},4D_{E},Ds_{N}$&&$100$&$100$&$100$&$87.50$&$100$&&$100$&$67.55$\\
					&&$3D_{C}$&&$-$&$83.99$&$-$&$100$&$-$&&$-$&$-$\\
					&&$Ds_{C}$&&$58.72$&$80.32$&$58.72$&$37.75$&$58.72$&&$22.94$&$100$\\
					\hline
					\multicolumn{2}{c}{Note: $-$ Singular designs}	&&&&&&&&&&\\
				\end{tabular}
			\end{small}
		}
	\end{table}
\end{center}
To make sure that the $D$- and $D_s$-optimal designs of the log case in table \ref{table.D,Ds-designs-efficiencies} are actually optimum, we plotted the sensitivity functions, Eq. (\ref{G-criterion}) and Eq. (\ref{Ds-variance}) for them, respectively. Figure \ref{fig:variance(equivalence)D-optim} shows that all the $D$-optimal designs have the same \E{maximal value} equal to the number of their respective parameters being three or four and for all the other points in the design region the value of \E{the sensitivity function} is less than the maximum. Figure \ref{fig:variance(equivalence)Ds-optim} similarly shows that the \E{sensitivity} function for $D_s$-optimal designs have the same maximum equal to one. Note that the red dots in the figures represent the values of sensitivity functions for the optimal designs in each case.

\section{Optimal designs for model discrimination}\label{sec-discrimination}

In the previous part we found optimal designs for estimation of parameters of each model. If there exists more than one model (like the case we have) and there is uncertainty in which model to choose, we need to perform experiments to find optimal designs for discrimination as what also was discussed in section \ref{subsec:likelihood ratio test}.
Note that the $D_s$-optimal designs presented in the previous section can be used for model discrimination. As the encompassing model discriminates the competitive and the non-competitive model completely by the respective value of the parameter $\lambda$, it is natural that good estimation of $\lambda$ ensures good discriminability. However, note that there is actually a great range of possible encompassing model specification and that the chosen one is subject to considerable arbitrariness.  

\subsection{T-optimal designs}
Another widely used discrimination criterion is  $T$-optimality introduced by \cite{atkinson1975design}. Here, we maximize the non-centrality parameter of the $F$-test for departures from the wrong model when the assumption is to know which model is the true one with all its parameters to be known so that the resulting optimal design depends on the fixed (or true) parameters $\bar{\thetav}_0$ in the assumed true model and therefore will be locally optimum as well. In this context, we denote those two models as $\eta_0(\thetav_0,\xv)$ and $\eta_1(\thetav_1,\xv)$.  \E{Note that the subscripts zero and one here are just suggesting the assumed true and wrong models for which we will use $C$ and $N$ for competitive and noncompetitive models in computations exchangeably.} Therefore, by assuming the first model to be true, a design $\xi_{T0}^{*}$ would be called $T$-optimal if it maximizes the lack of fit sum of squares for the second model being defined as
\begin{align} \label{formula-Toptimal-delta0}
	\Delta_0(\xi)&=\sum_{i=1}^{n}\omega_i\left( \eta_0(\bar{\thetav}_0,\xv_i)- \eta_1(\hat{\thetav}_1,\xv_i)\right)^2 \nonumber\\
	&=\inf_{\thetav_1\in\Thetav_1}\sum_{i=1}^{n}\omega_i\left( \eta_0(\bar{\thetav}_0,\xv_i)- \eta_1(\thetav_1,\xv_i)\right)^2,
\end{align}
\E{where $\hat{\thetav}_1$ is the estimate derived from minimization of (\ref{formula-Toptimal-delta0}).} Let $\Xi$ be a set of all approximate designs. Then, the design $\xi_{T0}^{*}\in \Xi$ will be called $T$-optimal, if
\begin{equation}
	\xi_{T0}^{*}\in \arg\max_{\xi\in \Xi}\Delta_0(\xi),
\end{equation}

In order to compare any design to a $T$-optimum design $\xi_{T0}^{*}$ (when $\eta_0$ is assumed true), $T$-efficiency is defined as
\begin{equation} \label{T-efficiency}
	\text{Eff}_{T0}(\xi)=\dfrac{\Delta_0(\xi)}{\Delta_0(\xi_{T0}^{*})}.
\end{equation}
The same definitions hold when the $\eta_1$ is assumed to be the true model with the only difference that the indices in Eqs. (\ref{formula-Toptimal-delta0})-(\ref{T-efficiency}) are interchanged. \cite{atkinson2012optimum} introduced the so-called Compound $T$-optimal ($CT$-optimal) designs to discriminate between both models which maximize a weighed product of efficiencies as
\begin{equation} \label{formula-weighted-T-efficiency}
	\left\lbrace \text{Eff}_{T0}\right\rbrace ^{1-\nu }\left\lbrace \text{Eff}_{T1}\right\rbrace ^{\nu }=\left\lbrace\dfrac{\Delta_0(\xi)}{\Delta_0(\xi_{T0}^{*})} \right\rbrace ^{1-\nu }\left\lbrace\dfrac{\Delta_1(\xi)}{\Delta_1(\xi_{T1}^{*})} \right\rbrace ^{\nu }, 0\leq \nu \leq 1.
\end{equation}
Here $\nu$ is a weighting coefficient such that when $\nu=0$ we obtain $T$-optimal designs when $\eta_0$ in assumed true and $\nu=1$ for $\eta_1$, similarly. By taking the logarithms of the right hand side of Eq. (\ref{formula-weighted-T-efficiency}) and omitting the constant values the $CT$-criterion
is
\begin{equation}
	\Phi_{CT}(\xi)=(1-\nu)\ln\Delta_0(\xi)+\nu\ln\Delta_1(\xi),
\end{equation}
\E{which is a convex combination of two design criteria, each of which is the logarithm of that for $T$-optimality. Further, since $\ln\Delta_0(\xi)$ is a concave function of a concave design criterion, $CT$-criterion satisfies the conditions of convex optimum design theory and therefore the equivalence theorem applies (\citealp{atkinson2012optimum}).} \cite{atkinson1975design} obtained an analogous of $D$-equivalence theorem to provide a check of $T$-optimal designs. Here we represents the general case \E{for $CT$-optimal designs} which is taken similarly from the results of \cite{atkinson2008dt} and works for any value of $\nu$ including $T$-optimal designs for $\nu=0$ and $\nu=1$ as
\begin{enumerate}
	\item \E{A necessary and sufficient condition for a design $\xi_{CT}^{*}$ to be $CT$-optimal is fulfillment of the inequality}
	$$\Psi_{CT}(\xv,\xi_{CT}^{*}) \leq 1, \quad \xv \in \mathfrak{X}, $$ with the sensitivity functions
	$\Psi_{CT}(\xv,\xi)=\left( 1-\nu\right) \dfrac{\Psi_{0}(\xv,\xi)}{\Delta_0(\xi)}+\nu \dfrac{\Psi_{1}(\xv,\xi)}{\Delta_1(\xi)}$,\\
	$\Psi_{0}(\xv,\xi)=\left( \eta_0(\bar{\thetav}_0,\xv)- \eta_1(\hat{\thetav}_1,\xv)\right)^2, \Psi_{1}(\xv,\xi)=\left( \eta_1(\bar{\thetav}_1,\xv)- \eta_0(\hat{\thetav}_0,\xv)\right)^2 $;
	\item at the points of the optimum design $\Psi_{CT}(\xv,\xi_{CT}^{*})$ achives its upper bound that is $\Psi_{CT}(\xv_{i}^{*},\xi_{CT}^{*})$;
	\item for any non-optimum design $\xi$, that is a design for which $\Phi_{CT}(\xi)<\Phi_{CT}(\xi_{CT}^{*})$,\\\\ $\sup_{\xv\in\mathfrak{X}}\Psi_{CT}(\xv,\xi)>1.$ 
\end{enumerate}

\E{Similar to} \cite{atkinson2012optimum}, we computed four approximate discriminating designs denoted here by $A_1$-$A_4$ also for the log case which are presented in the left hand part of table \ref{table.4optimaldes-log}. $A_1$ corresponds to a $T$-optimal design when the non-competitive model (\ref{formula-noncomp-model}) is assumed to be true. The estimates of parameters in the log case of the right hand side of table \ref{table.est-se-log1-2} are used as nominal parameter values. $A_2$ corresponds to a $CT$-optimal designs for $\nu=0.5$. We used the corresponding estimates in the table \ref{table.est-se-log1-2} as nominal  parameter values in each section of the compound criterion. $A_3$ is the $D_s$ optimal design for the discrimination parameter $\lambda$ in model (\ref{formula-combined-model}) at a nominal value of $\lambda=0.8737$. The estimates of parameters in table \ref{table.est-se-log3} are used as nominal values for the linearization. The last design $A_4$ refers to a $T$-optimal design when the competitive model (\ref{formula-comp-model}) is assumed to be the true one. The estimates of parameters in the log case of left hand side of table \ref{table.est-se-log1-2} are used as nominal values. The right hand part of table \ref{table.4optimaldes-log} corresponds to recalculations of Atkinson's designs for the standard case. Again some discrepancies are observed in the optimal designs of the standard case here, compared to the values reported in \cite{atkinson2012optimum} due to the differences in the nominal values for the parameters and the design space and accordingly some differences have occurred in the $T$-efficiencies.
\begin{table}[htp] 
	\caption{{Some} optimal discriminating designs and their $T$-efficiencies }
	\label{table.4optimaldes-log}
	\begin{center}
		\resizebox{\textwidth}{!}{
			\begin{small}
				\begin{tabular}{lccccccccccccc}
					\hline
					\multicolumn{6}{c}{ Log-case (Eq. \ref{formula-logmodel} )}&&&\multicolumn{6}{c}{ Standard case (Eq. \ref{formula-generalmodel} ) }
					\\
					\cline{1-6} \cline{9-14}&&&&\multicolumn{2}{c}{{\small Eff$_T (\%)$}}&&&&&&&\multicolumn{2}{c}{{\small Eff$_T (\%)$}} \\
					\cline{5-6} \cline{13-14}
					{\rm Design}&$x_S$&$x_I$&$\omega$&$A_1$&$A_4$&&&{\rm Design}&$x_S$&$x_I$&$\omega$&$\nu=0$&$\nu=1$\\ 
					\hline 
					$A_1$&$\varepsilon$&  $0$ & $0.0095$&$100$ &$76.50$&&&$\nu=0$&$30.000$&$0.000$&$0.063$&$100$&$57.40$ \\      
					&$30$ &  $0$ & $0.1402$&&&&&&$3.214$&$0.000$&$0.063$  \\  
					&$\varepsilon$&  $60$& $0.3600$&&&&&&$30.000$&$21.413$&$0.310$  \\
					&$30$ &  $60$& $0.4903$ &&&&&&$5.625$&$11.152$&$0.564$ \\
					\hline
					
					$A_2$&$\varepsilon$&  $0$ & $0.1688$&$74.89$&$89.72$  &&&$\nu=0.5$&$30.000$&$0.000$&$0.058$&$86.52$&$80.52$\\     
					&$30$ &  $0$ & $0.1818$ &&&&&&$3.348$&$0.000$&$0.189$ \\  
					&$\varepsilon$&  $60$& $0.3002$&&&&&&$30.000$&$22.082$&$0.260$  \\
					&$30$ &  $60$& $0.3492$ &&&&&&$5.759$&$11.375$&$0.493$ \\
					\hline

					$A_3$&$\varepsilon$&  $0$ & $0.1633$ &$72.96$&$92.40$  &&&$\lambda=0.9636$&$30.000$&$0.000$&$0.052$&$90.37$&$77.01$\\     
					&$30$ &  $0$ & $0.2189$ &&&&&&$2.678$&$0.000$&$0.137$ \\  
					&$\varepsilon$&  $60$& $0.2811$ &&&&&&$30.000$&$25.874$&$0.330$ \\
					&$30$ &  $60$& $0.3367$ &&&&&&$6.562$&$8.922$&$0.481$ \\
					\hline
					
					$A_4$&$\varepsilon$&  $0$ & $0.2500$&$ 57.85$&$ 100$ &&&$\nu=1$&$30.000$&$0.000$&$0.060$&$76.93$&$100$ \\     
					&$30$ &  $0$ & $0.2502$&&&&&&$3.080$&$0.000$&$0.250$  \\  
					&$\varepsilon$&  $60$& $0.2500$&&&&&&$30.000$&$22.751$&$0.250$  \\
					&$30$ &  $60$& $0.2498$&&&&&&$5.491$&$11.598$&$0.440$  \\
					\hline
					
				\end{tabular}
			\end{small} 
		}
	\end{center}
\end{table}
As we can observe from table \ref{table.4optimaldes-log} again all the support points of the log case designs are the same and on the corners of the design space and the difference between them is only due to their corresponding weights $\omega$. We mention that we used the Fedorov-Wynn algorithm (\cite{atkinson2007optimum}) to find the optimal designs $A_1$, $A_2$ and $A_4$ and set the maximum iteration equal to a fixed number (sufficiently large to ensure convergence of designs) as the stopping rule of the algorithm for all designs. 
Efficiencies are relatively high in both the log and standard cases \E{whether we assume $A_1$ or $A_4$ as the reference designs in the denominator of the Eff$_T (\%)$}. So we may state that the product of efficiencies are high for all the computed designs specifically for both $A_2$ and $A_3$ in the log case and $\nu=0.5$ and $\lambda=0.9636$ in the standard case, regardless of which model holds in comparisons, if we exclude the cases where we require the assumption of knowing the true fixed models in discrimination. \E{This provides a better interpretation if we are not interested in assuming any of the competitive or noncompetitive models to be the true models and accordingly the designs $A_2$ and $A_3$ provide higher efficiencies with this interpretation/assumption. }

\begin{figure}[htp]
	
	\begin{minipage}{.5\linewidth}
		\centering
		\subfloat{\label{fig:sensitivityT:a}\includegraphics[scale=.5]{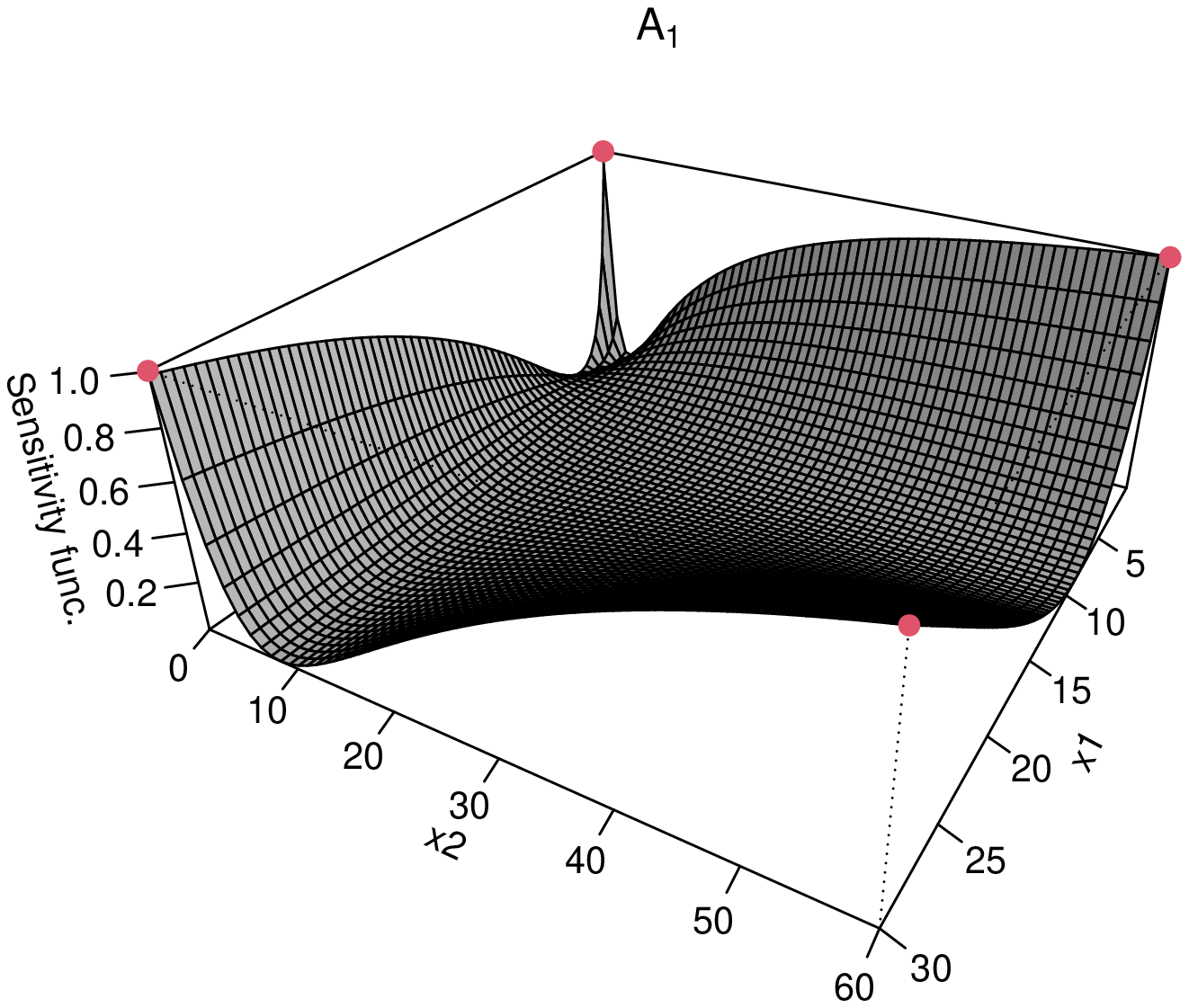}}
	\end{minipage}%
	\begin{minipage}{.5\linewidth}
		\centering
		\subfloat{\label{fig:sensitivityT:b}\includegraphics[scale=.5]{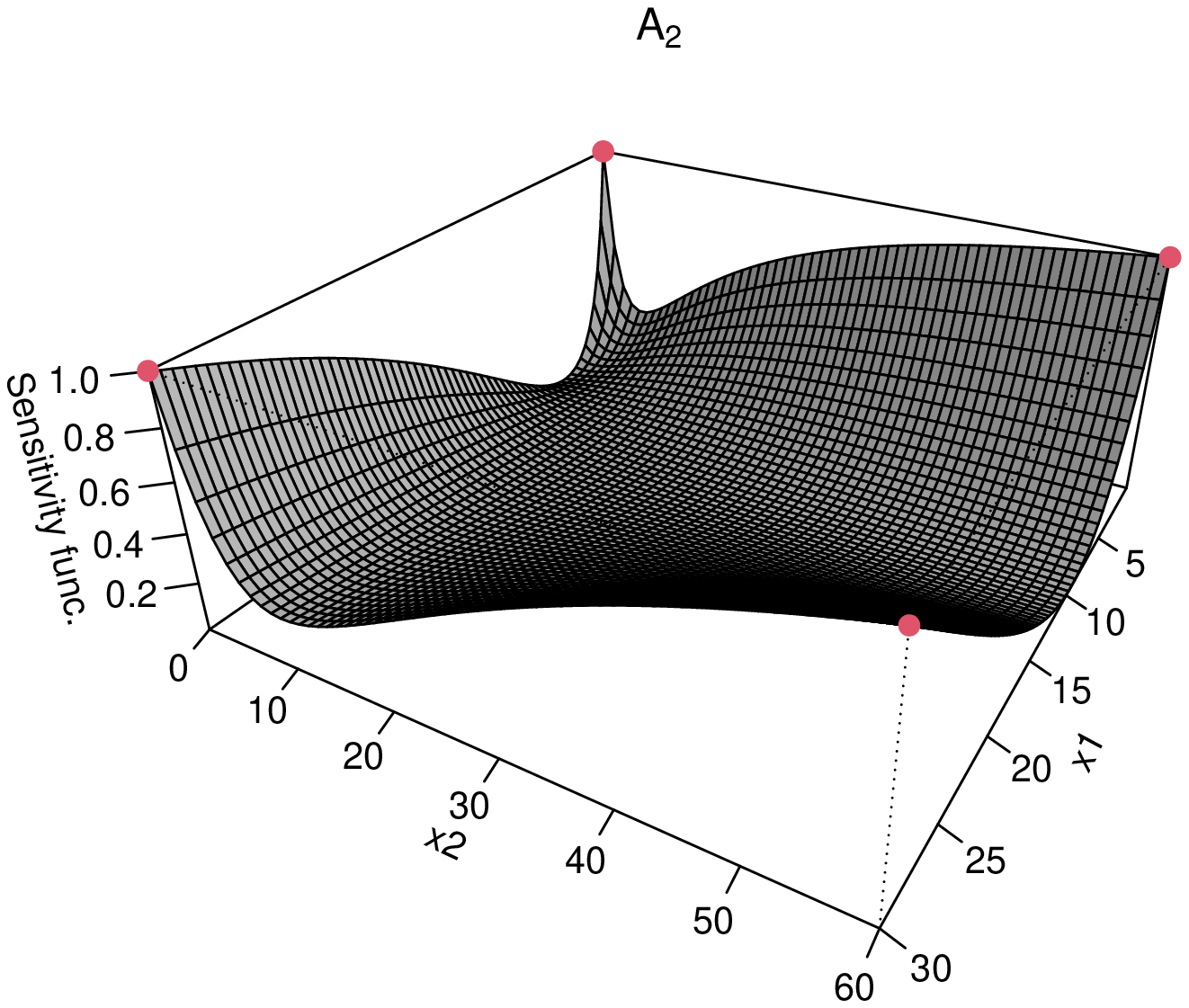}}
	\end{minipage}\par\medskip
	
	\begin{minipage}{.5\linewidth}
		\centering
		\subfloat{\label{fig:sensitivityT:c}\includegraphics[scale=.5]{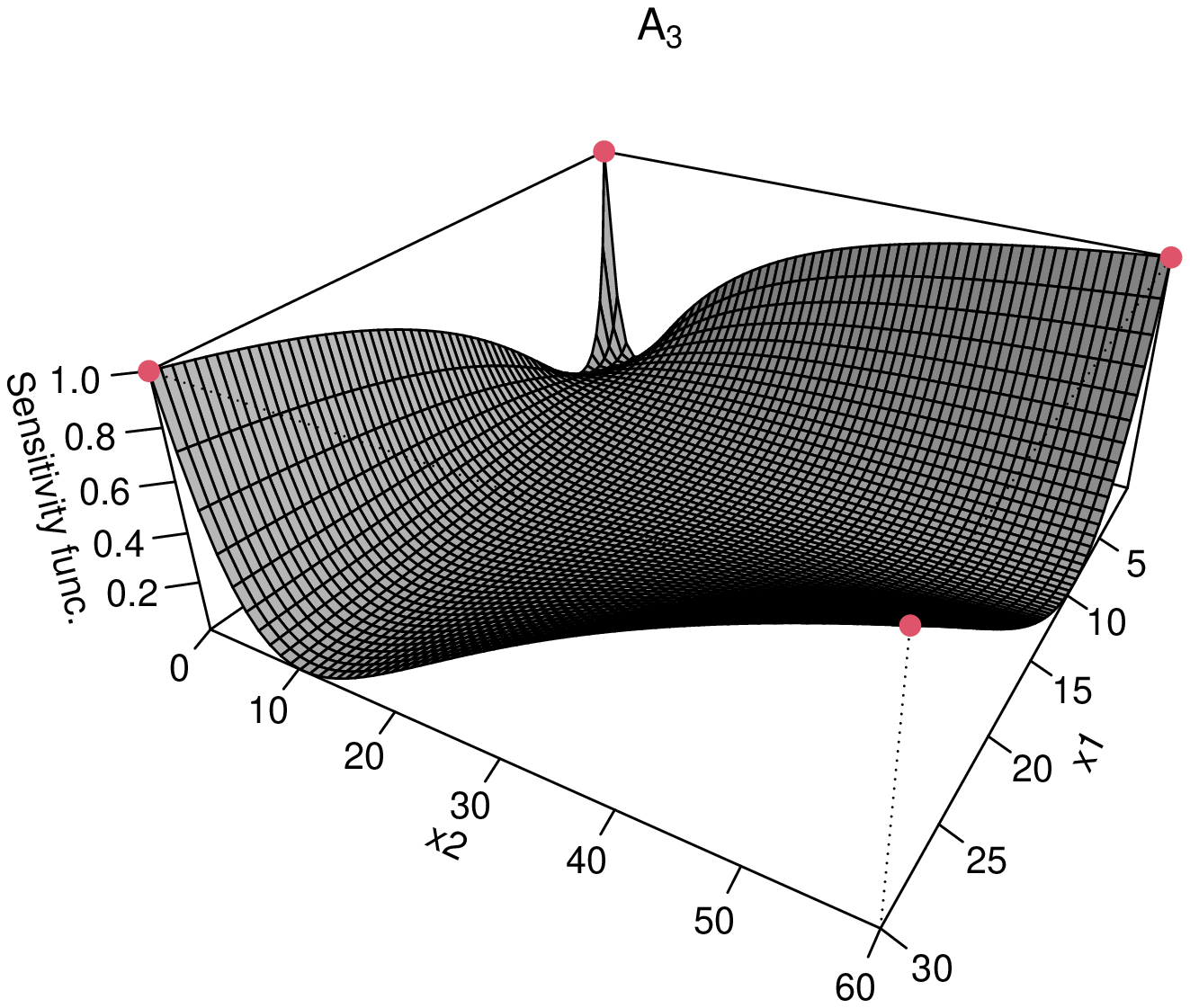}}
	\end{minipage}%
	\begin{minipage}{.5\linewidth}
		\centering
		\subfloat{\label{fig:sensitivityT:d}\includegraphics[scale=.5]{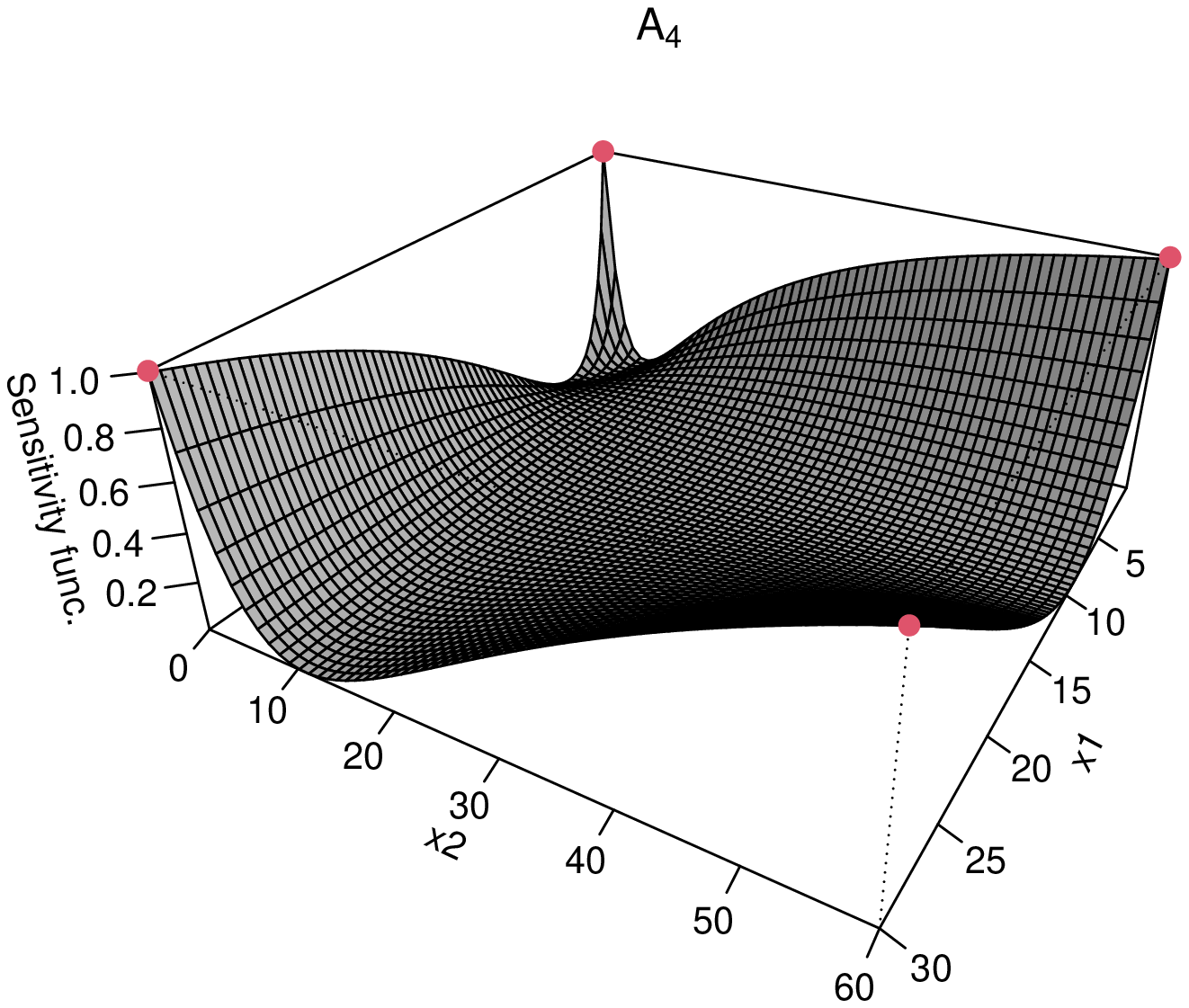}}
	\end{minipage}\par\medskip
	
	\caption{{Plot of sensitivity functions for $A_1$,$A_2$,$A_3$ and $A_4$}}
	\label{fig:sensitivityT}
\end{figure}

The sensitivity functions for $A_1$-$A_4$ are plotted in figure \ref{fig:sensitivityT} as an illustration of the equivalence theorems for $CT$- and $D_s$-optimal designs. As we can observe from the figure \ref{fig:sensitivityT} maximum value of the sensitivity functions are the same for all optimal designs and equal to one and for all the other non optimum designs in the design region, value of the sensitivity functions are below the maximum.

\subsection{$\delta$-optimal designs}
The last discrimination procedure used here is $\delta$-optimality introduced by \cite{harman2020design}. The method is a genuinely symmetric design criterion which is defined to discriminate between two statistical models of the form $\eta_u(\thetav_u,\xv_i)=\eta_{u}$ for $u=0,1$ \E{and $i=1,\dots,N$} with the same number of parameters $m$. \E{Note the we denote the size of exact designs here, by $N$, equal to the number of observations, since we allow replications in exact designs.} The idea of the method is to linearize both models at their respective nominal values, denoted by $\tilde{\thetav}_u$. Therefore the linearized models are 
\begin{equation*}
	(y_i)_{i=1}^{N}\approx \mathbf{F}_u(\mathcal{D})\thetav_{u}+\mathbf{a}_u(\mathcal{D})+\epsilon,\quad u=0,1,
\end{equation*}
where $\mathcal{D}=(\xv_1, \dots, \xv_N)$ is an exact design of size $N$ and $\mathbf{F}_u(\mathcal{D})$ is the $N\times m$ matrix with ith row $f^{T}(\xv_i,\tilde{\thetav}_{u})$ similarly computed from Eq. (\ref{formula-linearized model}). Further $\mathbf{a}_u(\mathcal{D})$ is the $N$ dimensional vector as
\begin{equation*}
	\mathbf{a}_u(\mathcal{D})=(\eta_u(\tilde{\thetav}_{u},\xv_i))_{i=1}^{N}-\mathbf{F}_u(\mathcal{D})\tilde{\thetav}_u.
\end{equation*}
According to above notations, the linearized distance criterion is (see \cite{harman2020design} for more details) 
\begin{align} \label{formula-linearized-distance-criterion}
	\delta(\mathcal{D})=\inf_{\thetav_{0}\in\tilde{\Thetav}_{0},\thetav_{1}\in\tilde{\Thetav}_{1}}\delta(\mathcal{D}\mid\thetav_{0},\thetav_{1}). \\
	\delta(\mathcal{D}\mid\thetav_{0},\thetav_{1})&=\lVert \mathbf{a}_0(\mathcal{D})+\mathbf{F}_0(\mathcal{D})\thetav_{0}-\left\lbrace \mathbf{a}_1(\mathcal{D})+\mathbf{F}_1(\mathcal{D})\thetav_{1}\right\rbrace \lVert , \nonumber
\end{align}
where $\tilde{\Thetav}_{0}\subseteq \mathbb{R}^{m},\tilde{\Thetav}_{1}\subseteq \mathbb{R}^{m}$ are called the flexible nominal sets which will not be considered fixed like the parameter spaces $\Thetav_{0}$ and $\Thetav_{1}$. Further the $\delta$-criterion, defined as a function of the exact design $\mathcal{D}$, is represented using the counting measure $\zeta$ on $\mathfrak{X}$ as 
\begin{equation*}
	{\zeta}\left( \left\lbrace \xv \right\rbrace \right):= \# \left\lbrace i\in\left\lbrace 1, \dots, N\right\rbrace : \xv_i=\xv \right\rbrace,  \xv \in \mathfrak{X}.
\end{equation*}
\E{where $\zeta$ here is the collection of exact designs of size $N$ with integer replications compared with $\xi$ where it refers to probability measures and continuous weights in the approximate case, as discussed in section \ref{sec-D-optimality}. Further, for the discussion on convexity of $\delta$-criterion see \cite{harman2020design}. } Finally for a set $\mathfrak{D}$ of all $N$-point designs, a design $\mathcal{D}^{*}\in \mathfrak{D}$ will be called $\delta$-optimal, if
\begin{equation}
	\mathcal{D}^{*}\in \arg\max_{\mathcal{D}\in\mathfrak{D}}\delta(\mathcal{D})
\end{equation} 
We need to emphasize that $\delta$-optimal designs are evaluated using the rapid and stable method for bounded variable least squares implemented in {\sf R} package {\sf bvls} 
\E{(see 
	\cite{stark1995bounded} and \cite{bvls2013mullen}
	)}. Therefore for implementation purposes  $\delta^{2}(\mathcal{D}\mid\thetav_0,\thetav_1)$ is used as
\begin{equation}
	\delta^{2}(D\mid\thetav_0,\thetav_1)=\lVert \left\lbrace \mathbf{a}_0(\mathcal{D})-\mathbf{a}_1(\mathcal{D})\right\rbrace -\left[-  \mathbf{F}_0(\mathcal{D}),\mathbf{F}_1(\mathcal{D})\right] \thetav \lVert^{2},
\end{equation}
\E{where $\thetav$ is the compound vector of unknown parameter vectors in both models.} For computation of $\delta$-optimal designs we used the standard KL-exchange heuristic \E{(\cite{atkinson2007optimum})}. The nominal values are chosen to be $\tilde{\thetav}_u=\hat{\thetav}_u$ and the nominal intervals are specifically chosen as $\tilde{\Thetav}_u=[ \tilde{\theta}_{u1}\pm r\tilde{\sigma}_{u1}] \times [ \tilde{\theta}_{u2}\pm r\tilde{\sigma}_{u2}] \times [ \tilde{\theta}_{u3}\pm r\tilde{\sigma}_{u3}]_{u=0,1}$ in which $\tilde{\theta}_{uv}=\hat{\theta}_{uv}$ and $\tilde{\sigma}_{uv}=\hat{\sigma}_{uv}$ for $u=0,1$ and $v=1,2,3$ \E{($\hat{\theta}_{uv}$ are basically the estimates of parameters of the models)}. Note that $r\geq0$ works as a tuning parameter which is specialized to change the size of nominal intervals and plays an important role in computation of the $\delta$-optimal designs. Therefore, we denote by $\delta_r$ a $\delta$-optimal design for a specific value of $r$.

Returning to our example, we would like to compute $\delta$-optimal designs for models (\ref{formula-comp-model}) and (\ref{formula-noncomp-model}) in the log case. According to Table \ref{table.est-se-log1-2} for initial estimates of the log cases, $r\in \left\lbrace 1,2,3,4 \right\rbrace $ higher values of which cause some or all values in the lower bounds of nominal intervals become negative. Therefore to fulfill this constraint, we used three alternatives to prevent having negative nominal intervals for values of $r$ more than $r>4$. The first alternative $a)$ was to increase $r$ and cut the lower bounds of the nominal intervals at zero wherever they are negative. The second $b)$ was to add the absolute values of negative lower bounds of the nominal intervals, cut at zero, into theirs upper bounds (shifting the upper bounds). For the third alternative $c)$, we used the remark below to arrive at positive intervals for the estimates of parameters.\\
\textbf{Remark 1:} Assume that the asymptotic distribution of the estimate of each parameter is normal $\sqrt{\E{N}}\left( \hat{\theta}-\theta\right) \sim \mathcal{N}(0,\sigma_{\theta}^2)$, under mild regularity conditions \E{in} \cite{lehmann1988theory}. Then implementing logarithmic transformations and the Delta method we can arrive at the asymptotic normal distribution of $\ln(\hat{\theta})$ as 
\begin{equation*}
	\sqrt{\E{N}}\left( \ln(\hat{\theta})-\ln(\theta)\right) \sim \mathcal{N}\left(0,\frac{\sigma_{\theta}^2}{\theta^2}\right).
\end{equation*}
Now, the asymptotic $100(1-\alpha)\%$ confidence interval for $\ln(\theta)$ is
\begin{equation*}
	\ln(\hat{\theta}) \pm z_{\frac{\alpha}{2}}\dfrac{\widehat{\sigma_{\theta}}}{\hat{\theta}}\equiv (L,U).
\end{equation*}
where $\widehat{\sigma_{\theta}}=\sqrt{\widehat{\text{Var}(\hat{\theta})}}$. Eventually using the inverse logarithmic transformation, an asymptotic $100(1-\alpha)\%$ confidence interval for each $\theta$ can be obtained as
\begin{equation*}
	({\rm e}^L, {\rm e}^U).
\end{equation*}
All above mentioned alternatives, denoted by the indices $a,b$ and $c$ respectively, are used to compute $\delta$-optimal designs for different values of $r$. 


\subsection{A simulation study of discriminating designs}
In this part we designed two experiments to compare discriminatory power of all discriminating methods of this section, first in a small scale and second in a large scale experiment. These experiments reflect the real discriminatory power of the designs resulting from different methods. The consequences would guide the experimenters, willing to work with log models of enzyme inhibitions, a path on which discriminating method to choose in practical situations.

\subsubsection{Exact designs, $N=6,7,8,9$}\label{subsubsec:exact:designs}
Since the designs of Table \ref{table.4optimaldes-log} have varying weights compared together and therefore they have different number of replications while rounding into exact ones, and also in order to observe how the designs will behave while their size changes, we designed experiments for $N=6,7,8,9$ in first part of the simulations. \E{Note that in all simulations studies of this part and the later parts, $N$ denoting the size of exact designs, is equal to the sample size (number of observations) at each step of Monte Carlo simulations and the goal of these parts is to compare the discriminatory power of all discriminating criteria using the exact designs resulted from section \ref{sec-discrimination} so far. In the case of $T$, $CT$ and $D_s$ criteria, the approximate designs of table \ref{table.4optimaldes-log} are rounded into their nearest integers depending on the size of exact design.} Therefore, we computed average values of correct classification (hit) rates when both models contribute equally in simulations presented in Table \ref{table.mean of hit rates n=6-9}. Note that the first support point of the design $A_1$ in Table \ref{table.4optimaldes-log} will not contain replications in its exact design for $N=6,7,8,9$ due to its very low weight, $\omega=0.0095$. Also in $\delta$-optimal designs, the tuning parameter is set to $r=\left\lbrace 1,2,3,4,5,10,15\right\rbrace$ to test discriminatory performance of $\delta$-optimal designs for different values of $r$.

\begin{figure}[htp]
	\centering
	\subfloat{\includegraphics[width= 16cm,height=6.8cm]{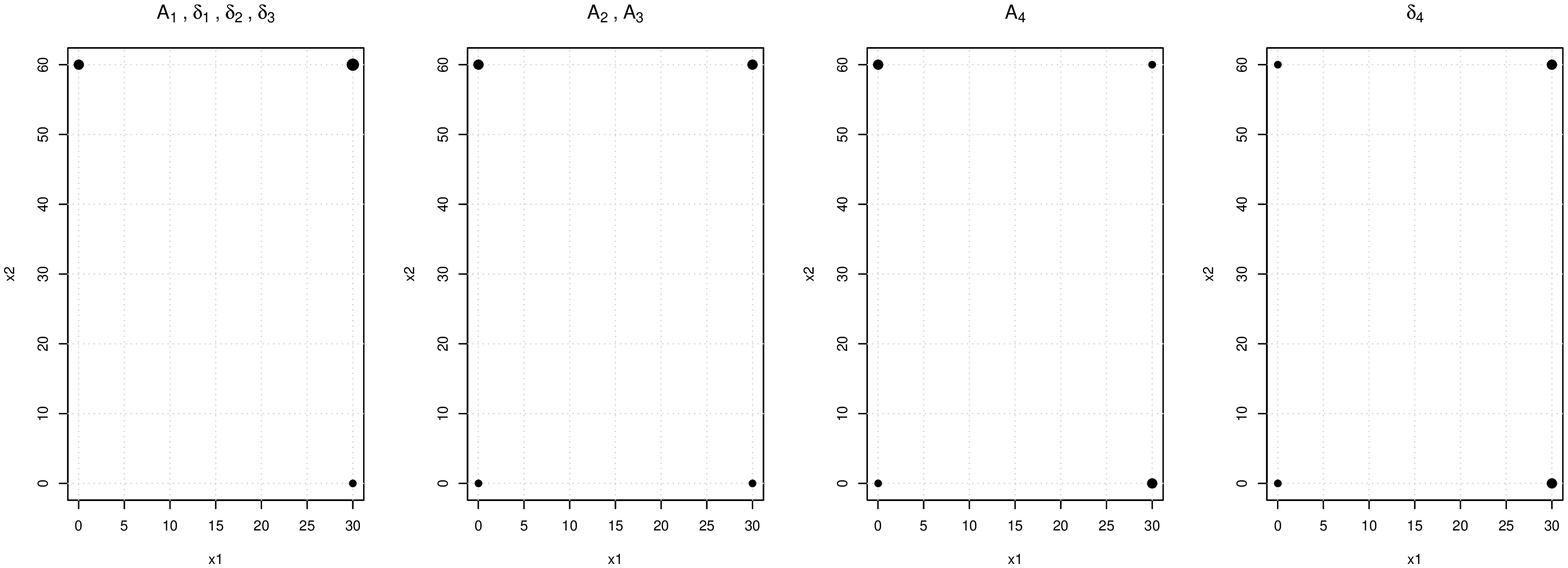}}
	
	\subfloat{\includegraphics[width= 16cm,height=6.8cm]{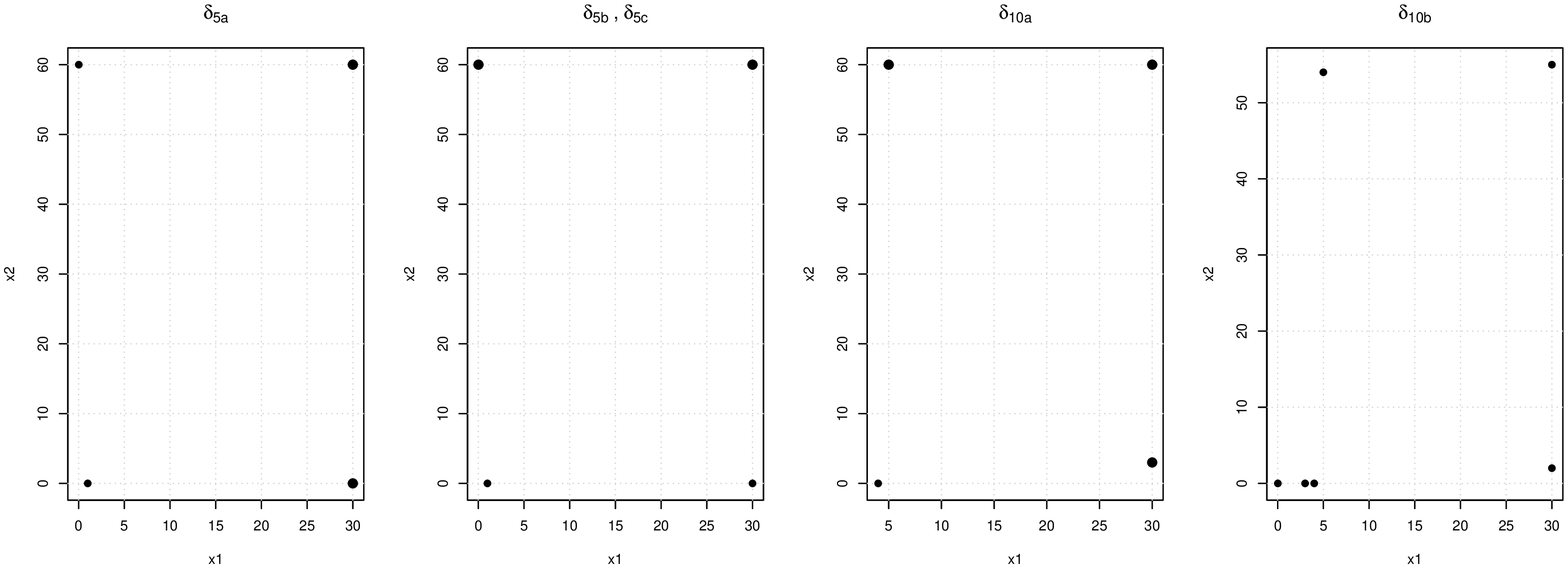}}
	
	\subfloat{\includegraphics[width= 16cm,height=6.8cm]{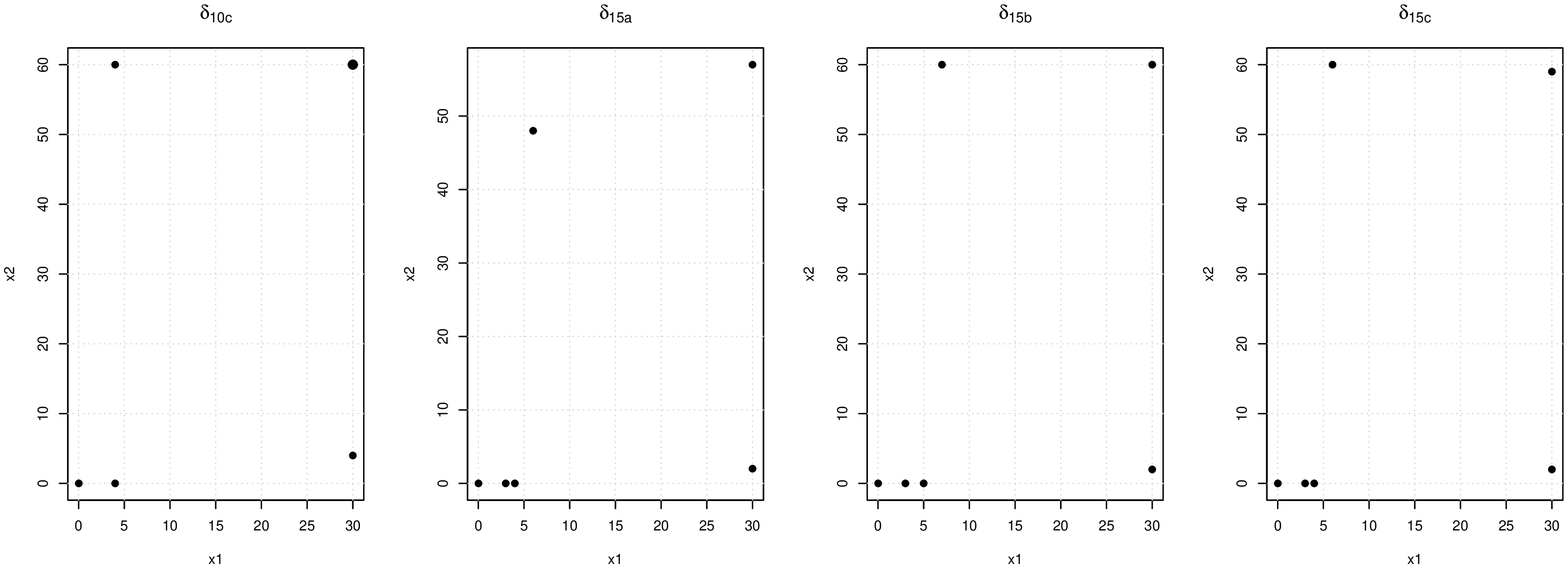}}

	\caption{All exact designs for $N=6$}
	\label{plot.alldes.$n=6$}
\end{figure}

Figures \E{\ref{plot.alldes.$n=6$}, \ref{plot.alldes.$n=7$}, \ref{plot.alldes.$n=8$} and \ref{plot.alldes.$n=9$} given in the supplementary material}, refer to plots of exact $A1$-$A4$ and different $\delta$-optimal designs for $N=6$, $N=7$, $N=8$ and $N=9$ \E{respectively}. As we can observe from the figures, in some cases the designs are equal to each other and therefore they will have the same average hit rates as we observe from Table \ref{table.mean of hit rates n=6-9}. Recall that in the figures \ref{plot.alldes.$n=6$}, \ref{plot.alldes.$n=7$}, \ref{plot.alldes.$n=8$} and \ref{plot.alldes.$n=9$}, $\delta_{5a}$, $\delta_{5b}$ and $\delta_{5c}$ refers to $r=5$ each of which is computed with the three alternatives to prevent negative nominal intervals, described before, respectively. The same description applies to $\delta_{10a}$, $\delta_{10b}$ and $\delta_{10c}$ for $r=10$ and $\delta_{15a}$, $\delta_{15b}$ and $\delta_{15c}$ for $r=15$. For these part of simulations we are using the estimate for the error standard deviation equal to $\hat{\sigma}=0.5128$ from the encompassing model in the log case as a base value for the simulation error standard deviation.

As we can observe from the table \ref{table.mean of hit rates n=6-9}, $A_2$, $A_3$ and $\delta_4$ and more specifically $A_2$ have the best performance for all number of exact designs when both models contribute equally in simulations. This result would be of high importance to those who seek to implement a tested method for discriminating between log models of enzyme inhibition.

\begin{table}[htp]
	\caption{Average values of hit rates (AvHr) for $B=100$ and $N=6$-$9$}
	\label{table.mean of hit rates n=6-9}
	\begin{center}
		\resizebox{\textwidth}{!}{
			\begin{tabular}{lccclccclccclc}
				\hline
				\multicolumn{2}{c}{$N=6$}&&&\multicolumn{2}{c}{$N=7$}&&&
				\multicolumn{2}{c}{$N=8$}&&&\multicolumn{2}{c}{$N=9$}\\
				\cline{1-2} \cline{5-6}\cline{9-10}\cline{13-14}Designs&AvHr&&&Designs&AvHr&&&
				Designs&AvHr&&&Designs&AvHr\\ 
				\hline
				$A_1,\delta_{1},\delta_{2},\delta_{3}$&$70.470$&&&
				$A_1$&$71.980$&&&
				$A_1,\delta_{2}$&$71.200$&&&
				$A_1$&$71.995$\\
				$A_2,A_3$&$\textbf{89.665}$&&&
				$A_2,\delta_{4}$&$\textbf{90.925}$&&&
				$A_2,A_3,\delta_{3},\delta_{4}$&$\textbf{92.595}$&&&
				$A_2,A_3$&$\textbf{93.635}$\\
				$A_4$&$87.525$&&&
				$A_3$&$90.675$&&&
				$A_4$&$91.640$&&&
				$A_4$&$92.715$\\
				$\delta_4$&$87.865$&&&
				$A_4$&$87.365$&&&
				$\delta_1$&$71.275$&&&
				$\delta_{1},\delta_{2},\delta_{3}$&$70.910$\\
				$\delta_{5a}$&$86.750$&&&
				$\delta_1,\delta_{2},\delta_{3}$&$70.680$&&&
				$\delta_{5a},\delta_{5b}$&$91.375$&&&
				$\delta_{4}$&$93.230$\\
				$\delta_{5b},\delta_{5c}$&$88.430$&&&
				$\delta_{5a}$&$89.470$&&&
				$\delta_{5c}$&$90.585$&&&
				$\delta_{5a},\delta_{5b}$&$91.870$\\
				$\delta_{10a}$&$75.590$&&&
				$\delta_{5b},\delta_{5c}$&$89.690$&&&
				$\delta_{10a}$&$78.710$&&&
				$\delta_{5c}$&$91.115$\\
				$\delta_{10b}$&$74.685$&&&
				$\delta_{10a}$&$78.185$&&&
				$\delta_{10b},\delta_{15a},\delta_{15b}$&$77.130$&&&
				$\delta_{10a}$&$78.935$\\
				$\delta_{10c}$&$79.065$&&&
				$\delta_{10b}$&$76.985$&&&
				$\delta_{10c}$&$80.070$&&&
				$\delta_{10b},\delta_{15a},\delta_{15b},\delta_{15c}$&$78.995$\\
				$\delta_{15a}$&$73.430$&&&
				$\delta_{10c}$&$79.005$&&&
				$\delta_{15c}$&$81.530$&&&
				$\delta_{10c}$&$83.825$\\
				$\delta_{15b}$&$72.725$&&&
				$\delta_{15a},\delta_{15b},\delta_{15c}$&$75.020$&&&
				&
				&\\
				$\delta_{15c}$&$73.555$&&&
				&&&&
				&&&&
				&\\
				\hline
			\end{tabular}
		}
	\end{center}
\end{table}
{In this research, we are trying not to have estimates beyond the boundaries of the parameter spaces, in order to have \E{$Q=100$} admissible experiments (\E{ $Q$} denotes the number of experiments done at each step of simulations), for each of the designs and at each step of the simulations. Such cases may be called inadmissible and \E{discarded} 
until we have $Q=100$ admissible cases in each scenario .} The average rates for inadmissible cases \E{amongst the total number of simulation steps (repeating until we have $Q=100$ valid experiments in each of $B=100$ Monte Carlo iterations)}, for $N=8$ is given in Table \ref{table.mean of inadmissibla-n=8}. As we can observe from the table, these rates are rather high in cases in which we have low averages of hit rates for the part of $N=8$ in Table \ref{table.mean of hit rates n=6-9}. The same table for the other $N=6,7,9$ could be provided. Since they represent the same consequences, we avoid to report them here. 
\begin{table}[htp]
	\caption{Average rates of inadmissible cases (ArIc) \E{amongst the total number of simulation steps}  for $N=8$}
	\label{table.mean of inadmissibla-n=8}
	\begin{center}
		\begin{footnotesize}
			\begin{tabular}{lcc}
				\hline
				Designs& ArIc(in \%)& \\ 
				\hline
				$A_1,\delta_{2}$&$41.44$&\\
				$A_2,A_3,\delta_{3},\delta_{4}$&$1.57$&\\
				$A_4$&$0.19$&\\
				$\delta_1$&$45.45$&\\
				$\delta_{5a},\delta_{5b}$&$3.00$&\\
				$\delta_{5c}$&$2.28$&\\
				$\delta_{10a}$&$3.31$&\\
				$\delta_{10b},\delta_{15a},\delta_{15b}$&$3.94$&\\
				$\delta_{10c}$&$2.38$&\\
				$\delta_{15c}$&$1.04$&\\
				\hline
			\end{tabular}
		\end{footnotesize}
	\end{center}
\end{table}

\subsubsection{A large scale experiment, $N=60$}
As a large scale experiment, we designed an experiment to compute total correct classification rates and also the average classification rates of all designs for $N=60$ in the second part of simulations. Figure A.\ref{plot.alldes.$n=60$} given in the Appendix refers to plots of different designs for $N=60$. Since the discriminatory power of all the designs for $N=60$ is perfect and rather the same when the estimated error standard deviation $\hat{\sigma}=0.5128$ from the encompassing model (\ref{formula-combined-model}) is used, we are required to inflate it. Therefore, the error standard deviation used was $4\times \hat{\sigma}$. The number of Monte Carlo simulations done for this part is $B=1000$. We need to mention that here, the tuning parameter is set to contain also $r=6$, beside the values used for the last part of the  low scale experiment.

\begin{figure}[htp]
	\centering	
	\begin{minipage}{\linewidth}
		\centering
		\subfloat[Total correct classification rates, white under $\eta_0$, grey under $\eta_1$,]{\label{all(28box)-60:a}\includegraphics[width=13cm,height=8cm]{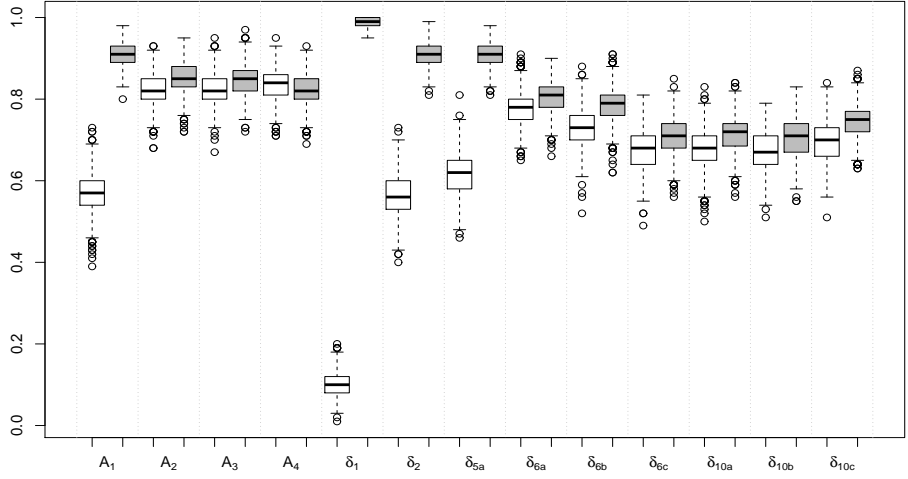}}
	\end{minipage}%
	
	\begin{minipage}{\linewidth}
		\centering
		\subfloat[Averages of classification rates]{\label{all(28box)-60:b}\includegraphics[width=13cm,height=8cm]{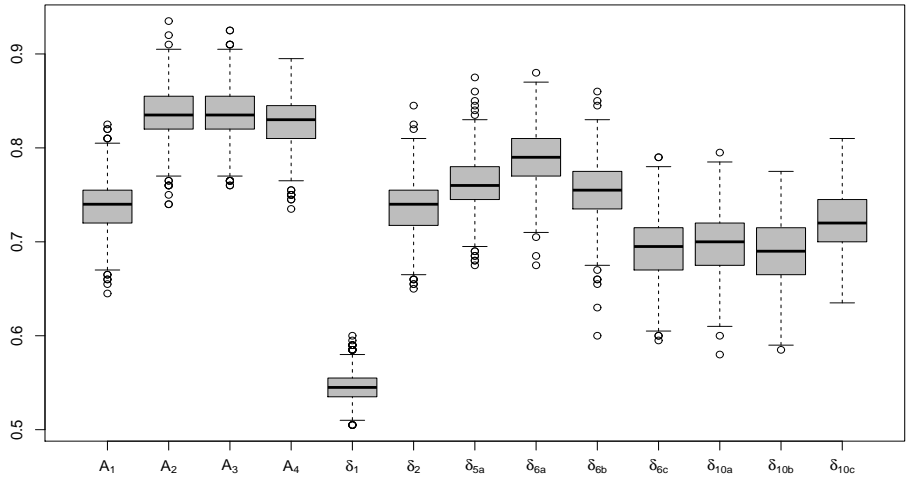}}
	\end{minipage}%
	\caption{{Boxplots for the correct classification rates of all designs $r=\left\lbrace1,2,3,4,5,6,10,15 \right\rbrace $, $\delta_{2}$ stands for all: $\delta_{2},\delta_{3},\delta_{4},\delta_{5b},\delta_{5c},$ and $\delta_{10b}$ stands for all: $\delta_{10b},\delta_{15a},\delta_{15b},\delta_{15c}$}. }		
	\label{all(28box)-60}
\end{figure}

The corresponding box plots of the total and average correct classification rates are given in Figure \ref{all(28box)-60}. All designs have reasonably high performances except the design $\delta_{1}$. Designs $A_2$ and $A_3$ and to be more specific $A_2$ is performing highly well according to both Figures \ref{all(28box)-60:a} and \ref{all(28box)-60:b} which confirm the results \E{presented in Table \ref{table.mean of hit rates n=6-9}}. Note that $A_1$ and $A_4$ are excluded from our comparisons, since the methods they are resulted from are inherently asymmetric. However, note that the average hit rate value for $A_1$  is the lowest, somehow also reflecting the rejection of  the noncompetitive model according to likelihood ratio tests from section \ref{subsec:likelihood ratio test}. Among the designs resulting from the symmetric method $\delta$-optimality, $\delta_{6a}$ is also performing well suggesting that $r=6$ is a good choice for the tuning parameter.

The total rates of inadmissible cases in each scenario within the total number of simulation steps, \E{(repeating until we have $Q=100$ valid experiments in each of $B=1000$ Monte Carlo iterations)} under each model and the average rates of them, when both models contribute equally in simulations, is given in the first two columns and the third column of Table \ref{table.ratios-inadmissible$n=60$} for $N=60$. As we can observe from the table, these rates are higher in cases in which we have low classification rates in Figure \ref{all(28box)-60}. These rates are observed specifically much higher when model (\ref{formula-comp-model}) is assumed true in simulations, a similar reflection we had from Table \ref{table.mean of inadmissibla-n=8}.

\begin{table}[htp]
	\caption{Rates of inadmissible cases \E{amongst the total number of simulation steps} for $N=60$}
	\label{table.ratios-inadmissible$n=60$}
	\begin{center}
		\begin{footnotesize}
			\begin{tabular}{lcccc}
				\hline
				
				True model&&$\eta_0$(in \%)&$\eta_1$(in \%)& ArIc(in \%)\\
				\hline				
				$A_1$&&$109.09$&$3.27$&$ 56.18$\\
				$A_2$&&$12.41$&$1.73$&$7.07$\\
				$A_3$&&$12.31$&$ 1.77$&$ 7.04$\\
				$A_4$&&$5.86$&$ 1.78 $&$3.82$\\
				$\delta_{1}$&&$176.39$&$ 3.13 $&$89.76$\\ 
				$\delta_{2},\delta_{3},\delta_{4},\delta_{5b},\delta_{5c}$&&$113.25$&$ 2.89$&$ 58.07$\\
				$\delta_{5a}$&&$91.02$&$ 2.79 $&$46.91$\\  
				$\delta_{6a}$&&$16.46$&$ 2.30 $&$ 9.38$\\ 
				$\delta_{6b}$&&$22.95$&$ 4.01$&$ 13.48$\\ 
				$\delta_{6c}$&&$24.43$&$ 3.35$&$ 13.89$\\   
				$\delta_{10a}$&&$23.67$&$ 3.31 $&$13.49$\\
				$\delta_{10b},\delta_{15a},\delta_{15b},\delta_{15c}$&&$24.70$&$ 3.52 $&$14.11$\\
				$\delta_{10c}$&&$21.16 $&$2.91 $&$12.04$\\
				\hline
			\end{tabular}
		\end{footnotesize}
	\end{center}
\end{table}

\subsection{A final discriminating design case, $N=10,100$}
In the last parts, we have discussed aspects of how the discriminating designs look like in the case of log models or how the discriminatory power of designs changes using different discriminating criteria in the log case. Still there may remain the question of how the discriminating designs look like when we want to differentiate between the models with different error structures, i.e. normal and log normal errors. For this purpose, we computed the discriminating designs using $\delta$-optimality criterion for the encompassing model using the two relations (\ref{formula-generalmodel}) and (\ref{formula-logmodel}). We do not prefer to set any assumption on which model is the true one. That's why $\delta$-optimality is a proper option here. Note that for this part, we set the \E{size of exact designs equal to} $N=10$ and $N=100$ to observe the stability of designs while switching from a low to a large number of exact designs. The tuning parameter $r$ is equal to $r=\left\lbrace 10,15\right\rbrace $. Note that the nominal interval for $\lambda$ may lead to an interval that does not completely fall inside its assumed parameter space; i.e. the unit interval. For this reason, we used the logit transformation, $g(\lambda)=\frac{\lambda}{1-\lambda}$, and Delta method to arrive at the unit interval for $\lambda$. This transformation is presented as a remark, below.

\textbf{Remark 2:} Assume that the asymptotic distribution of the estimate of $\lambda$ is normal
$\sqrt{\E{N}}\left( \hat{\lambda}-\lambda\right) \sim \mathcal{N}(0,\sigma_{\lambda}^2)$. Then implementing the logit transformation and the Delta method \E{(\citealp{ghitany2015estimation})} we can arrive at the asymptotic normal distribution of $g(\hat{\lambda})=\frac{\hat{\lambda}}{1-\hat{\lambda}}$ as 
\begin{equation*}
	\sqrt{\E{N}}\left( g(\hat{\lambda})-g(\lambda)\right) \sim \mathcal{N}\left(0,\frac{\sigma_{\lambda}^2}{\lambda^2(1-\lambda)^2}\right)
\end{equation*}
Now, the asymptotic $100(1-\alpha)\%$ confidence interval for $g(\lambda)$ is
\begin{equation*}
	g(\hat{\lambda}) \pm z_{\frac{\alpha}{2}}\dfrac{\widehat{\sigma_{\lambda}}}{\hat{\lambda}(1-\hat{\lambda})}\equiv (L,U).
\end{equation*}
where $\widehat{\sigma_{\lambda}}=\sqrt{\widehat{\text{Var}(\hat{\lambda})}}$. Finally an asymptotic $100(1-\alpha)\%$ confidence intervals for $\lambda$, can be obtained as
\begin{equation*}
	\left( \dfrac{{\rm e}^L}{1+{\rm e}^L}, \dfrac{{\rm e}^U}{1+{\rm e}^U}\right) .
\end{equation*}

\begin{figure}[htp]
	
	\begin{minipage}{.5\linewidth}
		\centering
		\subfloat[$N=10$]{\label{plot.deltadesrev1.$n=10$:a}\includegraphics[scale=.34]{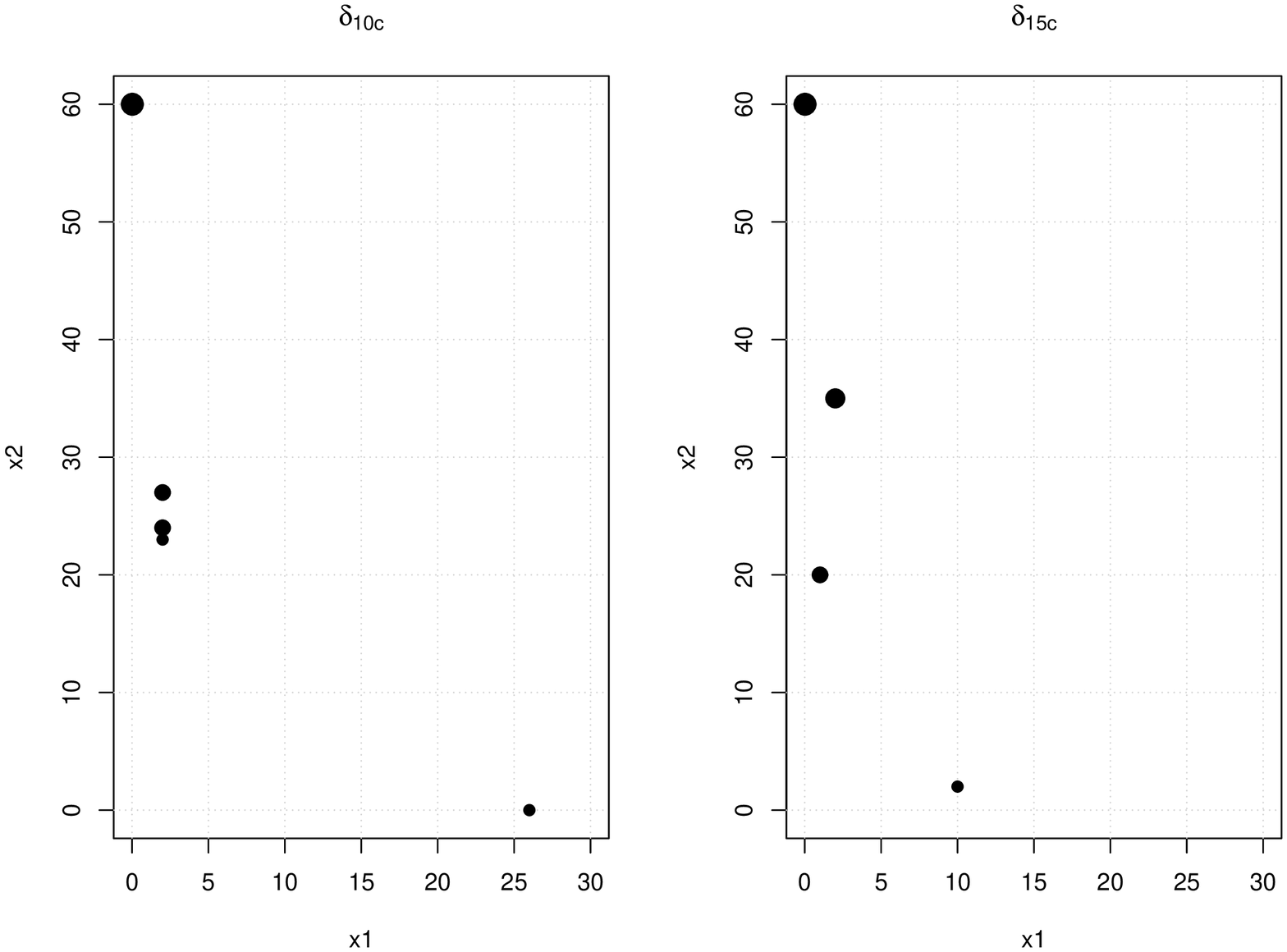}}
	\end{minipage}%
	\begin{minipage}{.5\linewidth}
		\centering
		\subfloat[$N=100$]{\label{plot.deltadesrev1.$n=10$:b}\includegraphics[scale=.34]{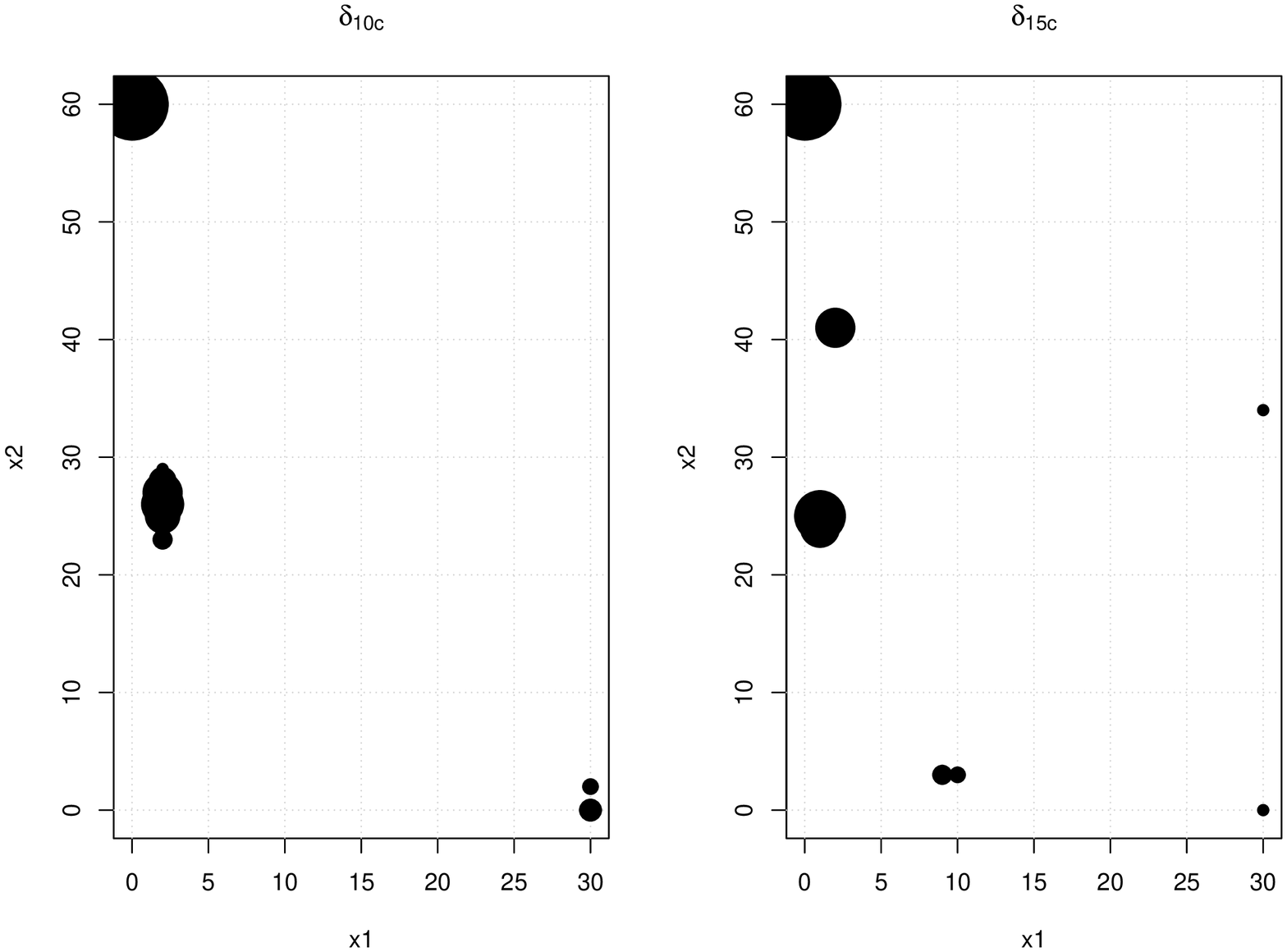}}
	\end{minipage}\par\medskip
	
	\caption{{Plots of $\delta$-optimal designs for discriminating between standard and log encompassing models}}
	\label{plot.deltadesrev1.$N=10$}
\end{figure}

Recall that for high values of $r$, the first remark is used to prevent negative lower bounds of nominal intervals for the parameters. Therefore, implementing both the first and the second remark here, $\delta$-optimal designs are plotted in Figure \ref{plot.deltadesrev1.$N=10$} which shows that switching from $N=10$ to $N=100$ does not change structure of the designs too much and the designs are quite stable.\\

\section{Conclusions}\label{sec-conclusions}

This paper provided optimal designs with high efficiencies either for estimation of the parameters of interest or for discrimination between enzyme kinetics log models whichever model holds. One should be careful which error structure to choose since the resulting designs showed considerably differing patterns. In the standard case, optimal designs are spread over the rectangular design region while the ones resulted from the log case are typically concentrated on the corners of that region. 

This means that optimal choices of the designs for estimation or discrimination should contain the most extreme pair concentrations of the substrates and inhibitors with different replications and therefore it is important to be aware of how to choose the maximum and minimum concentrations in the experiment. Misspecification of those concentrations may lead to irrecoverable results in producing Dextrometorphan-Sertaline and other similar biochemical products. There is no such sensitivity on how to choose the design region in the standard case as the designs are typically not dependent upon where the boundaries are set. On the other hand those designs are much more sensitive to the nominal values chosen, which is not the case for the log transformed model.

Both these observations for the log model are in accordance with the behaviour of linear models. So while those transformed models are not intrinsically linear \E{cf. \cite{pronzato2013design}}, this still points to the suspicion that their curvature must be flat for a wide range in the parameter space. The resulting robustifying effect on the designs may then be a desired quality for the experimenter.  

One other interesting result is that the optimal designs for discriminating between the nonlinear log models are similar to optimal designs for precise estimation of parameters of each model, in all but one case, with the only difference in their corresponding weights. 

Finally it was observed that in such transformed models - despite a firm theoretical grounding -  both $A_2$ and $A_3$ provided high relative efficiencies (or product of relative efficiencies), when the interest is to solely compare design methods avoiding asymmetries. 
In particular, due to the fact that we are more concerned with real discriminatory power in practical situations, comparisons are more straightforward using the results from the subsection 4.3. In particular, $A_2$ and $A_3$ have the best performances according to high average values of hit rates they have in Table \ref{table.mean of hit rates n=6-9}, irrespective of the number of exact designs required. They also perform well in parameter estimation as their D-efficiencies (in the encompassing model) are above $95\%$.  We note that since $D_s$-optimal designs are easy to calculate in comparison with $CT$ or $\delta$-optimal designs, this makes the design $A_3$ particularly attractive. If there is no such constraint about which method to choose due to complexity of the method or the time required for calculations, 
$A_2$ resulted from the $CT$-optimal design criterion at $\nu=0.5$ would \E{also be a recommendable choice (best choice when compared to the competitors in the present manuscript)} in the context of discrimination between log models of enzyme inhibition due to gaining the highest average rates of classification while still being sufficiently efficient for parameter estimation.

\begin{figure}[htp]
	\begin{minipage}{.5\linewidth}
		\centering
		\begin{tikzpicture}
			\draw[>=latex,step=0.5cm,gray,very thin] (0,0) grid (4,6);
			\draw[thin,->] (0,-1) node [anchor= east] {\footnotesize$x_{S}/\theta_{M}$} -- (4,-1) ;
			\draw[anchor=center] (0.25,6.2)  node{\footnotesize A};
			\draw[anchor=center] (0.75,6.2)  node{\footnotesize B};
			\draw[anchor=center] (1.25,6.2)  node{\footnotesize C};
			\draw[anchor=center] (1.75,6.2)  node{\footnotesize D};
			\draw[anchor=center] (2.25,6.2)  node{\footnotesize E};
			\draw[anchor=center] (2.75,6.2)  node{\footnotesize F};
			\draw[anchor=center] (3.25,6.2)  node{\footnotesize G};
			\draw[anchor=center] (3.75,6.2)  node{\footnotesize H};
			
			\draw[anchor=center] (0.25,-0.5)  node[rotate=-90]{\footnotesize 0.02};
			\draw[anchor=center] (0.75,-0.5)  node[rotate=-90]{\footnotesize 0.16};
			\draw[anchor=center] (1.25,-0.5)  node[rotate=-90]{\footnotesize 0.31};
			\draw[anchor=center] (1.75,-0.5)  node[rotate=-90]{\footnotesize 0.63};
			\draw[anchor=center] (2.25,-0.5)  node[rotate=-90]{\footnotesize 1.25};
			\draw[anchor=center] (2.75,-0.5)  node[rotate=-90]{\footnotesize 2.50};
			\draw[anchor=center] (3.25,-0.5)  node[rotate=-90]{\footnotesize 5.00};
			\draw[anchor=center] (3.75,-0.5)  node[rotate=-90]{\footnotesize 7.50};
			
			\draw[thin,->] (-1.5,0) node [anchor= east] {\footnotesize$x_I/IC_{50}$}-- (-1.5,6) ;
			\draw[anchor=center] (4.3,0.25)  node{\footnotesize 1};
			\draw[anchor=center] (4.3,0.75)  node{\footnotesize 2};
			\draw[anchor=center] (4.3,1.25)  node{\footnotesize 3};
			\draw[anchor=center] (4.3,1.75)  node{\footnotesize 4};
			\draw[anchor=center] (4.3,2.25)  node{\footnotesize 5};
			\draw[anchor=center] (4.3,2.75)  node{\footnotesize 6};
			\draw[anchor=center] (4.3,3.25)  node{\footnotesize 7};
			\draw[anchor=center] (4.3,3.75)  node{\footnotesize 8};
			\draw[anchor=center] (4.3,4.25)  node{\footnotesize 9};
			\draw[anchor=center] (4.3,4.75)  node{\footnotesize 10};
			\draw[anchor=center] (4.3,5.25)  node{\footnotesize 11};
			\draw[anchor=center] (4.3,5.75)  node{\footnotesize 12};
			
			\draw [decorate,decoration={brace,amplitude=10pt},xshift=0pt,yshift=0pt]
			(0.0,0.0) -- (0.0,1.5) node [black,midway,xshift=-0.7cm] 
			{\footnotesize $0.00$};
			\draw [decorate,decoration={brace,amplitude=10pt},xshift=0pt,yshift=0pt]
			(0.0,1.5) -- (0.0,3) node [black,midway,xshift=-0.7cm] 
			{\footnotesize $1.00$};
			\draw [decorate,decoration={brace,amplitude=10pt},xshift=0pt,yshift=0pt]
			(0.0,3.0) -- (0.0,4.5) node [black,midway,xshift=-0.7cm] 
			{\footnotesize $3.00$};
			\draw [decorate,decoration={brace,amplitude=10pt},xshift=0pt,yshift=0pt]
			(0.0,4.5) -- (0.0,6) node [black,midway,xshift=-0.7cm] 
			{\footnotesize $9.00$};

			\draw (0,0) rectangle (4,1.5);
			\draw[pattern=mydots1] (0,1.5) rectangle (4,3);
			\draw[pattern=mydots2] (0,3) rectangle (4,4.5);
			\draw[pattern=mydots3] (0,4.5) rectangle (4,6);
			
		\end{tikzpicture}
	\end{minipage}
	\begin{minipage}{.5\linewidth}
		\centering
		\begin{tikzpicture}
			\draw[>=latex,step=0.5cm,gray,very thin] (0,0) grid (4,6);
			\draw[thin,->] (0,-1) node [anchor= east] {\footnotesize$x_{S}/\theta_{M}$} -- (4,-1) ;
			\draw[anchor=center] (0.25,6.2)  node{\footnotesize A};
			\draw[anchor=center] (0.75,6.2)  node{\footnotesize B};
			\draw[anchor=center] (1.25,6.2)  node{\footnotesize C};
			\draw[anchor=center] (1.75,6.2)  node{\footnotesize D};
			\draw[anchor=center] (2.25,6.2)  node{\footnotesize E};
			\draw[anchor=center] (2.75,6.2)  node{\footnotesize F};
			\draw[anchor=center] (3.25,6.2)  node{\footnotesize G};
			\draw[anchor=center] (3.75,6.2)  node{\footnotesize H};
			
			\draw[anchor=center] (0.25,-0.5)  node[rotate=-90]{\footnotesize 0.02};
			\draw[anchor=center] (0.75,-0.5)  node[rotate=-90]{\footnotesize 0.02};
			\draw[anchor=center] (1.25,-0.5)  node[rotate=-90]{\footnotesize 0.02};
			\draw[anchor=center] (1.75,-0.5)  node[rotate=-90]{\footnotesize 7.50};
			\draw[anchor=center] (2.25,-0.5)  node[rotate=-90]{\footnotesize 7.50};
			\draw[anchor=center] (2.75,-0.5)  node[rotate=-90]{\footnotesize 7.50};
			\draw[anchor=center] (3.25,-0.5)  node[rotate=-90]{\footnotesize 7.50};
			\draw[anchor=center] (3.75,-0.5)  node[rotate=-90]{\footnotesize 7.50};
			
			\draw[thin,->] (-1.5,0) node [anchor= east] {\footnotesize$x_I/IC_{50}$}-- (-1.5,6) ;
			\draw[anchor=center] (4.3,0.25)  node{\footnotesize 1};
			\draw[anchor=center] (4.3,0.75)  node{\footnotesize 2};
			\draw[anchor=center] (4.3,1.25)  node{\footnotesize 3};
			\draw[anchor=center] (4.3,1.75)  node{\footnotesize 4};
			\draw[anchor=center] (4.3,2.25)  node{\footnotesize 5};
			\draw[anchor=center] (4.3,2.75)  node{\footnotesize 6};
			\draw[anchor=center] (4.3,3.25)  node{\footnotesize 7};
			\draw[anchor=center] (4.3,3.75)  node{\footnotesize 8};
			\draw[anchor=center] (4.3,4.25)  node{\footnotesize 9};
			\draw[anchor=center] (4.3,4.75)  node{\footnotesize 10};
			\draw[anchor=center] (4.3,5.25)  node{\footnotesize 11};
			\draw[anchor=center] (4.3,5.75)  node{\footnotesize 12};
			\draw[red, ultra thick] (1.5,0) -- (1.5,6) ;
			\draw[red, ultra thick] (0,2.5) -- (4,2.5) ;
			
			\draw[red, thick] (-1.4,-1.4) node [anchor= east] {\footnotesize$0.45\approx3/8$} -- (1.5,-1.4) ;
			\draw[red, thick, ->] (1.5,-1.4) -- (1.5,-0.1);
			\draw[red, thick, ->] (-1.4,2.5) node [anchor= east] {\footnotesize$0.38\approx5/12$} -- (-0.1,2.5) ;
			\filldraw [red] (1.5,0) circle (2pt);
			\filldraw [red] (0,2.5) circle (2pt);
			
			\draw [decorate,decoration={brace,amplitude=10pt},xshift=0pt,yshift=0pt]
			(0.0,0.0) -- (0.0,2.5) node [black,midway,xshift=-0.7cm] 
			{\footnotesize $0.00$};
			\draw [decorate,decoration={brace,amplitude=10pt},xshift=0pt,yshift=0pt]
			(0.0,2.5) -- (0.0,6) node [black,midway,xshift=-0.7cm] 
			{\footnotesize $9.00$};

			\draw[pattern=dots] (0,0) rectangle (1.5,2.5);
			\draw[pattern=mydots1] (1.5,0) rectangle (4,2.5);
			\draw[pattern=mydots2] (0,2.5) rectangle (1.5,6);
			\draw[pattern=mydots3] (1.5,2.5) rectangle (4,6);
			
		\end{tikzpicture}	
	\end{minipage}\par\medskip
	
	\caption{{96-well plate format for inhibitor modality studies. left: usual format, right: adapted for $A_3$ optimal design. }}
	\label{fig-96well-plate}
\end{figure}
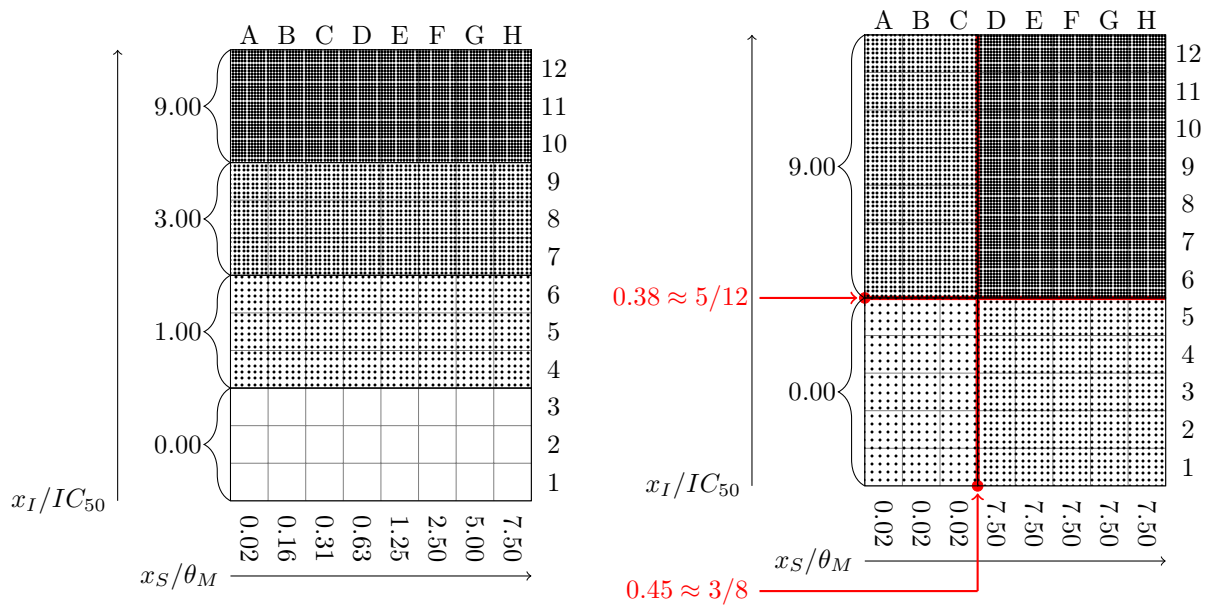

\E{According to the estimated parameters in Table \ref{table.est-se-log3}, $IC_{50}$ is equal to  $IC_{50}=6.638$ for the encompassing model in the log case using (\ref{IC50}). Determination of the reversible inhibition modality of a compound is of high importance in biopharmaceutical studies to observe whether an inhibitor may be detached from the enzyme complex, after the trace of its effect as a ligand is perceived (basically to reduce the side effects of using a ligand). For this purpose different $\%$ inhibition, defined in (\ref{Inhi-modality}) below, need to be determined. Together with different substrate concentrations, concentrations of both must simultaneously vary to determine the effect of these changes (forming a 96-well plate or other typical ones) on the reaction rate of the target enzyme. Here using different optimality criteria, when determination of the enzyme type is not possible or hardly possible (i.e. for discrimination purposes) we have computed these compounds (simultaneous concentrations of both the substrate and inhibitor) in an optimal way presented in Table \ref{table.4optimaldes-log}. Further, the substrate titration and inhibitor concentration ranges are slightly changed compared to Copelands suggestions  \cite{copeland2005evaluation} chapter 5, to match with our assumed design region and to make the results of these experimentation more feasible. Therefore the concentrations of inhibitor relative to $IC_{50}$ (for the encompassing model) for four different inhibitor concentrations each evaluated in triplicate, are used to form a similar of a 96-well plate format using optimal design $A_3$ to help visualization of concentration-response plots and other similar interpretations for an interested investigator, using the following equation
	\begin{equation} \label{Inhi-modality}
		x_{I}=IC_{50}\left( \dfrac{E_0[y]}{E_i[y]} -1 \right), 
	\end{equation}
	for the Hill coefficient being equal to one (which basically suggest a well-behaved concentration-response  relationship). The following Table \ref{table-I rel to IC50} for different $\%$ inhibition (here, $0,50,75$ and $90\%$ inhibition which have been chosen relative to $IC_{50}=6.638$ computed for the encompassing model and taking into account the assumed upper bound of $[x_I]_{\max}=60$ in the design region) helps to provide a convenient scheme for simultaneous inhibitor and substrate titration in a 96-well plate plotted next in Figure \ref{fig-96well-plate}.
	
	\begin{table}[htp]
		\caption{Concentrations of inhibitor relative to $IC_{50}=6.638$ for different inhibition levels}
		\label{table-I rel to IC50}
		\begin{footnotesize}
			\hfill{\renewcommand{\arraystretch}{1.4}
				\begin{center}
					\begin{tabular}{cccc}
						\hline 
						$\%$ Inhibition&Fractional activity $(E_i[y]/E_0[y])$&$E_0[y]/E_i[y]$&$x_I$\\
						\hline
						$0$&$1$&$1$&$0$\\
						$50$&$0.50$&$2$&$IC_{50}$\\
						$75$&$0.25$&$4$&$3 IC_{50}$\\
						$90$&$0.10$&$10$&$9 IC_{50}$\\
						\hline
					\end{tabular}
				\end{center}
			}
		\end{footnotesize}
	\end{table}
	
	A similar visualized result, compatible with the information in Table \ref{table.4optimaldes-log}, observed from the 96-plate (Figure \ref{fig-96well-plate}, the right one) is that using the optimal designs for discriminating between the enzyme log models ($A_3$ here), one do not need to simultaneously vary multiple pair concentrations for further investigation of velocity equations and curve fitting to the entire data set. Instead for example the suggested design $A_3$ require only two substrate and inhibition titration which require only one level change in each of substrate and inhibition concentrations (drawn in red thick vertical and horizontal lines, respectively) as apposed to wide titration ranges which are usually used for both concentrations (Figure \ref{fig-96well-plate}, the left one) in curve fitting and similar applications. The shading relates to the resulted weights for the design $A_3$ (see Table \ref{table.4optimaldes-log}). A similar procedure could be applied to provide 96-well plates for other optimal designs computed in this work for investigators having interest in other calculated designs either for estimation or discrimination (i.e. Tables \ref{table.D,Ds-designs-efficiencies} and \ref{table.4optimaldes-log}).
}

All these findings clearly point out how careful the experimenter needs to be in her/his decisions not only about the models used, but also the error structure and the form and boundaries of the design region. As usual, however, any effort invested in the experimental design pays off, if those choices stay within reasonable ranges.

%

\bibliographystyle{Chicago}

\bibliography{Bibliography-MM-MC}

\newpage


\begin{figure}[htp]
	\centering
	\subfloat{\includegraphics[width= 16cm,height=6.8cm]{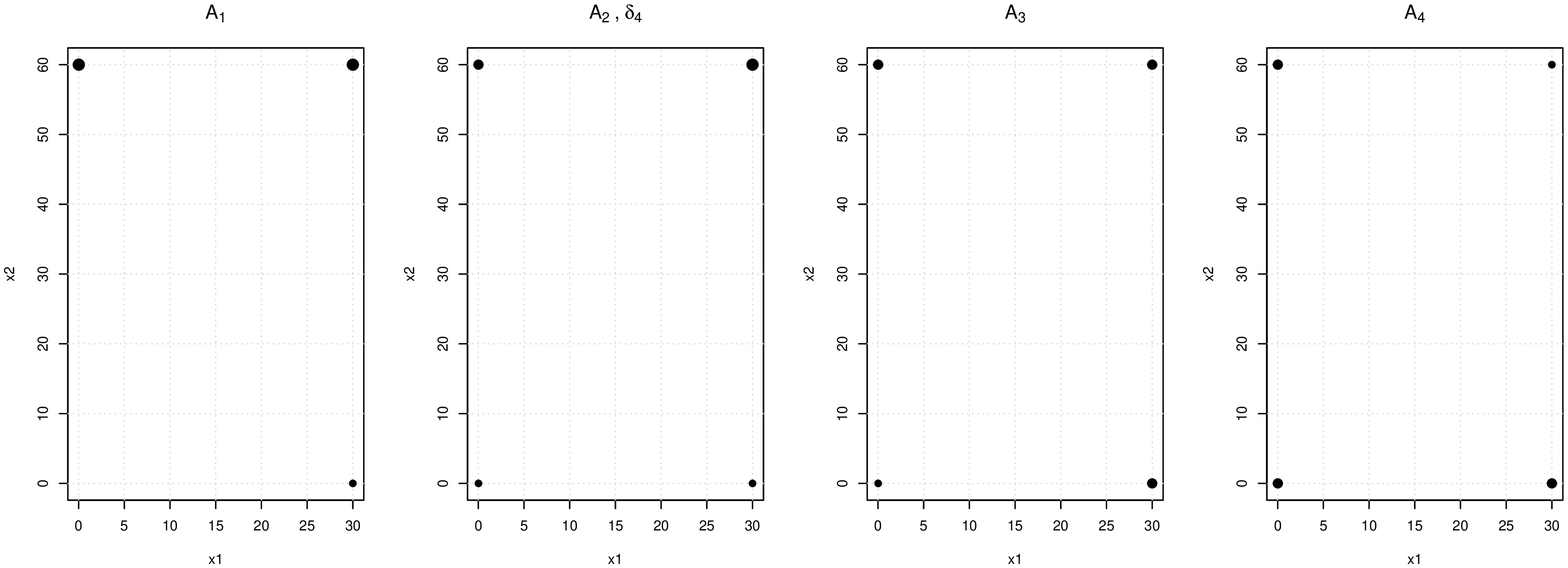}}
	
	\subfloat{\includegraphics[width= 16cm,height=6.8cm]{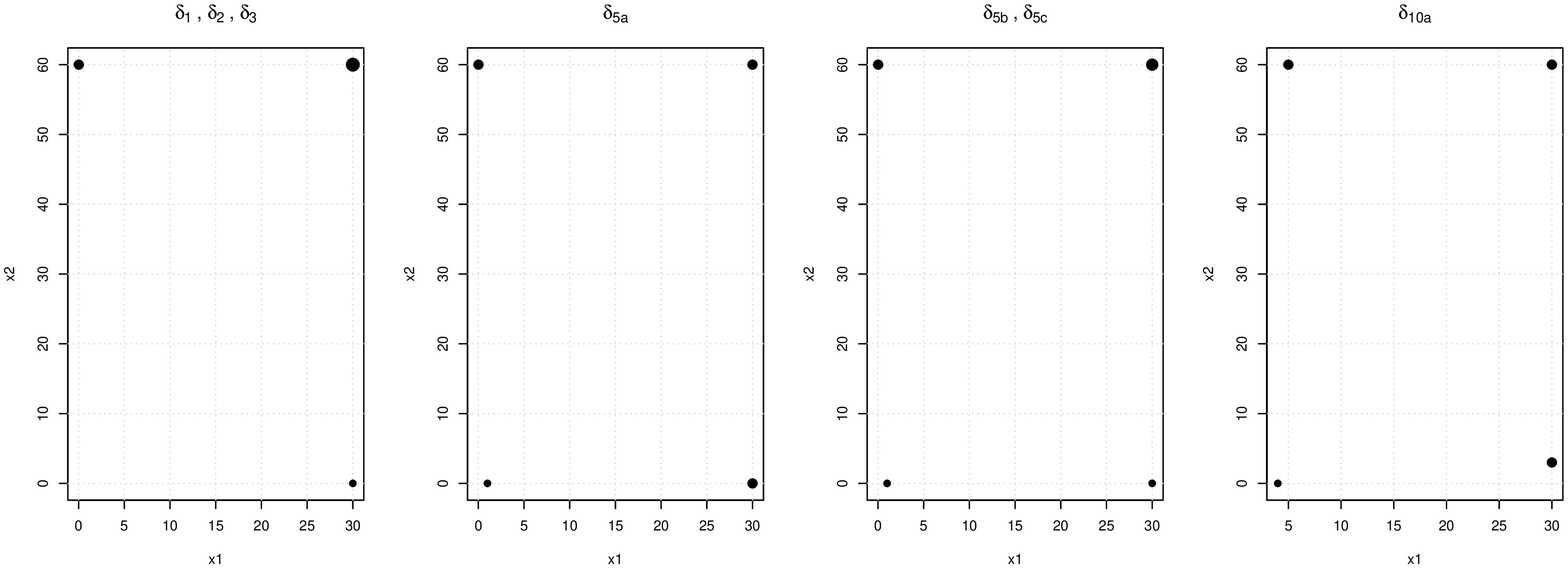}}
	
	\subfloat{\includegraphics[width= 16cm,height=6.8cm]{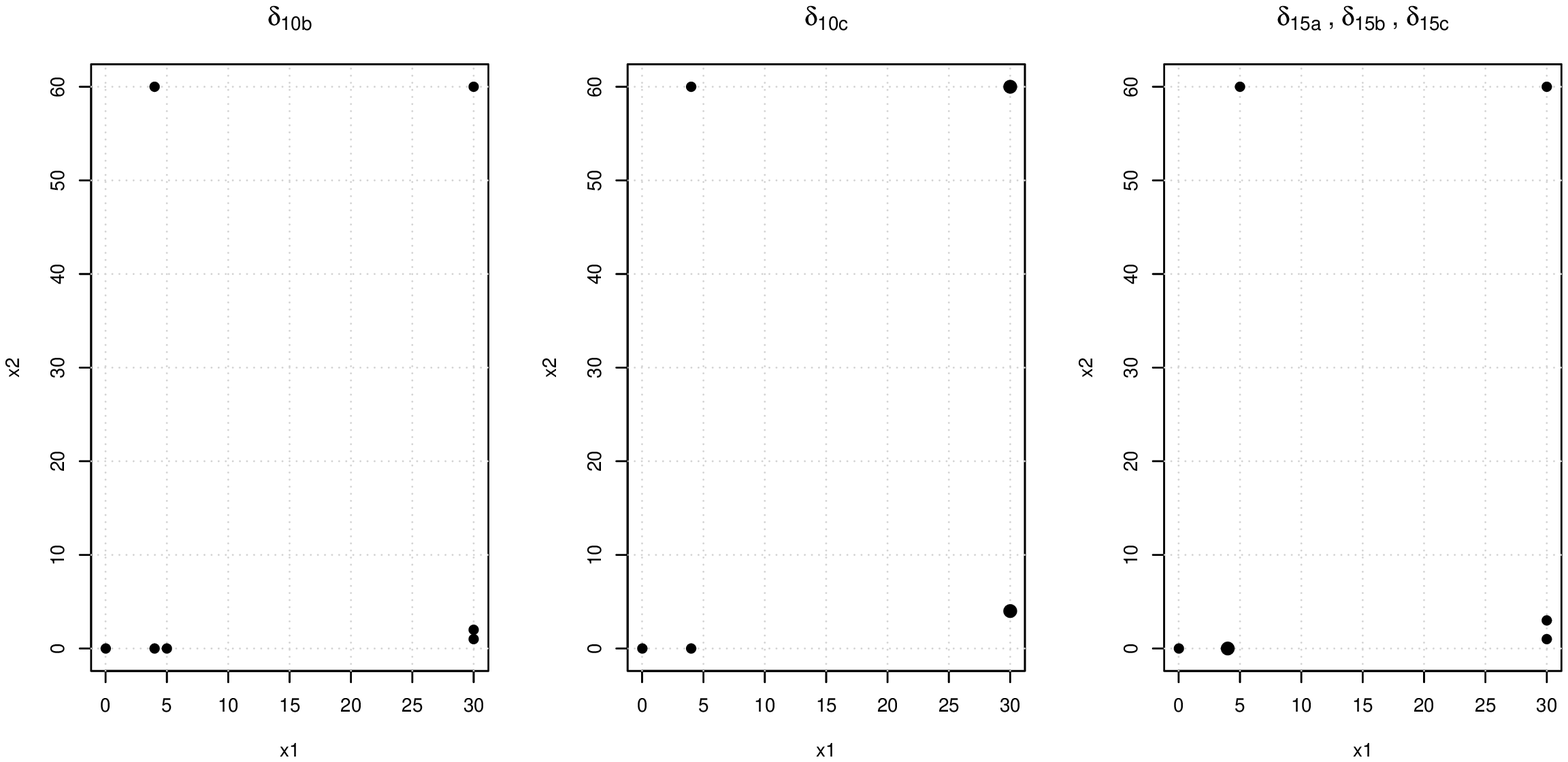}}

	\caption{All exact designs for $N=7$}
	\label{plot.alldes.$n=7$}
\end{figure}

\begin{figure}[htp]
	\centering
	\subfloat{\includegraphics[width= 16cm,height=6.8cm]{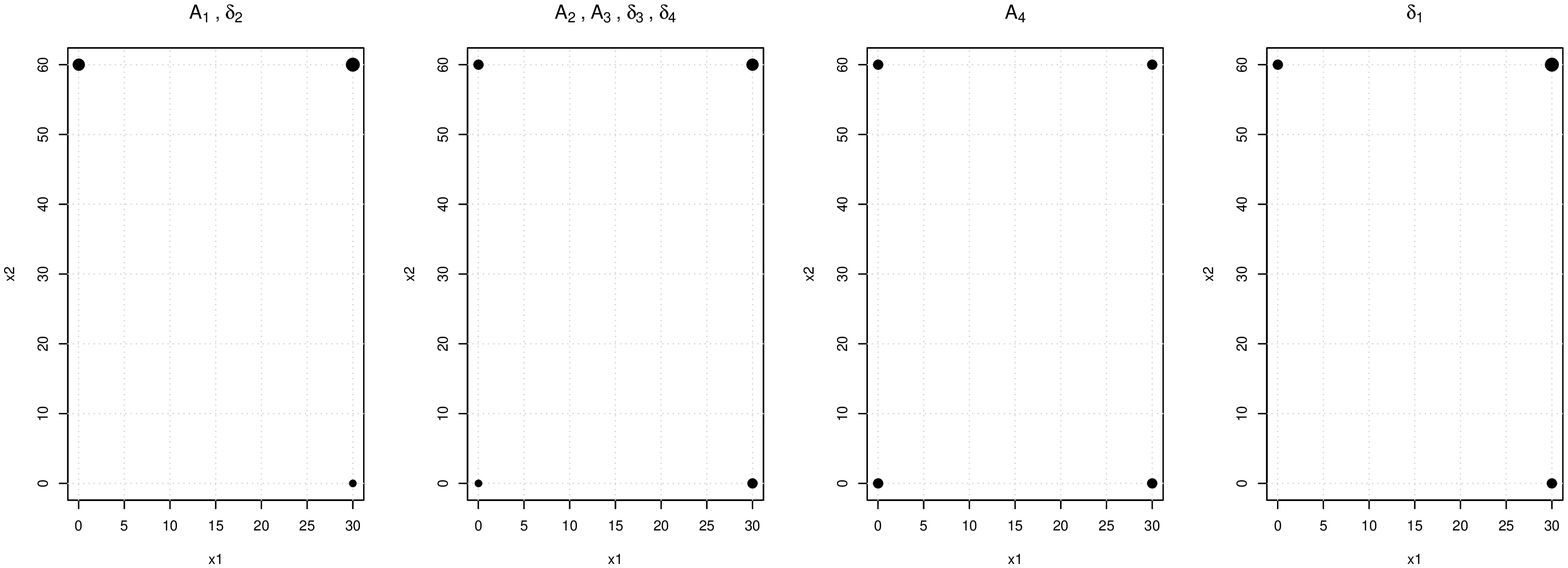}}
	
	\subfloat{\includegraphics[width= 16cm,height=6.8cm]{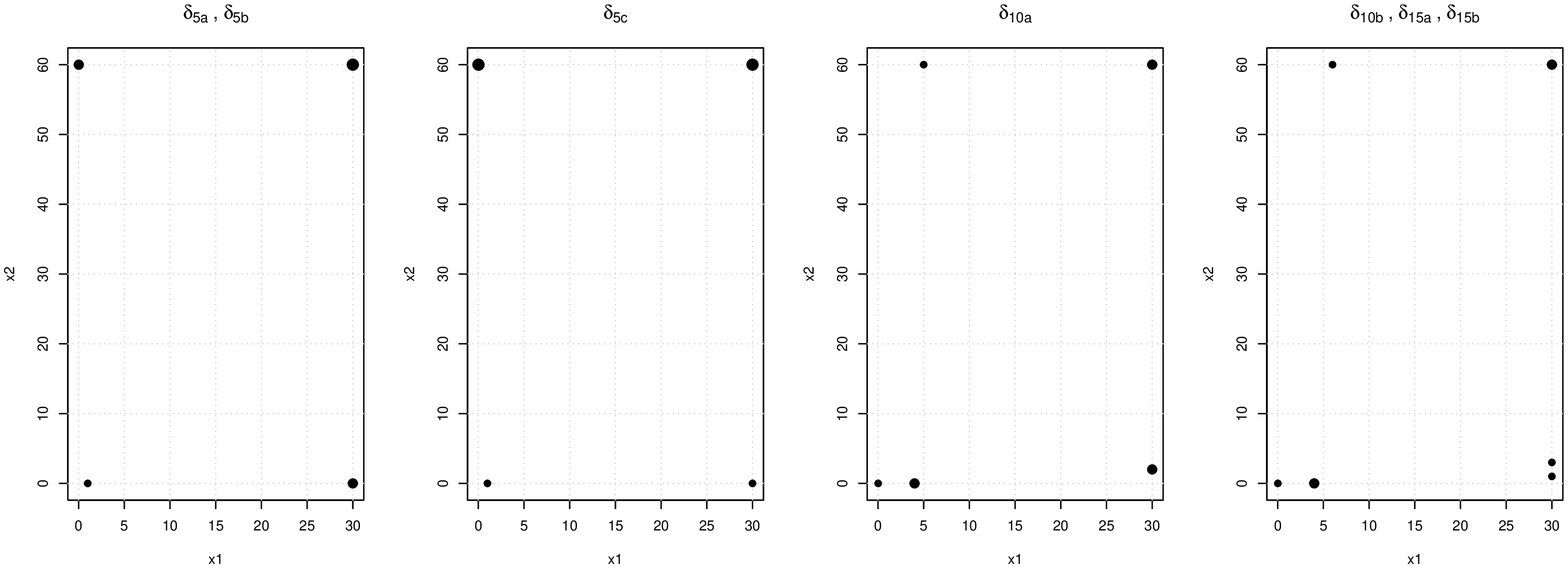}}
	
	\subfloat{\includegraphics[width= 16cm,height=6.8cm]{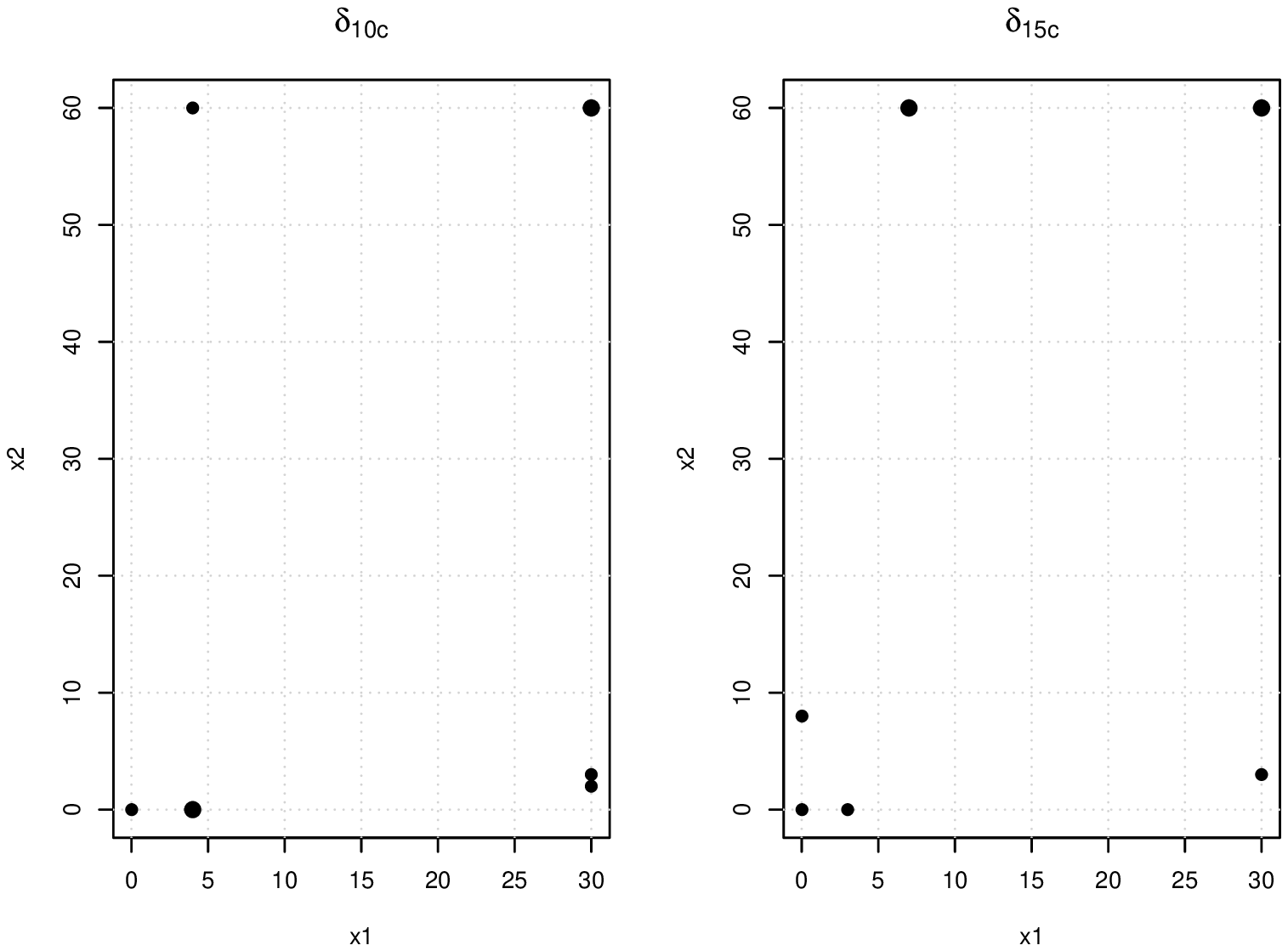}}

	\caption{All exact designs for $N=8$}
	\label{plot.alldes.$n=8$}
\end{figure}

\begin{figure}[htp]
	\centering
	\subfloat{\includegraphics[width= 16cm,height=6.8cm]{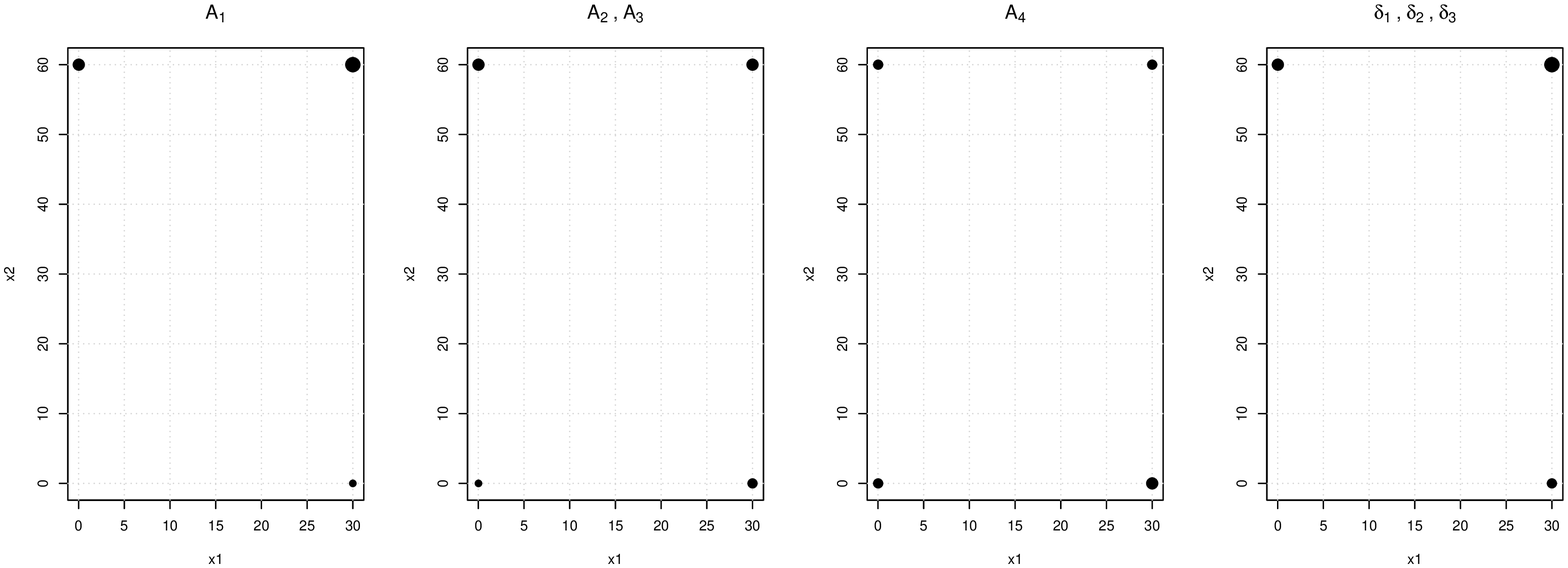}}
	
	\subfloat{\includegraphics[width= 16cm,height=6.8cm]{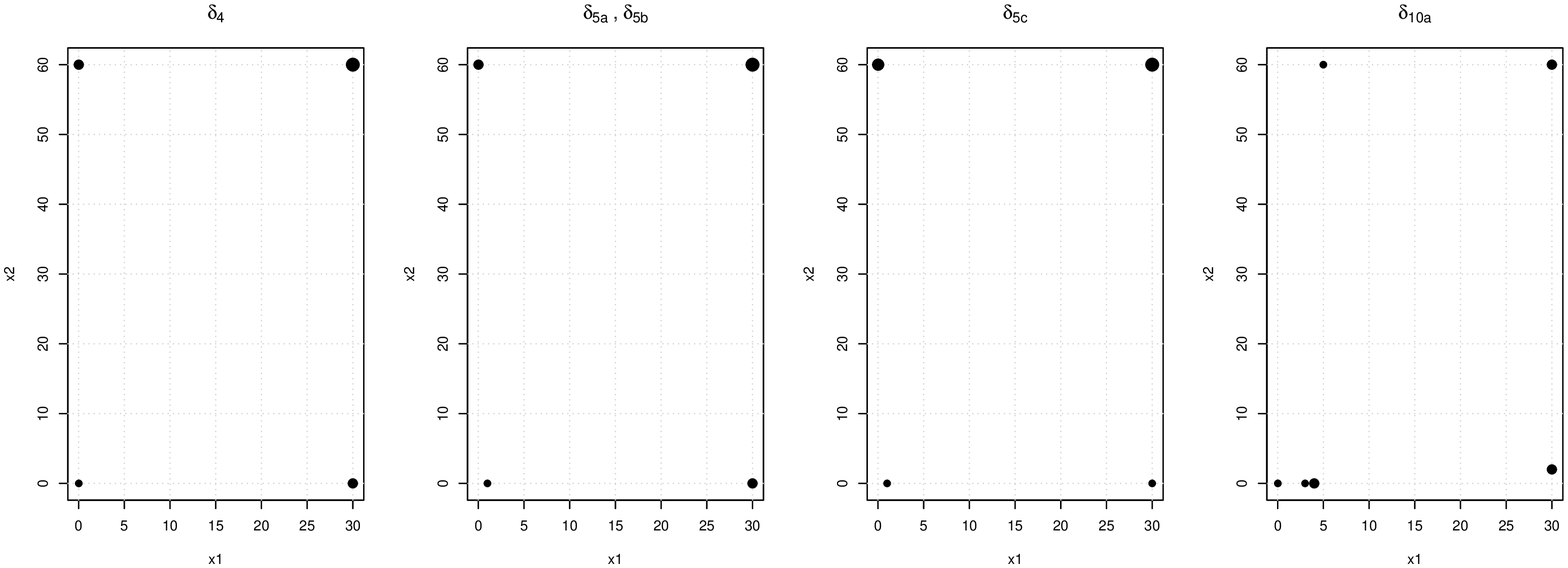}}
	
	\subfloat{\includegraphics[width= 16cm,height=6.8cm]{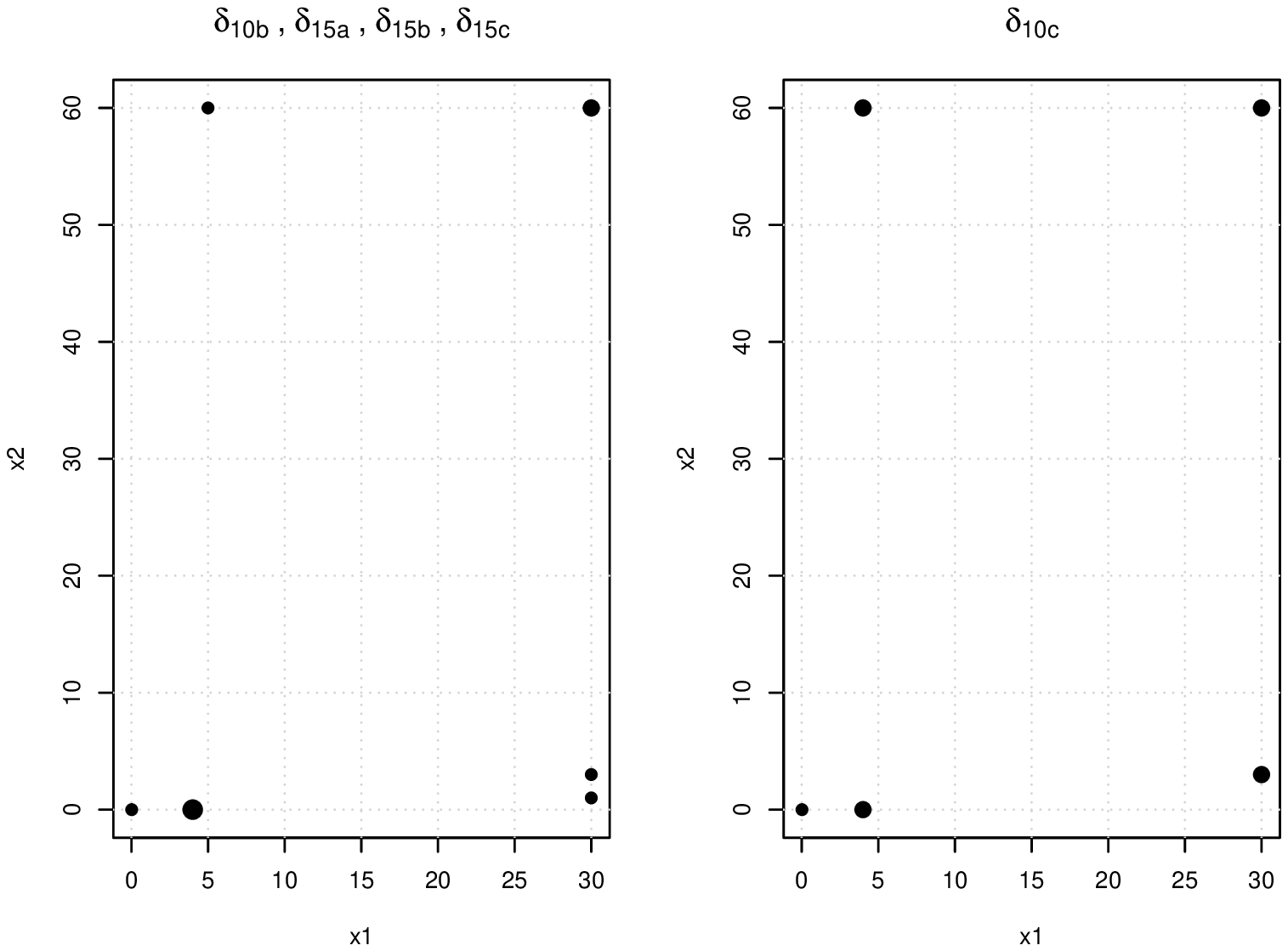}}

	\caption{All exact designs for $N=9$}
	\label{plot.alldes.$n=9$}
\end{figure}

\begin{figure}[htp]
	\centering
	\subfloat{\includegraphics[width= 16cm,height=5.6cm]{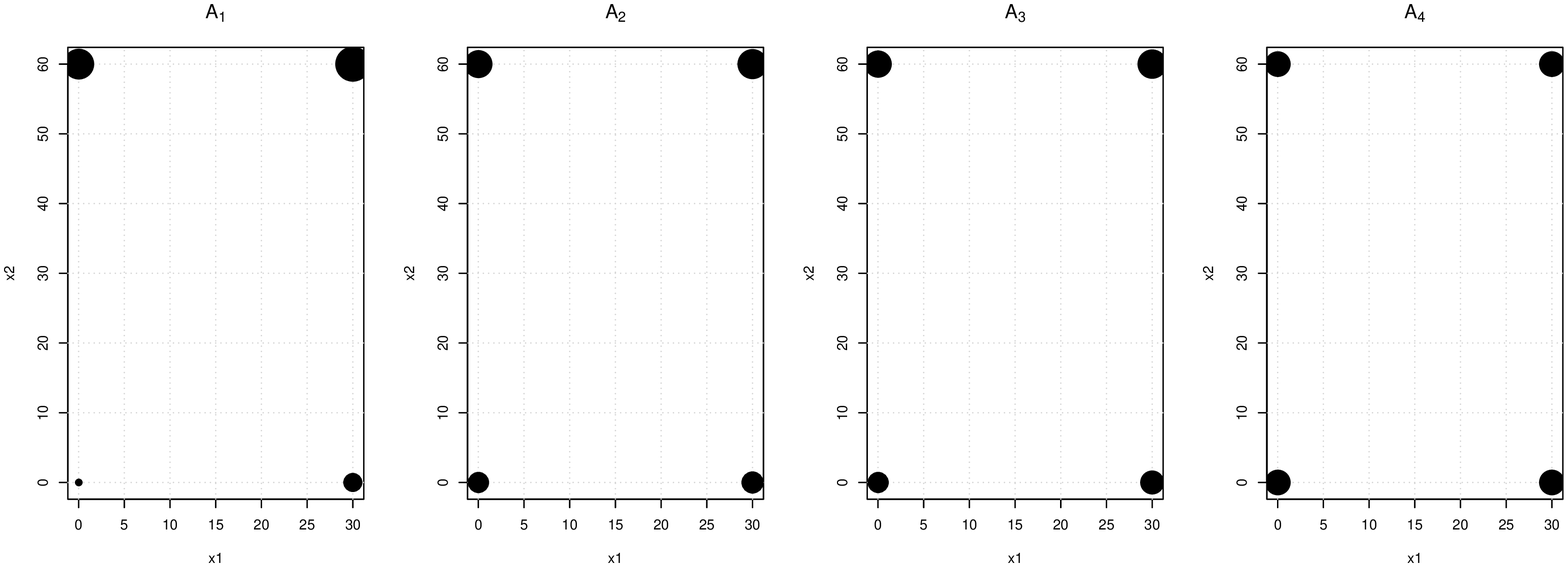}}
	
	\subfloat{\includegraphics[width= 16cm,height=5.6cm]{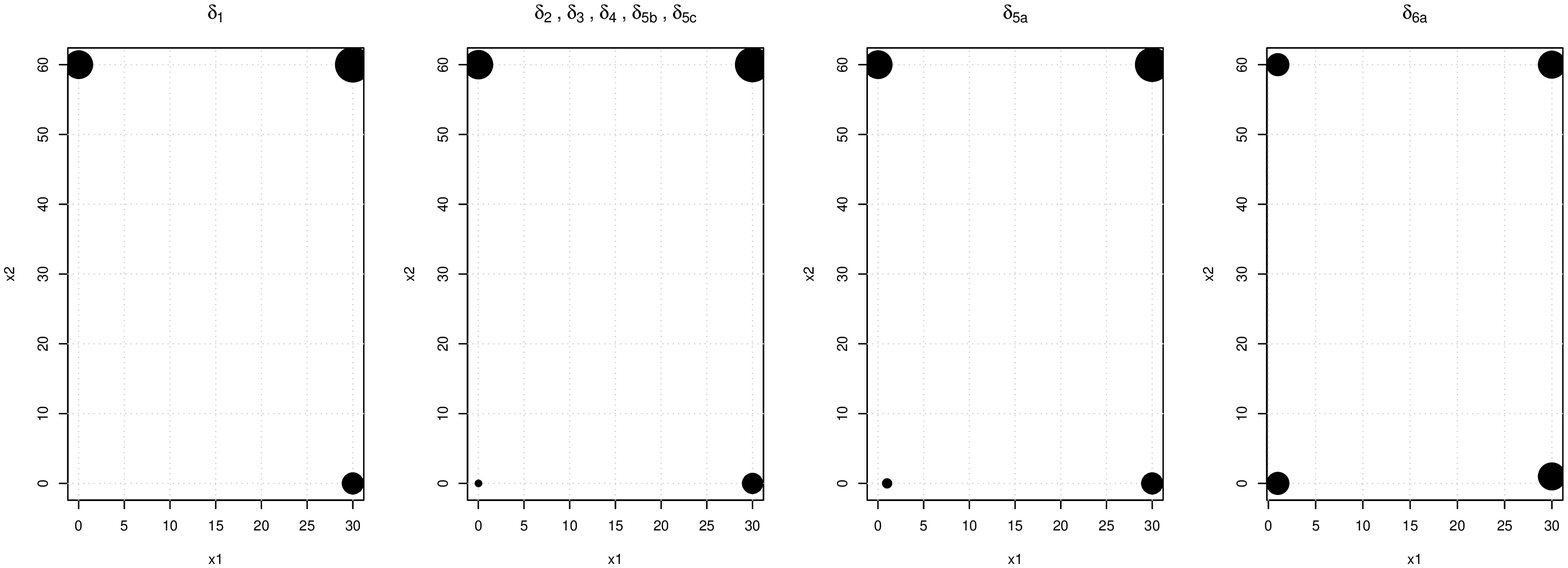}}
	
	\subfloat{\includegraphics[width= 16cm,height=5.6cm]{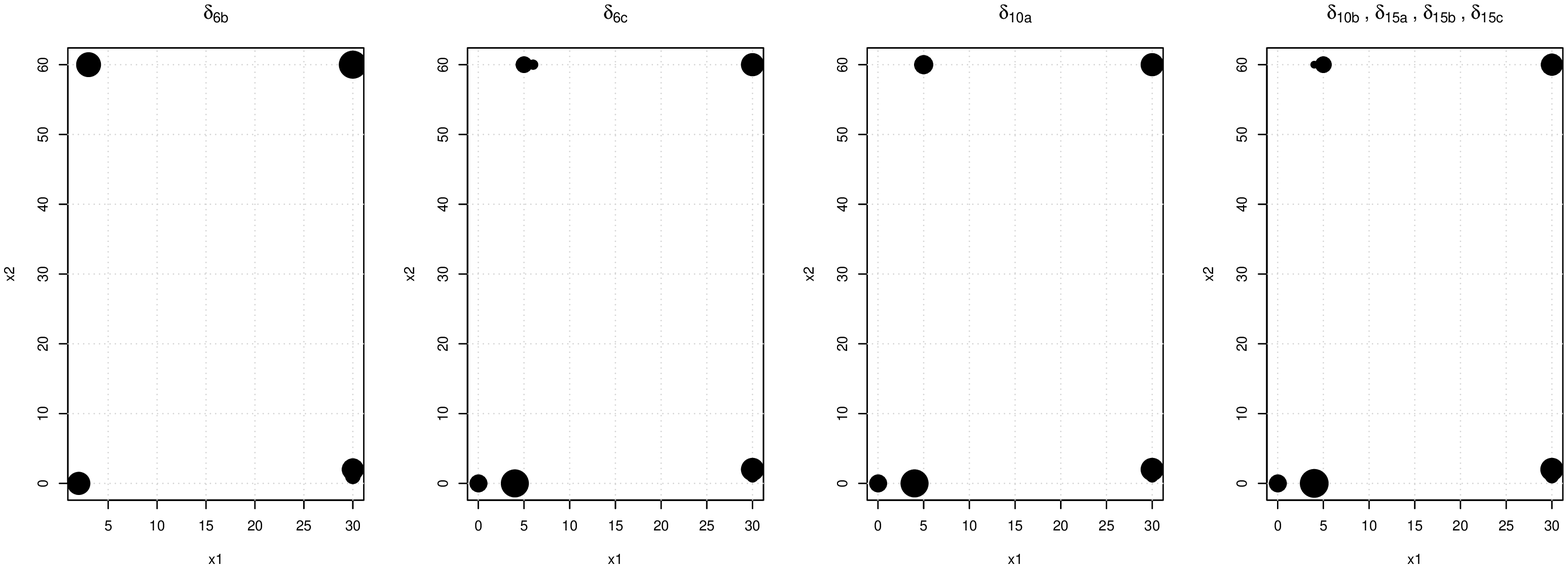}}
	
	\subfloat{\includegraphics[width= 16cm,height=5.6cm]{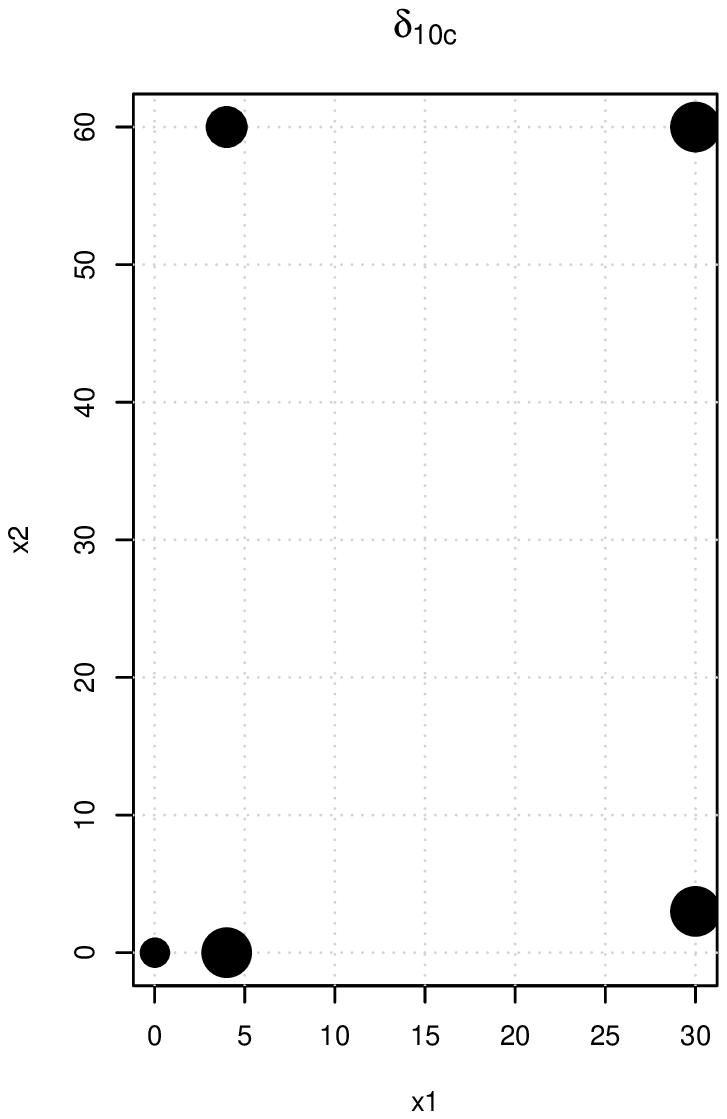}}
	
	\caption{All exact designs for $N=60$}
	\label{plot.alldes.$n=60$}
\end{figure}


\bigskip
\begin{center}
{\large\bf SUPPLEMENTARY MATERIAL}
\end{center}

\section*{Plots of exact designs\label{app1-plotdesigns}}
The plots of all exact designs for $N=6,7,8,9$ and $N=60$ are presented here for a better illustration about the location of design points in the design region. Note that the area of the circles is proportional to the number of replications.
\begin{description}

\item[Acknowledgments] We are grateful to Barbara Bogacka for providing the dataset used in this paper. The data that support the findings of this study are available from the corresponding author upon reasonable request.

\item[Author contributions] Elham Yousefi has performed all calculations and provided a first version of the text. Werner G. M{\"u}ller has conceptualized this work and edited the manuscript.

\item[Financial disclosure] None reported.

\item[Conflict of interest] The authors declare no potential conflict of interests.

\item[Supporting information] {EYs research was fully supported and WMs research was partially supported by project grants LIT-2017-4-SEE-001 funded by the Upper Austrian Government, and Austrian Science Fund (FWF): I 3903-N32}.

\end{description}

\end{document}